\documentclass[11pt]{article}

\usepackage{graphicx}
\usepackage{epsfig,cite}
\usepackage{amssymb}
\usepackage{amsmath}
\usepackage{dsfont}
\usepackage{multirow}

\usepackage{url}

\def\smallfrac#1#2{\hbox{${{#1}\over {#2}}$}}
\newcommand{\be}{\begin{equation}}
\newcommand{\ee}{\end{equation}}
\newcommand{\bea}{\begin{eqnarray}}
\newcommand{\eea}{\end{eqnarray}}
\newcommand{\bi}{\begin{itemize}}
\newcommand{\ei}{\end{itemize}}
\newcommand{\ben}{\begin{enumerate}}
\newcommand{\een}{\end{enumerate}}
\newcommand{\la}{\left\langle}
\newcommand{\ra}{\right\rangle}
\newcommand{\lc}{\left[}
\newcommand{\rc}{\right]}
\newcommand{\lp}{\left(}
\newcommand{\rp}{\right)}

\newcommand{\aq}{\alpha_s\left( Q^2 \right)}

\def\frac#1#2{{{#1}\over {#2}}}
\def\gsim{\mathrel{\rlap{\lower4pt\hbox{\hskip1pt$\sim$}}
    \raise1pt\hbox{$>$}}}         %greater than or approx. symbol
\def\lsim{\mathrel{\rlap{\lower4pt\hbox{\hskip1pt$\sim$}}
    \raise1pt\hbox{$<$}}}         %less than or approx. symbol

\newcommand{\dat}{\mathrm{dat}}

\newcommand{\rep}{\mathrm{rep}}
\newcommand{\net}{\mathrm{net}}

\newcommand{\tot}{\mathrm{tot}}

\newcommand{\draft}[1]{}

\begin{document}
\begin{flushright}
Edinburgh 2009/06\\
IFUM-941-FT\\
FREIBURG-PHENO-09/03
\end{flushright}
\begin{center}
{\Large \bf
Precision
 determination of electroweak parameters\\  
and the strange content of
 the proton\\\medskip
from neutrino deep--inelastic scattering}
\vspace{0.8cm}

{\bf  The NNPDF Collaboration:}\\
Richard~D.~Ball$^{1}$,
 Luigi~Del~Debbio$^1$, Stefano~Forte$^2$, Alberto~Guffanti$^3$, 
Jos\'e~I.~Latorre$^4$, Andrea~Piccione$^{2}$, 
Juan~Rojo$^2$ and Maria~Ubiali$^1$.

\vspace{1.cm}
{\it ~$^1$ School of Physics and Astronomy, University of Edinburgh,\\
JCMB, KB, Mayfield Rd, Edinburgh EH9 3JZ, Scotland\\
~$^2$ Dipartimento di Fisica, Universit\`a di Milano and
INFN, Sezione di Milano,\\ Via Celoria 16, I-20133 Milano, Italy\\
~$^3$  Physikalisches Institut, Albert-Ludwigs-Universit\"at Freiburg
\\ Hermann-Herder-Stra\ss e 3, D-79104 Freiburg i. B., Germany  \\
~$^4$ Departament d'Estructura i Constituents de la Mat\`eria, 
Universitat de Barcelona,\\ Diagonal 647, E-08028 Barcelona, Spain\\}
%\end{center}
\bigskip
\bigskip
{\it This paper is dedicated to the memory of Wu-Ki Tung}
\bigskip
\bigskip

%\begin{center}
{\bf \large Abstract:}
\end{center}
We use recent neutrino dimuon production data combined with a global
deep-inelastic parton fit to construct a new parton set, NNPDF1.2,
which includes a determination of the strange
and antistrange distributions of the nucleon. The result is
characterized by a faithful estimation of uncertainties thanks to the
use of the NNPDF methodology, and is free of model or theoretical 
assumptions other than the use of NLO perturbative QCD and exact sum rules. 
Better control of the uncertainties of the strange and antistrange parton 
distributions allows us to reassess the determination of 
electroweak parameters from the NuTeV dimuon data.  
We perform a direct determination of the $|V_{cd}|$ and 
$|V_{cs}|$ CKM matrix elements, obtaining central values in 
agreement with the current global CKM fit: specifically we find
$|V_{cd}|=0.244\pm 0.019$ and $|V_{cs}|=0.96\pm 0.07$. 
Our result for $|V_{cs}|$ is more precise 
than any previous direct determination. We also reassess the 
uncertainty on the NuTeV determination of $\sin^2\theta_W$ through 
the Paschos-Wolfenstein relation: we find that the very large 
uncertainties in the strange valence momentum fraction 
are sufficient to bring the NuTeV result into complete agreement with the
results from precision electroweak data. 

\clearpage

\tableofcontents

\clearpage

%----------------------------
%
% \section{Introduction}
%
%---------------------------------
%------------------------------------------------

\section{The strange content of the nucleon}
\label{sec-intro}

The determination of the strange and antistrange quark distributions
of the nucleon is of considerable phenomenological interest, because
many final states in the standard model and beyond couple directly
to strangeness. A notable example is the determination of the
electroweak mixing angle by the NuTeV
collaboration~\cite{Zeller:2001hh}, which might provide evidence for
physics beyond the standard model, and which is very 
sensitive~\cite{Davidson:2001ji} to the strange content of the nucleon.

Unfortunately, the bulk of the data which are used for
parton determination, namely neutral-current 
deep-inelastic scattering, have minimal
sensitivity to flavour separation, and no sensitivity at all to the
separation of quark and antiquark contributions. As a consequence,
until very recently in standard parton fits such as
CTEQ6.5~\cite{Tung:2006tb} and MRST2006~\cite{Martin:2007bv}, the strange 
and antistrange quark distributions were not determined directly:
rather, they were assumed to be equal, and then 
proportional to the total light antiquark sea distribution. The
only available attempt at a determination 
of the strange and antistrange distributions~\cite{Barone:1999yv} was 
based on a re-analysis of old (mostly bubble-chamber) charged-current
neutrino-nucleon scattering data: unfortunately, the quality of 
these old data was insufficient for a reliable determination. 

This situation has changed
recently, due to the availability of a wider set of inclusive neutrino
deep-inelastic scattering data~\cite{Tzanov:2005kr,Onengut:2005kv}
and, more importantly, of data for deep-inelastic neutrino and
anti-neutrino 
production
of charm~\cite{Goncharov:2001qe,chorus-dimuon,Mason:2007zz}
(``dimuon'' data, henceforth), 
which is directly sensitive to the strange and antistrange
parton distributions. As a consequence, dedicated analyses of the
strange quark distribution have been
performed~\cite{Olness:2003wz,Kretzer:2003wy,Lai:2007dq,Alekhin:2008mb},
and independent parametrizations of the strange and antistrange
distributions are included in most recent parton
fits~\cite{Martin:2009iq}. 
However, the standard method of parton determination used in all
these references, which is based on fitting the parameters of a fixed
functional form, is known to be hard to handle when the 
experiments are relatively unconstraining. Indeed, it is not uncommon that the
addition of new experimental information to a parton fit of this kind, 
actually leads to an increase rather than a decrease of uncertainty bands (see
e.g.~\cite{Watt:2008hi}), because the new data require 
the use of a more general parametrization. This hampers a
direct statistical interpretation of the uncertainty bands on 
parton distributions obtained
in this way: indeed, in some of these parton
determinations~\cite{Martin:2009iq,Nadolsky:2008zw}  
experimental 
uncertainties are inflated 
by suitable ``tolerance'' criteria. 
Precision measurements are thus very difficult to obtain whenever the results 
are significantly affected by parton uncertainties. This is clearly
the case
in the extraction of the electroweak mixing angle from
the NuTeV data of Ref.~\cite{Zeller:2001hh}, and it could be more
generally an issue for LHC observables which depend crucially on the
strange distribution, such as the ``standard candle''
$\sigma_Z/\sigma_W$~\cite{Nadolsky:2008zw}. 

A method of parton determination which is free of these difficulties
was developed by us in a series of
papers~\cite{f2ns,f2p,DelDebbio:2007ee}, and has led recently to the
construction of~ a full parton set based on a fit to a global set of
deep-inelastic scattering data: NNPDF1.0~\cite{Ball:2008by}. This
method is based on the use of neural networks for parton
parametrization, 
and a Monte Carlo method supplemented by a suitable training  and
stopping algorithm for the construction of the parton fit. In this approach, 
parton distributions are given as a Monte Carlo sample 
representing their probability distributions as inferred from
the data:
%whose
%statistical properties represent the available information: 
so, for instance, uncertainties can be obtained from the sample by
computing standard deviations, likelihood intervals by determining
frequency histograms, and so on.

It was 
shown that this methodology is largely free of bias related to parton
parametrization, and it handles in a satisfactory way incomplete
information, contradictory data, and the addition of new data within a
single framework. In particular, in Ref.~\cite{Ball:2008by} it was
explicitly verified that when data are removed by changing the
kinematic cuts, the uncertainty
bands widen in such a way that results 
before and after the cuts remain compatible, while results 
outside the data region directly affected by the cuts remain stable. 
In Ref.~\cite{Dittmar:2009ii} it was further
checked that the same behaviour is observed when the whole dataset is
altered, e.g. by removing all data from one or more experiments: 
a fit to a smaller dataset has wider uncertainties, but
remains compatible with the fit to the larger dataset. 

That these stability properties of the NNPDF approach apply also to the way
the strange distribution is treated was shown in a  dedicated
 study based on the same 
methodology~\cite{Rojo:2008ke}, leading to the NNPDF1.1 parton set. 
In NNPDF1.1, the strange parton distributions
$s^\pm=s\pm \bar s$ are parametrized by two independent neural
networks, 
instead of being
taken to be proportional to the light
antiquark distribution as in NNPDF1.0. However, the dataset is the
same as for NNPDF1.0: so the $s^+$ distribution is only very weakly
constrained, and the $s^-$ essentially unconstrained by the the
data. Nevertheless, when results of this pair of fits are compared,
they show remarkable stability, despite the fact that each neural
network is parametrized by a very redundant set of parameters (the
addition of two neural nets results in the addition of 74 extra free
parameters in the fit). Indeed,
 parton distributions which are
unaffected by the addition of independent strange degrees of
freedom (such as the gluon) are  unchanged, and the only marked
effect of the independent parametrization of strangeness is an
increase, by about a factor two,
of the uncertainty on the  total valence quark distribution ($u-\bar
u+d-\bar d+s-\bar s$ ). Remarkably, statistical analysis of the
NNPDF1.0 set alone was already sufficient to show~\cite{Ball:2008by} 
that the uncertainty on this combination was underestimated.

In this paper, by adding recent dimuon data
to the global deep-inelastic
scattering dataset on which the NNPDF1.0 and NNPDF1.1 fits were
based, we construct a new  parton set, NNPDF1.2, which includes
a determination of the strange and antistrange distributions. Furthermore,
we determine directly the $|V_{cs}|$ and $|V_{cd}|$ CKM matrix elements which
control the strength of the charged--current coupling to neutrinos
in dimuon production 
of the strange and down
quarks respectively, and we use our determination of the strange quark
distribution to compute the correction to the Paschos-Wolfenstein
ratio to be used in extractions of the electroweak mixing angle.

We find that the shape of the strange and antistrange distributions
which are compatible with data are rather more general than those
obtained in other recent
studies~\cite{Olness:2003wz,Kretzer:2003wy,Lai:2007dq,Alekhin:2008mb,
Martin:2009iq,Nadolsky:2008zw}. 
Our uncertainty on the ratio $K_S=\lc S^+\rc/\lc \bar U+\bar D\rc$
of strange to light sea momenta is rather more asymmetric than
hitherto assumed:  $K_S\lp Q^2=20 \,{\rm GeV}^2\rp = 
0.71 {{}^{+0.19}_{-0.31}}^{\rm stat}$. 
This may have nontrivial implications for 
LHC observables, such as the $Z/W$ cross section ratio
mentioned above.
Despite these increased uncertainties, we find that, perhaps 
surprisingly, the dimuon data are sufficient to determine 
$|V_{cs}|=0.96\pm 0.07^{\rm tot}$. 
This is one order of magnitude more precise than any
other direct determination from neutrino deep-inelastic scattering,
and is comparable to the current PDG best average of direct determinations from
$D$ meson decays, 
($|V_{cs}|=1.04\pm0.06$~\cite{Amsler:2008zzb}), though still two
orders of magnitude worse than the results of a global
CKM fit. The related CKM element $|V_{cd}|$ is also determined,
$|V_{cd}|=0.244\pm 0.019^{\rm tot}$, with a similar
accuracy to other determinations from
dimuon data. 

We further find that the $s-\bar s$ distribution, which must
change sign as a function of $x$ in order for the total nucleon
strangeness to vanish, can do so  in a wide variety of ways, and that
its sign at any given $x$ is not well determined. 
As a consequence,
the uncertainty in the strange valence momentum fraction, and thus in
the correction to the Paschos-Wolfenstein ratio, is much larger than
hitherto assumed, and is such that the NuTeV
measurement of $\sin^2\theta_W$ is actually in complete agreement 
with determinations from precision electroweak data once this 
uncertainty is taken into account. 

Many of the techniques and tools that we use in this paper are
part of the standard NNPDF methodology, already described in detail in
Refs.~\cite{f2p,DelDebbio:2007ee,Ball:2008by} and used there for the
construction of the NNPDF1.0 parton set.  Here we will focus on the
new aspects of the NNPDF1.2 set, and then discuss our main
results. Thus in Sect.~\ref{sec:expdata} we describe the dimuon
cross section and its available experimental determinations, and in
Sect.~\ref{sec:evolution} we will give its expression in perturbative
QCD and thus its relation to the strange distribution, and discuss the
way the strange and antistrange distributions are treated, as well as some
specific theoretical issues related to the treatment of this
observable, such as the treatment of the charm mass and of nuclear
corrections.  Full details of the hard kernels used to construct the
physical observables are given in Appendix~\ref{sec:kernels}.  In
Section~\ref{sec:results} we present our determination of the
strange and antistrange distributions, specifically their shape and
their contribution to the nucleon momentum, and compare them to results
obtained by other groups. In Section~\ref{sec:ewparam} we will
discuss in detail the implications of our results for precision
electroweak measurements, and discuss specifically the determination
of the CKM matrix elements $|V_{cs}|$ and $|V_{cd}|$ 
and the impact of our results on
the NuTeV determination of the electroweak mixing angle.

%------------------------------
%
%\section{Experimental NuTeV data}
%
%------------------------------
%----------------------------------------------------------

\section{Experimental data}
\label{sec:expdata}

The NNPDF1.2 parton determination is based on the same data set used for
NNPDF1.0, supplemented by data on deep-inelastic neutrino production
of charm from NuTeV~\cite{Goncharov:2001qe,masonPhD} 
which give us a handle on the strange distribution, whose
determination is the main goal of this paper. We also add to the data set
some recently published measurements of neutral current and charged current
deep-inelastic cross sections by the ZEUS experiment based on HERA-II
data~\cite{Chekanov:2009gm,Chekanov:2008aa}. 

An earlier 
measurement of the dimuon cross section using the same detector (but a
different beam-line) was performed by the CCFR
collaboration~\cite{Bazarko:1994tt}. This previous measurement is
significantly  less accurate and its compatibility with the NuTeV data is
debatable~\cite{Alekhin:2008mb,Goncharov:2001qe}; we will not include
it in our fit.
A recent measurement of the dimuon cross section has 
also been performed by the CHORUS collaboration~\cite{chorus-dimuon};
unfortunately, however, only the results of a leading--order QCD
analysis of this data have been published, and not the cross-section
data themselves, which therefore cannot be used in our analysis. 

%------------------------------------------------------------
\begin{figure}[t!]
\begin{center}
\epsfig{width=0.95\textwidth,figure=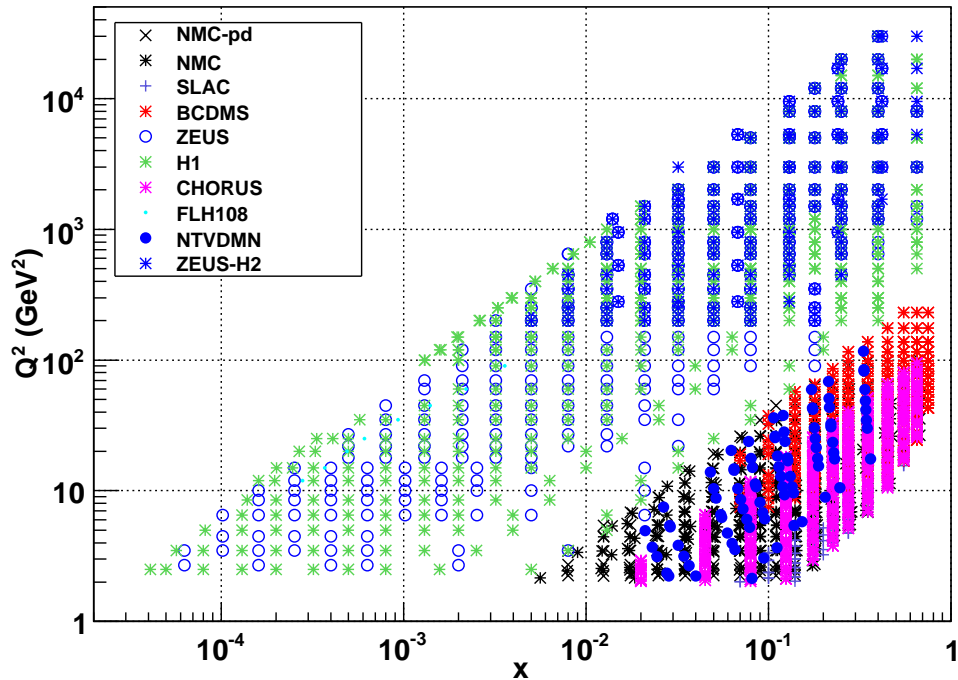} 
\caption{\small Experimental data in the $(x,Q^2)$ plane used
    in the  NNPDF1.2 analysis after kinematic cuts.}
\label{fig:dataplot}
\end{center}
\end{figure}
%------------------------------------------------------------------

The treatment of experimental data in the present fit follows
Ref.~\cite{Ball:2008by}.
In particular, all information on correlated systematics is included
in our fit, in that the full covariance matrix is computed including
all available correlated uncertainties (including normalization
uncertainties). 

Below we give more explicit details of the new data and corresponding
observables which have been included in the current fit.

\subsection{Data set, uncertainties and correlations}
 \label{sec:dataset}

The data set used for the present fit is obtained by supplementing the  
data set used for the NNPDF1.0 fit, as  summarized in Table~1 of
Ref.~\cite{Ball:2008by}, with the data summarized in
Table~\ref{tab:expsets}  given
here.  A scatter plot of the full data set is displayed in
Figure~\ref{fig:dataplot}.  Note that NuTeV dimuon 
data overlap with the rest of fixed target experiments, providing
information on the proton strangeness for $x \gsim 10^{-2}$.

 %------------------------------------------------------------
\begin{table}
\begin{center}

 \tiny
 \begin{tabular}{|ll|c|c|c|c|c|c|c|c|}
  \hline
  Experiment & Set & $N_{\rm dat}$ & $x_{\rm min}$ &  $x_{\rm max}$ 
 &  $Q^2_{\rm min}$ &  $Q^2_{\rm max}$ & $\sigma_{\rm tot}$ (\%) & $F$ & Ref.\\ \hline
 \hline
 \multicolumn{2}{|l|}{ZEUS-HERA-II} & & & & & & & &\\
 &  ZEUS06NC   &   90 (90) & .005 & 0.65 &    200.0 &    30000.0 &   2.6 & $\tilde{\sigma}^{NC,e^-}$ &\cite{Chekanov:2009gm}\\
 &  ZEUS06CC &   37 (37) & .015 & 0.65 &   280.0 &   30000.0 &   14 & $\tilde{\sigma}^{CC,e^-}$ &\cite{Chekanov:2008aa}\\
\hline
 \multicolumn{2}{|l|}{NUTEV Dimuon} & & & & & & & &\\
 &  NuTeV Dimuon $\nu$ &  45 (43) & .0267 & 0.37 &    1.1 &    116.5 &   19 & $\tilde{\sigma}^{\nu,c}$ &\cite{masonPhD}\\
 &  NuTeV Dimuon $\bar{\nu}$ &  45 (41) & .021 & 0.25 &    0.8 &
 68.3 &  23 & $\tilde{\sigma}^{\bar{\nu},c}$ &\cite{masonPhD}\\
\hline
\multicolumn{2}{|l|}{Total (including Tab.~1 of Ref.~\cite{Ball:2008by})}     &  4165 (3372) & \multicolumn{7}{|c}{}\\
\cline{1-3}
 \end{tabular}

\end{center}
\caption{\small Further experimental data included in the
  present analysis in addition to those given in Table~1 of
  Ref.~\cite{Ball:2008by}. We show the number of points before (after)
applying kinematic cuts, the kinematic range, the average total
uncertainty after cuts and the observable which is measured. Different sets
within an experiment are correlated with each other, while data from
different experiments are uncorrelated. The total number of data
points refers to the full dataset. }
\label{tab:expsets}
\end{table}
%------------------------------------------------------------------

The covariance matrix is computed for all the data included in the fit, as
discussed in Ref.~\cite{Ball:2008by}. The NuTeV dimuon data are
affected by a common normalization uncertainty of
2.1\%~\cite{Tzanov:2005kr}; 
eight
correlated systematics; and a statistical uncertainty.
The statistical uncertainty is around 15\% for neutrino and around 25\% for
anti-neutrino data, while correlated systematics are generally smaller
by a factor between three and five. This dominant statistical
uncertainty 
is
affected by a bin by bin correlation due to the  unfolding procedure
used in  extracting the dimuon cross section from the measured
observable. The covariance matrix which describes these
correlations is not available. Its effect has been summarized in
Ref.~\cite{masonPhD}  by
providing for each bin an ``effective number of degrees of freedom'', which
provides the expected value of the best-fit $\chi^2$ to the given data
bin, i.e., effectively, a rescaling for the statistical
error. These rescaling factors can be as low as 30\%, and are
typically around 50\%, indicating sizable correlations.

Rescaling of statistical errors in order to account for missing
correlations could bias the fit in an unpredictable way and it is a
dangerous procedure if the information on the covariance matrix is
lost. On the other hand, only including correlations for the
subdominant systematic errors could lead to an underestimate of the
relative impact of these uncertainties. Hence, because the
covariance matrix of the NuTeV data is unfortunately unavailable, 
 the only consistent procedure for the treatment of these data is to
 add all uncertainties in quadrature, and only consider
 normalizations as correlated uncertainties. This is the procedure
 that we shall follow.

\begin{table}[t!]
\begin{center}

%\begin{flushleft}
{\tiny
\begin{tabular}{|c|c|c|c|}
 \hline
Experiment & ZEUS-HERA-II & NuTeV Dimuon  & \multicolumn{1}{|c|}{Total}\\
\hline
 \hline
$\la PE\left [\la F^{\rm (art)} \ra_{\rm rep}\right]\ra_{\rm dat}$ &
-4.2 $\cdot 10^{-4}$  &
-5.4 $\cdot 10^{-4}$ & \multicolumn{1}{|c|}{-2.3 $\cdot 10^{-4}$}\\
$ r\left[ F^{\rm (art)} \right] $ & 0.999  & 0.999 & \multicolumn{1}{|c|}{0.999}\\
\hline
$\la PE\left [\la \sigma^{\rm (art)} \ra_{\rm rep}\right]\ra_{\rm dat}$ & 
6.5 $\cdot 10^{-3}$ & 
-2.6 $\cdot 10^{-3}$ &  \multicolumn{1}{|c|}{-6.1 $\cdot 10^{-4}$}\\
$ \la \sigma^{\rm (exp)} \ra_{\rm dat} $ & 13.79\%  & 21.23\% & \multicolumn{1}{|c|}{11.24\%}\\
$ \la \sigma^{\rm (art)} \ra_{\rm dat} $ & 13.88\%  & 21.17\% &\multicolumn{1}{|c|}{11.24\%}\\
$ r\left[ \sigma^{\rm (art)} \right] $ & 0.999 & 0.998 & \multicolumn{1}{|c|}{0.999}\\
\hline
$ \la \rho^{\rm (exp)} \ra_{\rm dat} $ & 0.287 & 0.034 & \multicolumn{1}{|c|}{0.146}\\
$ \la \rho^{\rm (art)} \ra_{\rm dat} $ & 0.294 & 0.034 & \multicolumn{1}{|c|}{0.146}\\
$ r\left[ \rho^{\rm (art)} \right] $ &  0.994 & 0.978 & \multicolumn{1}{|c|}{0.996}\\
\hline
$ \la {\rm cov}^{\rm (exp)} \ra_{\rm dat} $ & 
6.89 $\cdot 10^{-4}$ & 0.169 & \multicolumn{1}{|c|}{1.61 $\cdot 10^{-3}$}\\
$ \la {\rm cov}^{\rm (art)} \ra_{\rm dat} $ &
7.03 $\cdot 10^{-4}$ &  0.168 & \multicolumn{1}{|c|}{1.54 $\cdot 10^{-3}$}\\
$ r\left[ {\rm cov}^{\rm (art)} \right] $ & 0.997 & 0.988 & \multicolumn{1}{|c|}{0.988}\\
\hline
\end{tabular}
}
 
%\end{flushleft}
\end{center}
\caption{\small Statistical estimators for the
Monte Carlo artificial data
generation with $N_{\rep}=1000$, for
experiments not included in Ref.~\cite{Ball:2008by}. 
The definition of the statistical estimators 
is given in Appendix B of~\cite{DelDebbio:2007ee}. The faithfulness of
the Monte Carlo sampling of experimental data is assessed
quantitatively by these estimators.
}
\label{tab:mcest}
\end{table}

\subsection{Observables, kinematic cuts and pseudo-data sample}
\label{sec:physobs}

The set of observables considered in these fits consists of the
structure functions and reduced cross-sections considered in
Ref.~\cite{Ball:2008by} and summarized in Table~1 of that reference,
supplemented by the dimuon cross section. 
Neutrino dimuon production is induced by charm production through
charged current
interactions of neutrinos with the target nuclei, followed
 by the fragmentation
of the charm quark into a charmed hadron and its decay into
a muon. 
The corresponding cross section is given by
\begin{eqnarray}
\label{eq:nuxsecdimuon}
&&\tilde{\sigma}^{\nu (\bar{\nu}),c}(x,y,Q^2)\equiv 
\frac{1}{E_{\nu}}\frac{d^2\sigma^{\nu(\bar{\nu}),c}}{dx\,dy}
(x,y,Q^2)\nonumber\\
&&\qquad =\frac{G_F^2M_N}{2\pi(1+Q^2/M_W^2)^2}
\Bigg[
\left( \lp Y_+ - \frac{2M^2_Nx^2y^2}{Q^2} -y^2\rp \lp
1+ \frac{m_c^2}{Q^2}\rp +y^2\right)
F_2^{\nu(\bar{\nu}),c}(x,Q^2) \nonumber\\ 
&&\qquad\qquad\qquad\qquad\qquad\qquad\qquad
 -y^2F_L^{\nu(\bar{\nu}),c}(x,Q^2)\pm 
\,Y_-\,xF_3^{\nu(\bar{\nu}),c}(x,Q^2)
\Bigg],
\end{eqnarray}
where 
\be
Q^2=2M_NE_{\nu}xy,\qquad Y_\pm = 1\pm(1-y)^2.
\label{eq:nukin}
\ee

The charm production cross section is obtained from the
published NuTeV
neutrino dimuon production cross sections~\cite{masonPhD} as
\be
\label{eq:dimuon}
\frac{1}{E_{\nu}}\frac{d^2\sigma^{\nu(\bar{\nu}),c}}{dx\,dy}
(x,y,Q^2)  =\frac{1}{\la {\rm Br}\lp D\to \mu \rp\ra 
\cdot \mathcal{A}\lp x,y,E_{\nu}\rp} \frac{1}{E_{\nu}}\frac{d^2\sigma^{\nu(\bar{\nu}),2\mu}}{dx\,dy}
(x,y,Q^2) ,
\ee
where $\la {\rm Br}\lp D\to \mu \rp\ra $ is the average branching ratio of
charmed hadrons into muons and $\mathcal{A}\lp x,y,E_{\nu}\rp$ is a
bin-dependent experimental acceptance correction. Acceptances are 
provided by the NuTeV collaboration, based on a leading-order 
model~\cite{Masonpriv}; 
next-to-leading order 
acceptances~\cite{olnesspriv} (not publicly available) differ by
less than 3\% from the leading-order ones.
The branching ratio
used in the NuTeV analysis~\cite{Mason:2007zz}
comes from 
a reanalysis of the emulsion data of the
FNAL E531 experiment and turns out to be $\la {\rm Br}\lp D\to \mu
\rp\ra=0.099\pm0.012$, in agreement with other
determinations~\cite{Bolton:1997pq,chorus-dimuon}. A simultaneous
extraction of this parameter along with the determination of
strangeness in Ref.~\cite{Alekhin:2008mb} leads to a similar
result. In the determination of the dimuon cross section, the 
branching ratio will be set equal to the central value used
in the NuTeV analysis \cite{Mason:2007zz}. The associated 
uncertainty will then be included in our fit
as discussed in Section~\ref{sec:diphysobs} below.

Our data set is obtained by imposing on all the data listed in
Table~1 of Ref.~\cite{Ball:2008by}
and in Table~\ref{tab:expsets} the same kinematical cuts as in
NNPDF1.0, namely  $Q^2 > Q^2_{\rm cut}=2$ GeV$^2$ and $W^2>12.5$~GeV$^2$.
After these cuts, 84 out of the 90 
NuTeV dimuon data points are left. After cuts, the total number of
data points in the NNPDF1.2 analysis is $N_{\rm dat}=3372$.

Error propagation from the experimental data to the fit is performed
through a Monte Carlo procedure, described in detail in
Ref.~\cite{Ball:2008by}, by generating a set of 1000 pseudo-data
replicas, whose faithfulness can be verified by studying suitable
statistical estimators. The statistical estimators for the new data
sets included in the present fit, as well as for the global data set,
are summarized in Table~\ref{tab:mcest}.

%--------------------------------------------------------

% ------------------------------------------------------
%
%\section{Theoretical treatment}
%
% ----------------------------------------------------
%----------------------------------------------------------------------

\section{Neural networks,  parton distributions and physical observables}
\label{sec:evolution}

Physical observables are determined from a set of PDFs given at a
reference scale, which are in turn parametrized in terms of neural
networks, according to  the formalism discussed in detail in Sect.~3-4
of Ref.~\cite{Ball:2008by}. Here we summarize the new features of this
determination: the use of an independent parametrization for the
strange and antistrange distribution and its construction in terms of
neural networks,
the new physical observables used
for dimuon data, and some issues that require
reconsidering when dealing with this observable, namely the
treatment of the charm mass and nuclear corrections.

\subsection{Parametrization of the strange PDF}
\label{sec:strangepar}

In the NNPDF1.0 fit of  Ref.~\cite{Ball:2008by}, parton distributions
were parametrized using five independent neural networks: four
independent linear combinations of  the two light
flavours and anti-flavours, and the gluon. The strange and
antistrange quark distributions were assumed to be given by
$s=\bar{s}=\kappa \lp \bar{u}+  \bar{d}\rp/2$ with
$\kappa=0.5$, and heavy quarks were generated dynamically, using a
zero-mass variable flavour number scheme (ZM-VFN). 
In the subsequent NNPDF1.1 fit~\cite{Rojo:2008ke}, two further neural
networks were introduced to parametrize
the strange and antistrange quark distributions. Here, as in Ref.~\cite{Rojo:2008ke} we parametrize parton
distributions in terms of seven independent neural networks, as we now
discuss.

The primary partonic quantities out of which all physical observables
are built up are the gluon, the singlet quark distribution, the
total valence quark distribution, and ten nonsinglet
combinations of the valence $(q_i-\bar q_i$) or total ($q_i+\bar q_i$)
quark and antiquark distribution for the $i$-th quark
flavor. These are constructed as in Ref.~\cite{Ball:2008by}, to which
we refer for more details. The
starting scale is chosen at the charm threshold, where the charm
distributions are assumed to vanish, and the remaining six light quark
distributions and the gluon distribution are parametrized 
in terms of independent neural
networks. The possibility of introducing an intrinsic charm distribution
will not be studied in the present fit, though there is no obstacle to
including it in future studies. 

The four light non-strange distributions and the gluon distribution
are parametrized in terms of neural networks as in
Ref.~\cite{Ball:2008by}, by letting
\be
f(x,Q_0^2) = A_f \lp 1-x\rp^{m_f}x^{-n_f}
 {\mathrm NN}_{f}(x),
\label{gennn}
\ee
where $f(x,Q_0^2)$ is a linear combination of parton distributions, and
NN$_f(x)$ is a multi-layer feed-forward neural network with two
intermediate layers and architecture 2-5-3-1, parametrized by 37
free parameters (weights and thresholds). The constants $A_f$ are
either simply set to one, or else used to enforce the valence and 
momentum sum rules. 

The preprocessing function $\lp
1-x\rp^{m_f}x^{-n_f}$ is included in order to speed up the
convergence of the fit: the neural network only has to fit 
the deviation from the behaviour of the
preprocessing function, whose exponents are thus  fixed to values which
absorb some of the gross behaviour of the function $f(x,Q_0^2)$
without biasing the result (i.e. without imposing a steep growth or
fall which ${\mathrm NN}(x)$ would have trouble in reabsorbing). 
Independence of
the results on the choice of the preprocessing exponents was verified in
Ref.~\cite{Ball:2008by} by varying them within a reasonable stability
range. This stability range is identified in Ref.~\cite{Ball:2008by} by
requiring the quality of the fit to be unchanged as the exponents are varied.
A small residual dependence on the preprocessing exponents  was found
in
Ref.~\cite{Ball:2008by}  for  the triplet and total
valence quark distributions. In order to be able to disentangle 
accurately the strange contribution it is important that uncertainties
on all light quark flavours are estimated as precisely as possible:
for this purpose, in the NNPDF1.1 fit of Ref.~\cite{Rojo:2008ke} and
in the present fit all preprocessing exponents are randomized: a
different value is taken for each Monte Carlo replica, uniformly
distributed within the stability range. 

\begin{table}
\small
  \begin{center}
    \begin{tabular}{|c|c|c|}
      \hline PDF & $m$ & $n$ \\
      \hline
      $\Sigma(x,Q_0^2)$  & $\lc 2.7,3.3\rc$ & 
      $\lc 1.1,1.3\rc$ \\
      \hline
      $g(x,Q_0^2)$  & $\lc 3.7,4.3\rc$ & 
      $\lc 1.1,1.3\rc$ \\
      \hline
      $T_3(x,Q_0^2)$  & $\lc 2.7,3.3\rc$ & 
      $\lc 0.1,0.4\rc$ \\
      \hline
      $V(x,Q_0^2)$  & $\lc 2.7,3.3\rc$ & 
      $\lc 0.1,0.4\rc$ \\
      \hline
      $\Delta_S(x,Q_0^2)$  & $\lc 2.7,3.3\rc$ & 
      $\lc 0,0.01\rc$ \\
     \hline
       $s_+(x,Q_0^2)$  & $\lc 2.7,3.3\rc$ & 
      $\lc 1.1,1.3\rc$ \\
      \hline
      $s_-(x,Q_0^2)$  & $\lc 2.7,3.3\rc$ & 
      $\lc 0.1,0.4\rc$ \\
      \hline
    \end{tabular}
    \caption{\small \label{tab:prepexps} The range
of variation of the randomized preprocessing exponents
      used in  the present NNPDF1.2 fit. }
  \end{center}
\end{table}

The choice of linear combinations of the two lightest flavours
which are parametrized independently
according to
Eq.~(\ref{gennn}) is the same in the present fit as in NNPDF1.0. On top
of them, we add two independent neural networks in the strange sector,
in order to parametrize
\be
\label{spmdef}
s^\pm(x,Q^2)\equiv s(x,Q^2)\pm \bar s(x,Q^2)
\ee
according to
\bea
s^+(x,Q_0^2) &=& \lp 1-x\rp^{m_{s^+}}x^{-n_{s^+}}
 {\rm NN}_{s^+}(x)  \ , \label{eq:strangePDFsop} \\
s^-(x,Q_0^2) &=& \lp 1-x\rp^{m_{s^-}}x^{-n_{s^-}}
 {\rm NN}_{s^-}(x) -  s_{\rm aux}(x,Q_0^2), 
\label{eq:strangePDFsom}
\eea
where
\be
s_{\rm aux}(x,Q_0^2)=A_{s^-}\lc x^{r_{s^-}}\lp 1-x\rp^{t_{s^-}}\rc .
\ee
The exponents $m$, $n$ of the preprocessing functions are randomized
as discussed above, and their ranges are also listed in
Table~\ref{tab:prepexps}. 

The contribution $s_{\rm aux}(x,Q_0^2)$ in Eq.~(\ref{eq:strangePDFsom}) 
is introduced in order 
to enforce the strange valence sum rule: the constant 
$A_{s^-}$ is fixed by requiring
\be
\label{eq:sr}
\int_0^1\! dx \, s^-(x) = 0,
\ee
which gives the condition
\be
\label{eq:asbar2}
A_{s^-} = \frac{\Gamma\lp r_{s^-}+ t_{s^-}+2\rp}{
\Gamma\lp r_{s^-}+1\rp\Gamma\lp t_{s^-}+1\rp}
\int_{0}^1   \lp 1-x\rp^{m_{s^-}}x^{-n_{s^-}}
 {\rm NN}_{s^-}(x)dx.
\ee
Clearly, the sum rules requires $s^-$ to change sign at least once. 
This way of implementing the sum rule is designed in order to ensure 
that this crossing happens naturally in the valence region, rather than 
in some contrived way outside the data region where the shape 
of $s^-$ is completely unconstrained. To this purpose, 
the exponents $r_{s^-},t_{s^-}$ are
chosen in such a way that $s_{\rm aux}(x,Q_0^2)$ peaks in the valence
region, and that the small $x$ and large $x$ behaviour of
$s^-(x,Q_0^2)$ are not controlled  by the 
$s_{\rm aux}(x,Q_0^2)$  contribution. In practice the latter condition 
is enforced by requiring
$r_{s^-} \ge  -n_{s^-}$ and $t_{s^-}\ge m_{s^-}$, while the former is
enforced by letting $r_{s^-}=t_{s^-}/k$, which sets the maximum of 
$s_{\rm aux}(x,Q_0^2)$ at $x=\frac{1}{k+1}$. We then choose $t_{s^-}=3.5$, and
take $k$ as a uniformly distributed random number in the range
$k\in\lc 1,3\rc$. The consequences of this very flexible
implementation of the strangeness valence sum rule will be discussed
in Sect.~\ref{sec:strange} below.

\subsection{The dimuon physical observable}
\label{sec:diphysobs}

The NNPDF1.2 data set, displayed in Fig.~\ref{fig:dataplot}, contains 
data for the same set of observables discussed in
Ref.~\cite{Ball:2008by}, with the addition of the dimuon  cross section
Eq.~(\ref{eq:nuxsecdimuon}). The latter is determined by the charm
structure functions $F_2^{\nu(\bar{\nu}),c}$,$F_L^{\nu(\bar{\nu}),c}$ and
$xF_3^{\nu(\bar{\nu}),c}$, which in the quark model are given by
\bea
\label{fnexpr}
  &&F_2^{\nu,p,c}(x,Q^2)=x F_3^{\nu,p,c}(x,Q^2)= 2x\,
  \big(|V_{cd}|^2\,d(x)\,+|V_{cs}|^2\,s(x)+|V_{cb}|^2\,b(x)\big), \\
\label{fnbexpr}
  &&F_2^{\bar\nu,p,c}(x,Q^2)=-xF_3^{\bar\nu,p,c}(x,Q^2)=
2x\,
\big(|V_{cd}|^2\,\bar d(x)\,+|V_{cs}|^2\,\bar s(x)
+|V_{cb}|^2\,\bar b(x)\big) \ ,
\eea
with $F_L^{\nu(\bar{\nu}),c}=0$.
Full expressions for these structure functions in perturbative QCD 
at any scale in terms of the basis of
PDFs used in our fits are given in
Appendix~\ref{sec:kernels}. 

Because they are not inclusive with respect to the final state quark
flavour, these structure functions depend on CKM matrix
elements. These are extremely well
determined by current global fits including 
unitarity constraints; for our global fits we will use the current
best-fit 
PDG~\cite{Amsler:2008zzb} values: uncertainties on them are tiny and
will be neglected. In Section~\ref{sec:results} we will
then study the quality of our fit as the parameters $|V_{cs}|$ and
$|V_{cs}|$ are varied without the unitarity constraint, and use 
this to provide a direct determination of these
parameters from the dimuon data. 

Also, as already discussed in Sect.~\ref{sec:expdata}, the dimuon 
cross-section Eq.~(\ref{eq:dimuon}) depends on the branching
ratio $\la {\rm Br}\lp D\to \mu \rp\ra$. The uncertainty in this is
actually rather significant: in previous
analyses~\cite{Mason:2007zz,Alekhin:2008mb} of dimuon data 
this turned out to be one of the dominant sources of uncertainty. To
take account of this uncertainty, the value of the branching ratio
used in the fit has been randomized about its central value, 
analogously to the procedure used for the
preprocessing exponents, with a Gaussian distribution of width equal to
the stated uncertainty $\la {\rm Br}\lp D\to \mu
\rp\ra=0.099\pm0.012$~\cite{Mason:2007zz}.

\subsection{Treatment of the charm mass}
\label{sec:hq}

%------------------------------------------------------------
\begin{figure}[t!]
\begin{center}
\epsfig{width=1.00\textwidth,figure=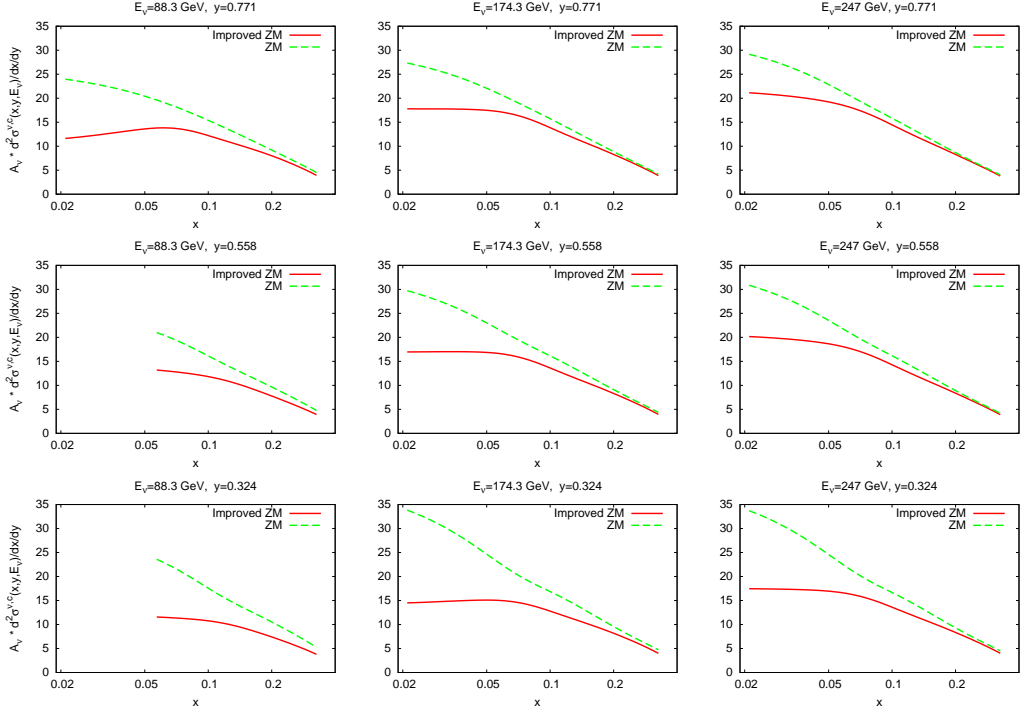} 
\end{center}
\caption{\small Comparison of the ZM and I-ZM computations of the
  dimuon cross section $d^2\sigma^{\nu(\bar{\nu}),c}/dx\,dy$
for typical NuTeV kinematics. All cross sections in the plots are
rescaled by a factor $A_{\nu}=100/G_F^2M_NE_{\nu}^2$. 
The neutrino kinematic parameters $(E_{\nu},y)$
are related to $x$ and $Q^2$ by Eq.~(\ref{eq:nukin}).
Results for anti-neutrinos are very similar.}
\label{fig:izmcomp}
\end{figure}
%------------------------------------------------------------------

In the previous NNPDF1.0 and NNPDF1.1 parton
determinations,  heavy quarks were treated in a zero mass variable
flavour number (ZM-VFN) scheme, as discussed in Sect.~3.4 of
Ref.~\cite{Ball:2008by}. Contributions suppressed by powers of the
heavy quark mass, i.e. of order 
$m_h^2/Q^2$, which are neglected in this scheme, have a small
but not entirely negligible impact~\cite{Pumplin:2007wg}, mostly through
the initial condition on the charm distribution, which
then affects all other PDFs due to the momentum sum rule. 
For the purpose of the present analysis, an improved treatment of the
charm mass is advisable at least for dimuon production, as the dimuon
cross section measures charm production, and a sizable 
fraction of the NuTeV dimuon data are thus at scales close to the charm mass.

To this purpose, we employ (for the  dimuon observable only)
the improved ZM-VFN (I-ZM-VFN) scheme, proposed in 
Ref.~\cite{Thorne:2008xf} and discussed in detail in
Ref.~\cite{Nadolsky:2009ge}. There, it was shown that the bulk of the
charm mass effects near threshold can be accounted for by requiring
that the threshold for the
inclusion of heavy quarks in the sum over final states be set at its
physical value $W^2=m_c^2$, and
that the phase-space constraint due to the heavy quark mass be
respected in convolution integrals. The latter requirement is
in practice implemented by replacing the Bjorken $x$ variable by
a rescaling variable
$\chi_c$ defined as
\be
\label{slowrescdef}
\chi_c \equiv x\lp 1+ \frac{m_c^2}{Q^2}\rp.
\ee
Whereas results obtained with this I-ZM-VFN scheme are in fair
agreement with those obtained with a full treatment of the quark mass
(so-called general mass, or GM scheme),
and in rather better agreement with the data, they may lead to an
excessive suppression of heavy quark production: to this purpose, in
Ref.~\cite{Nadolsky:2009ge} a one-parameter family of rescaling
variables has been constructed, such that the agreement with the GM
scheme 
can be optimized by tuning this parameter. It turns out, however, that
the simplest choice Eq.~(\ref{slowrescdef}) is actually very close to
the optimal one for charged current deep inelastic scattering.

Hence, in the present analysis we will use the ZM-VFN for all
inclusive observables, but for the dimuon cross section
Eq.~(\ref{eq:dimuon}) we will use the I-ZM-VFN of
Ref.~\cite{Thorne:2008xf}. In practice, this means that we will retain
the full $m_c$ dependence in Eq.~(\ref{eq:nuxsecdimuon}), and in the
expressions for the structure functions  $F_i^{\nu,c}$ Eq.~(\ref{fnu_charm_qcd})
all convolutions are defined as
\be
\lc C \otimes q \rc\lp x,Q^2\rp 
= \theta\lp W^2 - m_c^2\rp \int_{\chi_c}^1
\frac{dy}{y}C\lp  y,\aq \rp q\lp \frac{\chi_c}{y},Q^2\rp.
\label{eq:convchi}
\ee

The impact of this treatment of the charm mass is shown in
Fig.~\ref{fig:izmcomp}, where we compare a NLO determination of the 
dimuon cross section Eq.~(\ref{eq:dimuon}) within the ZM-VFN and
I-ZM-VFN schemes, based on our previous NNPDF1.0 parton set. The
suppression of the cross section at small  $x$ 
due to finite quark mass is apparent
from this plot.
Clearly, the inclusion of quark mass effects only in the 
determination of the dimuon cross section, and
then in the I-ZM-VFN scheme, is an approximation. This approximation
will lead to a systematic uncertainty in our determination of the
strange PDFs and of CKM matrix elements in the next sections. We will
estimate this uncertainty by comparing results obtained
in the ZM-VFN and I-ZM-VFN scheme: as the full GM scheme is actually
in between these two, this provides a rather conservative overestimate of
the associated uncertainty. We will then see that this systematic
uncertainty is actually small in comparison to the
statistical uncertainty on strangeness and associated observables.

\subsection{Nuclear Corrections}
\label{sec:nucorr}

%------------------------------------------------------------
\begin{figure}[t!]
\begin{center}
\epsfig{width=0.49\textwidth,figure=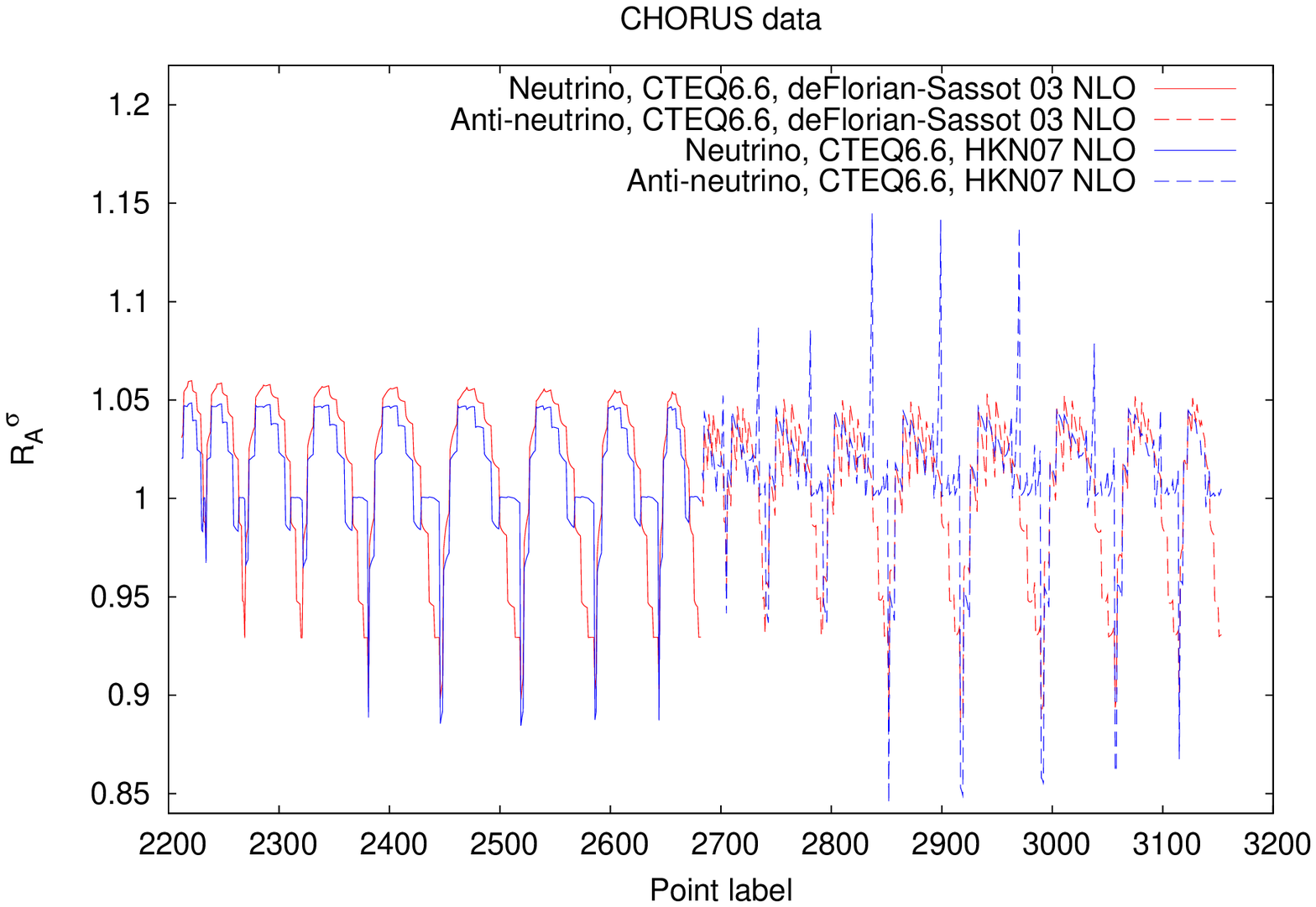} 
\epsfig{width=0.49\textwidth,figure=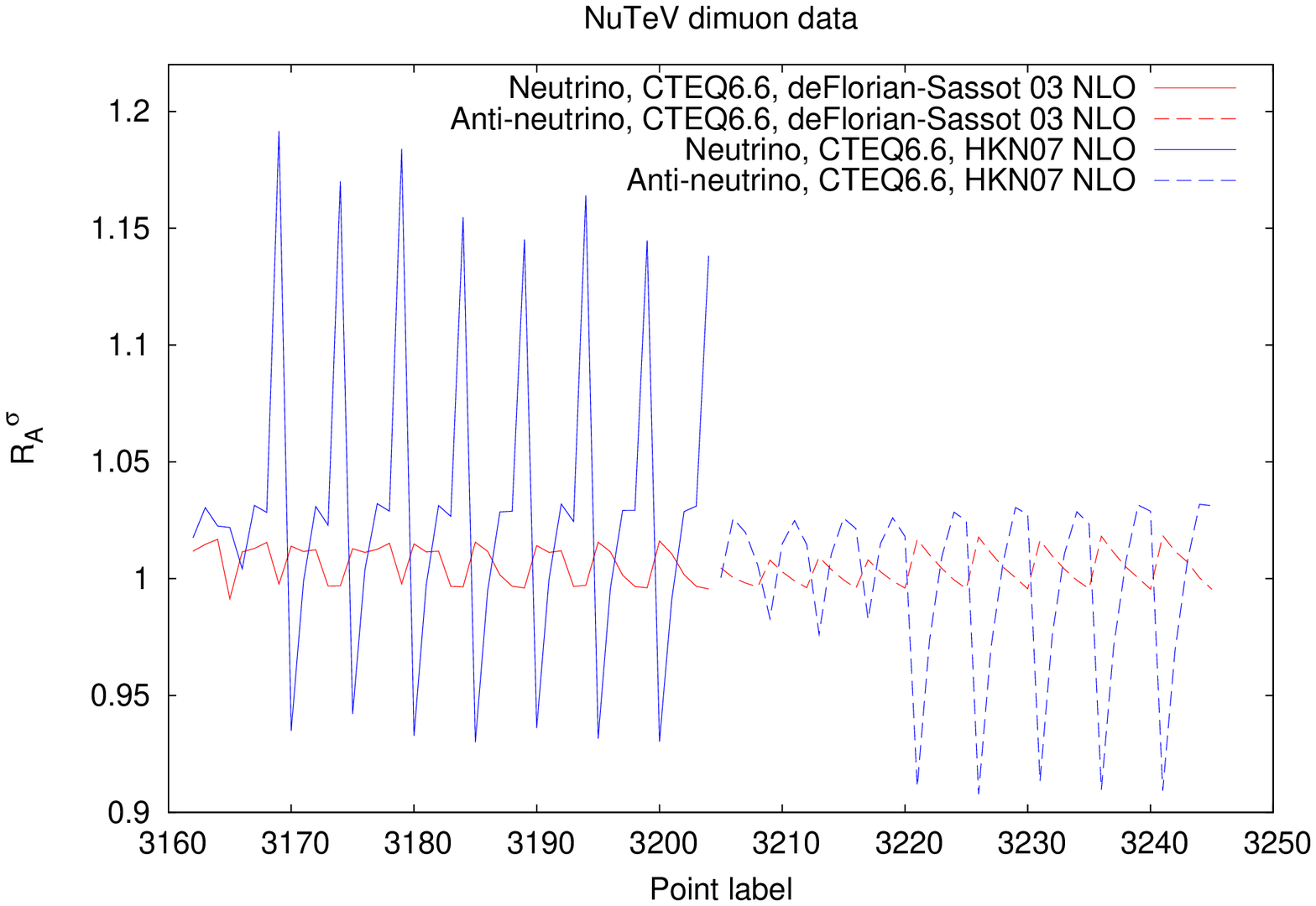} 
\end{center}
\caption{\small Comparison of the nuclear corrections to the
reduced cross sections
for inclusive CHORUS data (left) and for 
NuTeV charm production data (right)
from the  
de~Florian-Sassot~\cite{deFlorian:2003qf} and
HKN07~\cite{Hirai:2007sx}parametrization. The
correction is shown for individual experimental data points, versus
the point label (arbitrary
order).}
\label{fig:nucorr2}
\end{figure}
%------------------------------------------------------------------

%-----------------------------------------------------------------------
Neutrino data are obtained from deep-inelastic scattering off a
nuclear target: for NuTeV essentially Fe, $A_{\rm
  NuTeV}=49.6$~\cite{masonPhD}, and for CHORUS (whose inclusive
structure function measurements are also included in our data set) Pb,
$A_{\rm chorus}=207$,~\cite{Onengut:2005kv}.
Therefore, a suitable nuclear
correction should be introduced in order to obtain from these data a
determination of the PDFs of free nucleons. 

\begin{figure}[t!]
\begin{center}
\epsfig{width=0.49\textwidth,figure=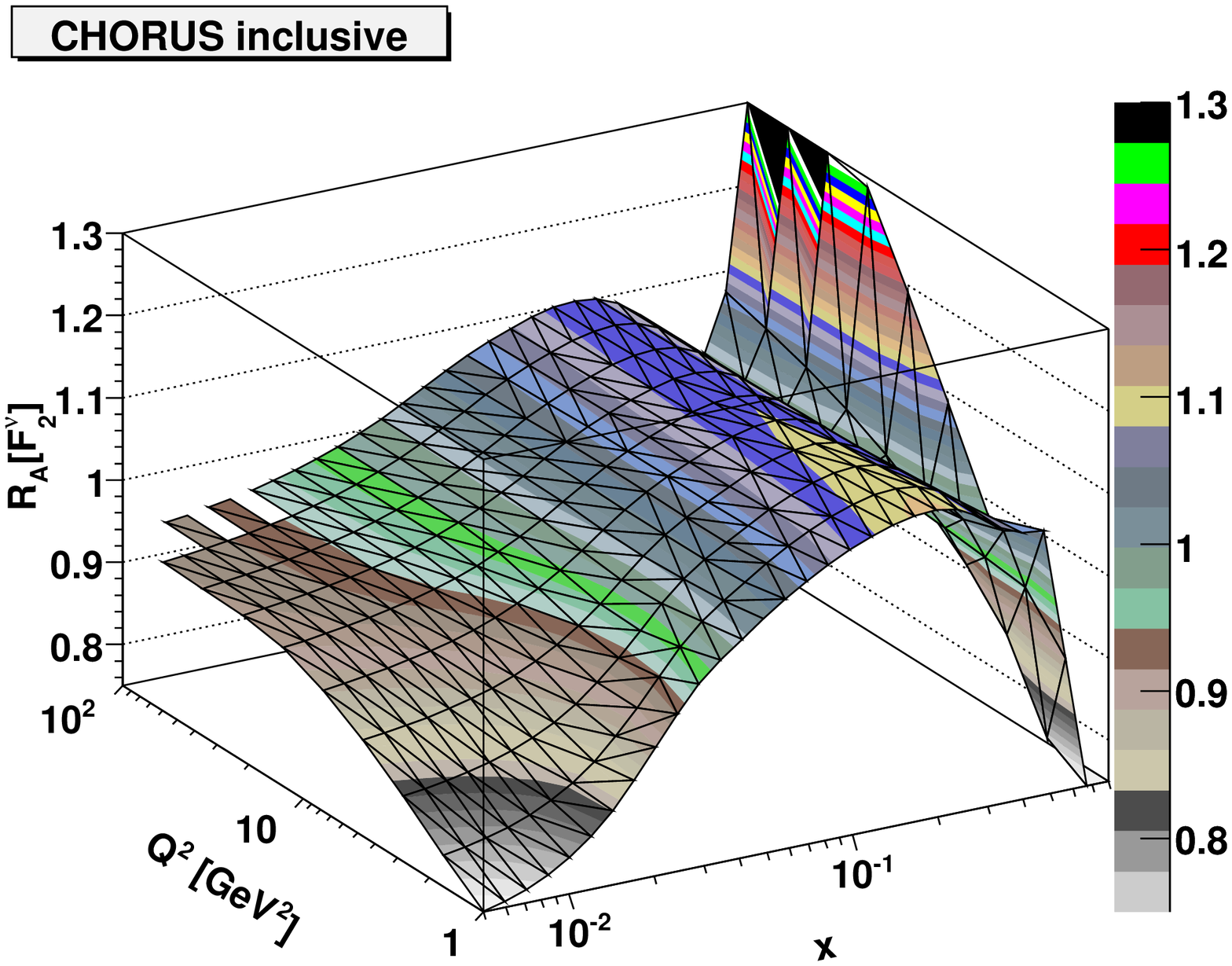} 
\epsfig{width=0.49\textwidth,figure=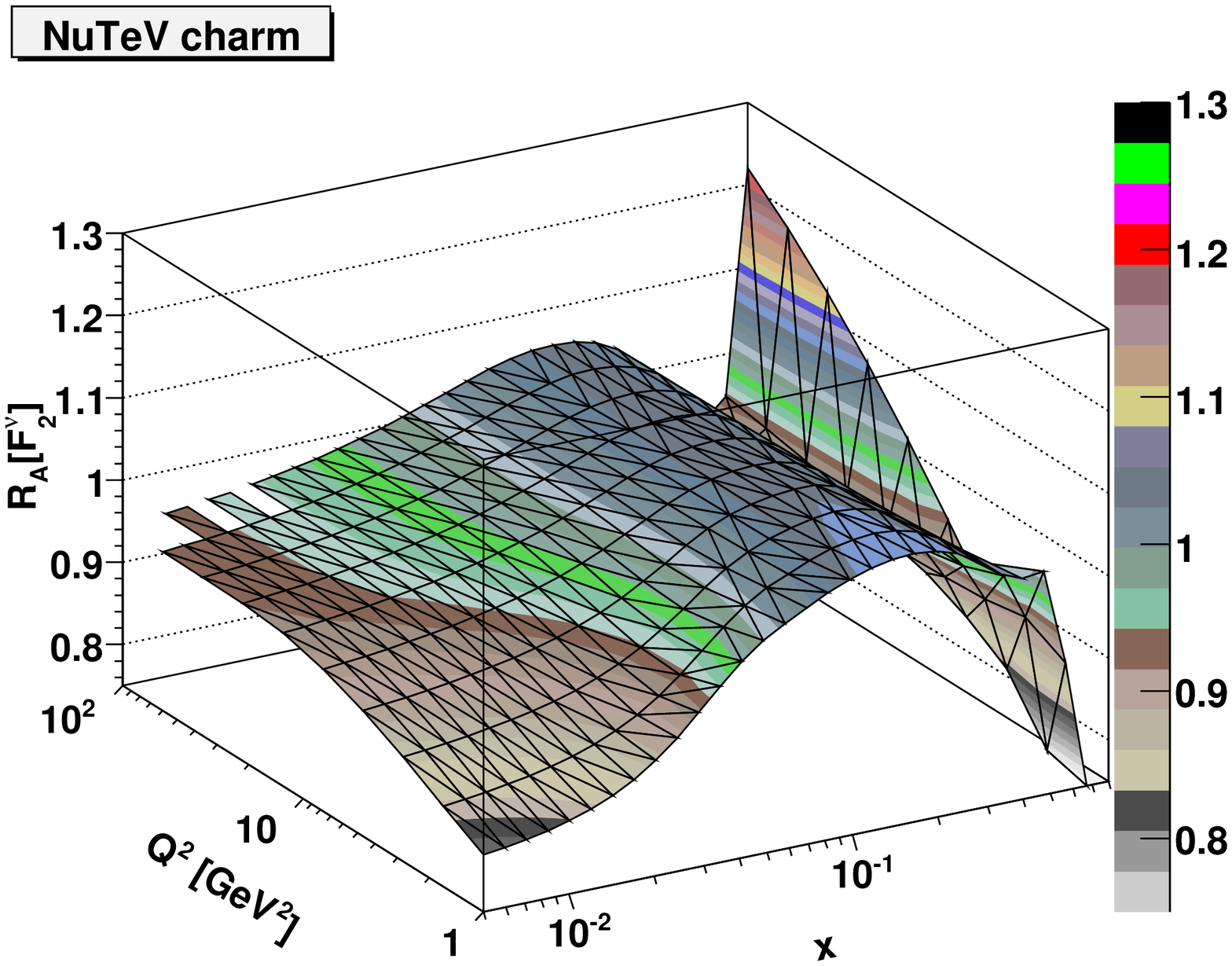} 
\end{center}
\caption{\small Nuclear corrections to the 
neutrino structure function
$F_2^{\nu}$ for inclusive CHORUS data (left) and for 
NuTeV charm production data (right)
from the Kulagin-Petti parametrization~\cite{Kulagin:2007ju}. The
correction is shown in the full kinematic region relevant for both experiments.}
\label{fig:nucorr}
\end{figure}

Nuclear corrections 
have been determined by various 
groups~\cite{deFlorian:2003qf,Hirai:2007sx,Kulagin:2007ju,Eskola:2009uj},
using models of nuclear structure.
The correction 
\be
R_A\lc F_2^{\nu}(x,Q^2)\rc \equiv \frac{F_2^{\nu,A}(x,Q^2)}{
AF_2^{\nu,p}(x,Q^2)}, 
\ee
to the reduced cross sections
$\sigma^{\nu(\bar{\nu})}$ and $\sigma^{\nu(\bar \nu),c}$,
 obtained
using the
parametrizations of Refs.~\cite{deFlorian:2003qf,Hirai:2007sx},
are displayed in
Figs.~\ref{fig:nucorr} for the experimental CHORUS inclusive and
dimuon NuTeV data.  
It is apparent that 
corrections obtained using different models can be significantly
different, but they are all quite small. For
this reason, nuclear corrections were not used in the NNPDF1.0
fit~\cite{Ball:2008by}. In the NNPDF1.2 fit presented here we will not
include nuclear corrections in  our baseline fit, but, in order to
determine the associated systematic uncertainty, we will 
repeat the fit with the nuclear corrections computed using the
models of Refs.~\cite{deFlorian:2003qf,Hirai:2007sx}, which provide
corrections to the parton distributions.
The dependence of the nuclear correction on the kinematic variables is shown 
in Fig.~\ref{fig:nucorr} in the kinematic region and for
$A$ values relevant for CHORUS and NuTeV data, using 
 the model of
Ref.~\cite{Kulagin:2007ju}, which instead  provides  directly a correction to the
structure function.

%------------------------------------------------------------
%------------------------------------------------------------------

% --------------------------------------------
%
%\section{Results}
%
%
%-------------------------------------------------
%%--------------------------------------------------------

\section{Results}
\label{sec:results}

In this section we present the NNPDF1.2 parton set. After discussing
the general features of the fit and its result, and comparing these to 
the previous NNPDF parton set, we discuss in detail the determination
of the strange and antistrange distributions, which are the novel
features of this fit. We finally compare results to experimental data,
including a comparison with the recent~\cite{Chekanov:2009gm}
determination
of the $F_3$ structure function.

\subsection{The NNPDF1.2 parton set: statistical features}
\label{sec:genres}

\begin{table}
\small
 \begin{center}
  \begin{tabular}{|c|c|c|c|c|c|c|c|c|c|c|c|}
    \hline 
    $\eta^{(0)}_{i,\mathrm{\Sigma}}$ & $\eta^{(0)}_{i,\mathrm{g}}$ 
    & $\eta^{(0)}_{i,T_3}$  & 
    $\eta^{(0)}_{i,\mathrm{V_T}}$  & $\eta^{(0)}_{i,\mathrm{\Delta_S}}$
& $\eta^{(0)}_{i,\mathrm{s_+}}$& $\eta^{(0)}_{i,\mathrm{s_-}}$
    & $N_{\rm ite}^{\rm max}$ & $r_{\eta}$ & $N_{\rm cop}$
    & $E_{\rm sets}$ &$N_{\rm update}$\\
    \hline
    $[10,1]$ & $[10,1]$ & $[1,0.1]$ & $[1,0.1]$ & $[1,0.1]$ & $[5,0.5]$ 
& $[1,0.1]$ & 5000
    & 1/3 & 120 & 3 &10 \\
    \hline
  \end{tabular}
\end{center}
  \caption{Parameters controlling the genetic algorithm minimization.
    Since we work with $N_\mathrm{mut}=2$ there are two entries in
    each column for the values of $\eta^{(0)}$.}
  \label{tab:ga_params}
\end{table}

We have produced a  set of
$N_{\rm rep}=1000$ replicas of seven PDFs, each determined as an
optimal fit to one of the Monte Carlo replicas obtained from the
data set of  Sect.~\ref{sec:expdata}. We have used 
the genetic algorithm minimization and a
cross-validation method for the determination of the optimal fit,
according to the method 
 presented in Sect.~4 of Ref.~\cite{Ball:2008by}. The parameters of
 the genetic algorithm are summarized in Table~\ref{tab:ga_params};
they coincide with those used in Ref.~\cite{Ball:2008by} for the five PDFs
already present in that fit.

The general statistical features of our final parton set are
summarized in Tables~\ref{tab:est-fin-tot}-\ref{tab:est-fin}, to be  
compared with the
corresponding tables (Tables~7-8) of Ref.~\cite{Ball:2008by}, where all
the relevant quantities are defined (note
that average uncertainties are now given in percentage value, while
they were given as absolute values in Ref.~\cite{Ball:2008by}).

The statistical features of the fit
 can be
summarized as follows:
\begin{itemize}
\item The general features of the total fit (Tab.~\ref{tab:est-fin-tot}) 
are essentially indistinguishable from those
  of Ref.~\cite{Ball:2008by}, and the comments we made then still apply. The
  same is true for the features of the fit to individual experiments
  (Tab.~\ref{tab:est-fin}) when these were already included in the dataset of
  Ref.~\cite{Ball:2008by}.
This stability upon the addition of two new independent PDFs (thus 74
extra free parameters) and a randomization of the preprocessing
exponents supports the reliability of the results obtained in NNPDF1.0
for all PDFs which were determined there.
\item The quality of the fit to the new HERA II data is comparable to
  that to the BCDMS data, and somewhat worse than that of the fit to
  other HERA data. These new data mostly probe the large $x$ region,
  like BCDMS and unlike other HERA data (see Fig.~\ref{fig:dataplot}), 
and are generally rather precise, also like BCDMS and unlike 
other HERA data (see
  Tab.~\ref{tab:expsets} and Tab.~1 of Ref.~\cite{Ball:2008by}). 
This somewhat larger value of the $\chi^2$ for large $x$ high
precision data, though compatible with statistical fluctuations and
with the theoretical error related to the use of NLO perturbation theory, may
suggest some minor data incompatibility in this region. 
\item The $\chi^2$ of the fit to dimuon data is rather smaller than
  one. This is a consequence of the fact that, as discussed in
  Sec.~\ref{sec:dataset}, correlations have not been included for
  these data because the covariance matrix is not available. The
  average value
  of the $\chi^2$ we obtain is in good agreement with that expected
  on the basis of the ``effective number of degrees of freedom''
  published in Ref.~\cite{masonPhD}, and with other fits to the
  same data~\cite{Alekhin:2008mb}.
\item The uncertainty of the fit to dimuon data, as measured by the
  average standard deviation $\langle \sigma\rangle$ is very close to
  the uncertainty of the data, unlike that of all other data sets
  (reflected by the results for the total fit), where the fit
  uncertainty is much smaller than the data uncertainty (4\% vs. 11\%
  for the total fit). This is a consequence of the fact that dimuon
  data have  little redundancy, and are sensitive to strangeness, 
to which other data are essentially
  insensitive; while all other data have a very large redundancy,
  especially low-$x$ HERA data which depend mainly on the quark
  singlet and gluon. This effect can also be observed in the
  comparison between experimental data and NNPDF1.2
predictions of Fig.~\ref{fig:dimuon}.
\item The average correlation is very low for the dimuon data, because
  the only correlated systematics is normalization. However, the fit
  to these data does display a correlation of the same order of
  magnitude as for other data, reflecting the underlying smoothness of
  parton distributions.
\end{itemize}
\begin{table}
\begin{center}
\begin{tabular}{|c|c|}
\hline 
$\chi^{2}_{\tot}$ &      1.31 \\
\hline
\hline
$\la E \ra $   &       2.80      \\
$\la E_{\rm tr} \ra $   &       2.75      \\
$\la E_{\rm val} \ra $   &       2.80      \\
$\la{\rm TL} \ra $   &     1024      \\
\hline
 $\la \sigma^{(\exp)}
\ra_{\dat}$  &  11.0\%\\
 $\la \sigma^{(\net)}
\ra_{\dat}$  &  4.0\% \\
\hline
 $\la \rho^{(\exp)}
\ra_{\dat}$ &  0.15\\
 $\la \rho^{(\net)}
\ra_{\dat}$&  0.32\\
\hline
 $\la {\rm cov}^{(\exp)}
\ra_{\dat}$ &  $1.6~10^{-3}$\\
 $\la  {\rm cov}^{(\net)}
\ra_{\dat}$&  $6.1~10^{-3}$\\
\hline
\end{tabular}

\end{center}
\caption{\small Statistical estimators for the
final PDF set 
 with $N_{\rep}=1000$ for the total data set. \label{tab:est-fin-tot}}
\end{table}
\begin{table}
\begin{center}
{
\tiny
\begin{tabular}{|c|c||c|c|c|c|c|c|c|}
\hline 
Experiment    & $\chi^{2}_{\tot}$  & $\la E\ra $   & $\la \sigma^{(\exp)}\ra_{\dat}$    & $\la \sigma^{(\net)}\ra_{\dat}$  & $\la \rho^{(\exp)}\ra_{\dat}$ & $\la \rho^{(\net)}\ra_{\dat}$ & $\la \rm cov^{(\exp)}\ra_{\dat}$ & $\la \rm cov^{(\net)}\ra_{\dat}$\\
\hline
SLAC     &    1.27&   3.32& 4.2\% & 2.6\% & 0.31& 0.63& 
$3.1~10^{-5}$& $2.7~10^{-5}$\\
\hline
BCDMS     &   1.57&   3.14& 5.7\% & 4.5\% & 0.47& 0.51&
$ 2.9~10^{-5}$& $1.0~10^{-5}$\\
\hline
NMC      &    1.70&   3.09& 4.9\% & 2.3\% & 0.16& 0.62&
 $4.4~10^{-4}$& $3.8~10^{-5}$\\
\hline 
NMC-pd   &    1.46&   3.12& 1.7\% & 1.7\% & $3.3~10^{-2}$& 0.36& 
$6.5~10^{-6}$& $6.0~10^{-5}$\\
\hline
ZEUS      &   1.07&   2.64& 13\% &  3.9\% & 
$7.9~10^{-2}$& 0.26& $1.5~10^{-4}$& $2.9~10^{-5}$\\
\hline
H1        &   1.03&   2.52&  12\% & 3.3\%  & $2.7~10^{-2}$& 0.25& 
$4.9~10^{-2}$& $2.7~10^{-5}$\\
\hline
CHORUS    &   1.37&   2.88& 15\% & 3.7\% & $9.4~10^{-2}$&0.27 &
 $2.2~10^{-3}$& $3.8~10^{-4}$\\
\hline
FLH108   &    1.67&   2.56& 72\% &  5.7\% & 0.65& 0.76& 
$2.0~10^{-2}$ & $2.5~10^{-4}$\\
\hline
NuTeV Dimuon   &    0.62 &   2.62& 21\% &  22\% & 0.03 & 0.50& 
$1.7~10^{-3}$ & $1.7~10^{-4}$\\
\hline
ZEUS-HERA-II   &    1.51&   2.90& 14\% &  2.5\% & 0.29 & 0.34& 
$6.9~10^{-4}$ & $3.2~10^{-5}$\\
\hline
\end{tabular}
}

\end{center}
\caption{\small Statistical estimators for the
final PDF set 
 with $N_{\rep}=1000$ for individual experiments. \label{tab:est-fin}}
\end{table}

%------------------------------------------------------------
\begin{figure}
\begin{center}
\epsfig{width=0.48\textwidth,figure=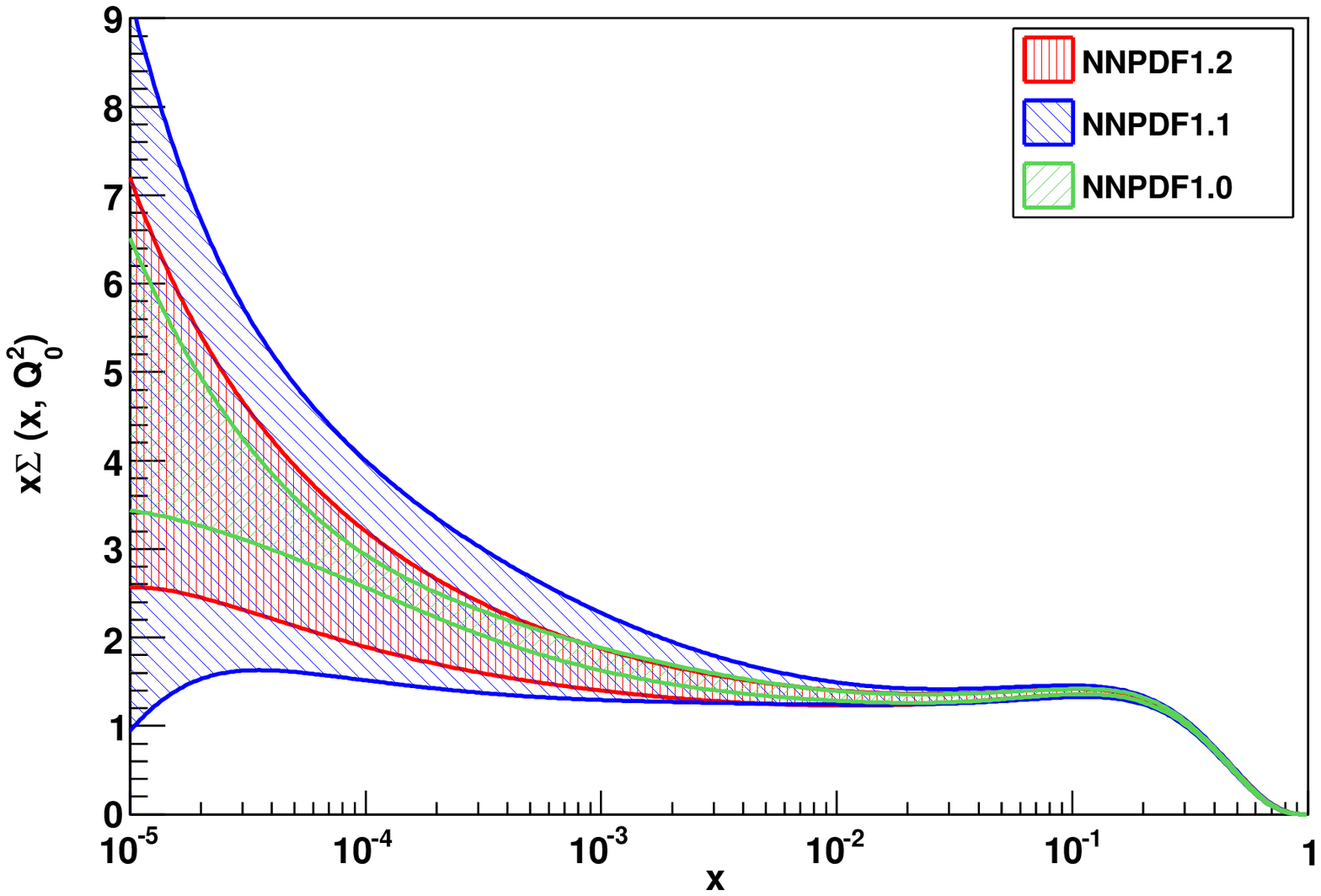}
\epsfig{width=0.48\textwidth,figure=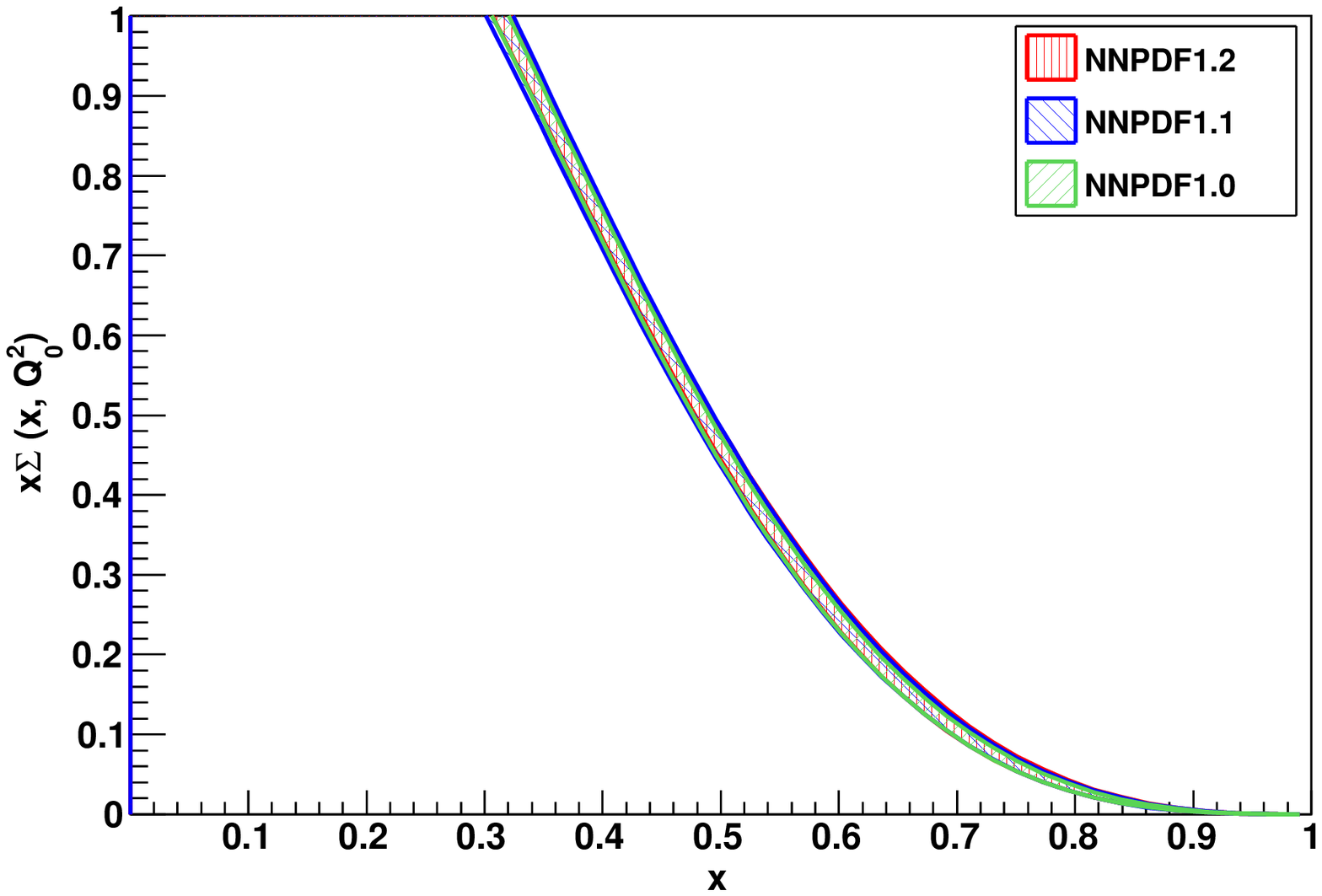}
\epsfig{width=0.48\textwidth,figure=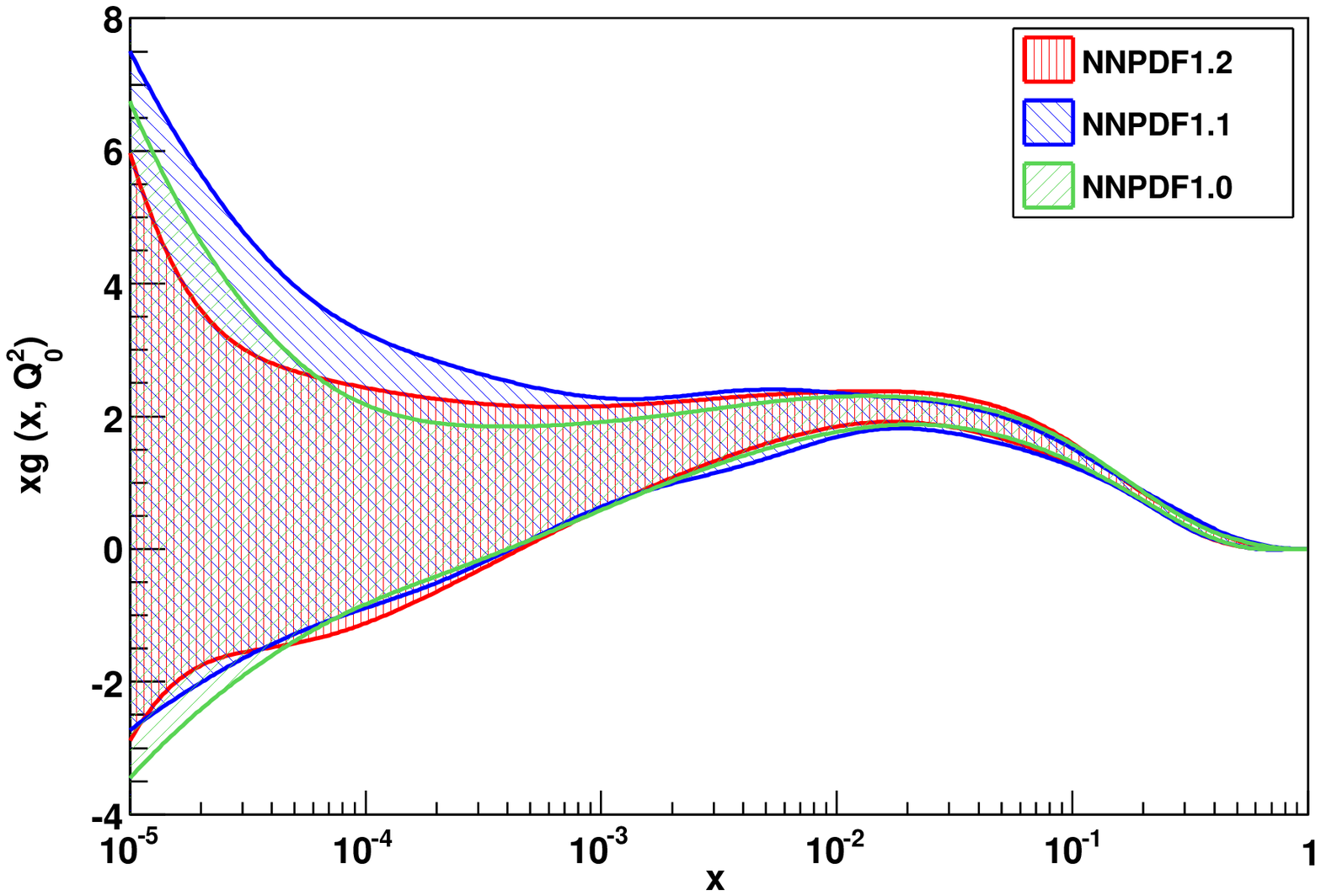}
\epsfig{width=0.48\textwidth,figure=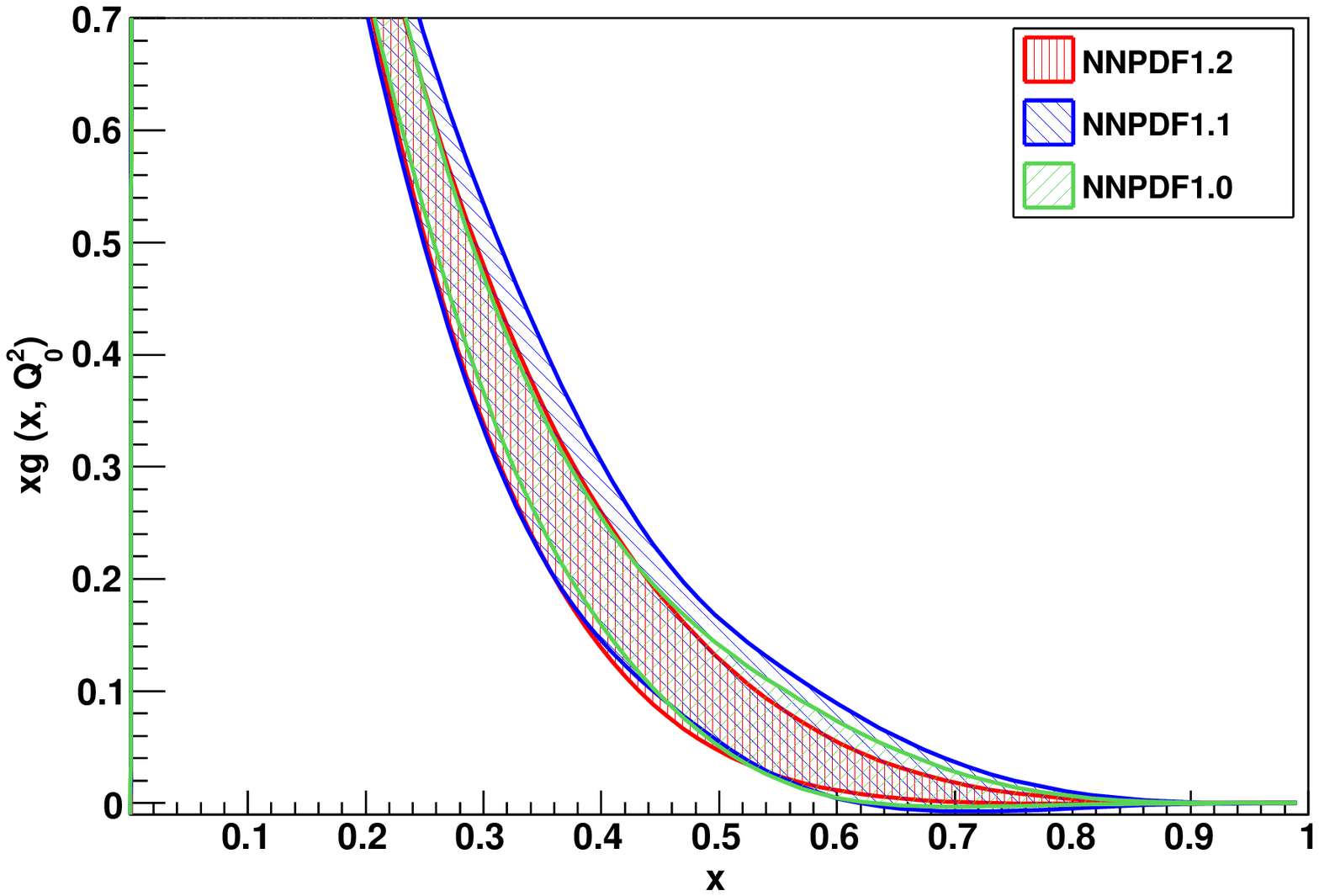}
\end{center}
\caption{\small \label{fig:1p2sing}
 The singlet and gluon PDF at the starting scale
  $Q^2_0=2$ GeV$^2$, plotted versus $x$ on a log (left) or linear
  (right) scale. The PDFs from the previous sets
  NNPDF1.0~\cite{Ball:2008by} and NNPDF1.1~\cite{Rojo:2008ke} are also
  shown for comparison. Note that while the PDFs from
NNPDF1.2 and NNPDF1.0 have been computed with $N_{\rep}=1000$,
those of NNPDF1.1 use $N_{\rep}=100$ only.}
\end{figure}
%------------------------------------------------------------

%------------------------------------------------------------
\begin{figure}
\begin{center}
\epsfig{width=0.48\textwidth,figure=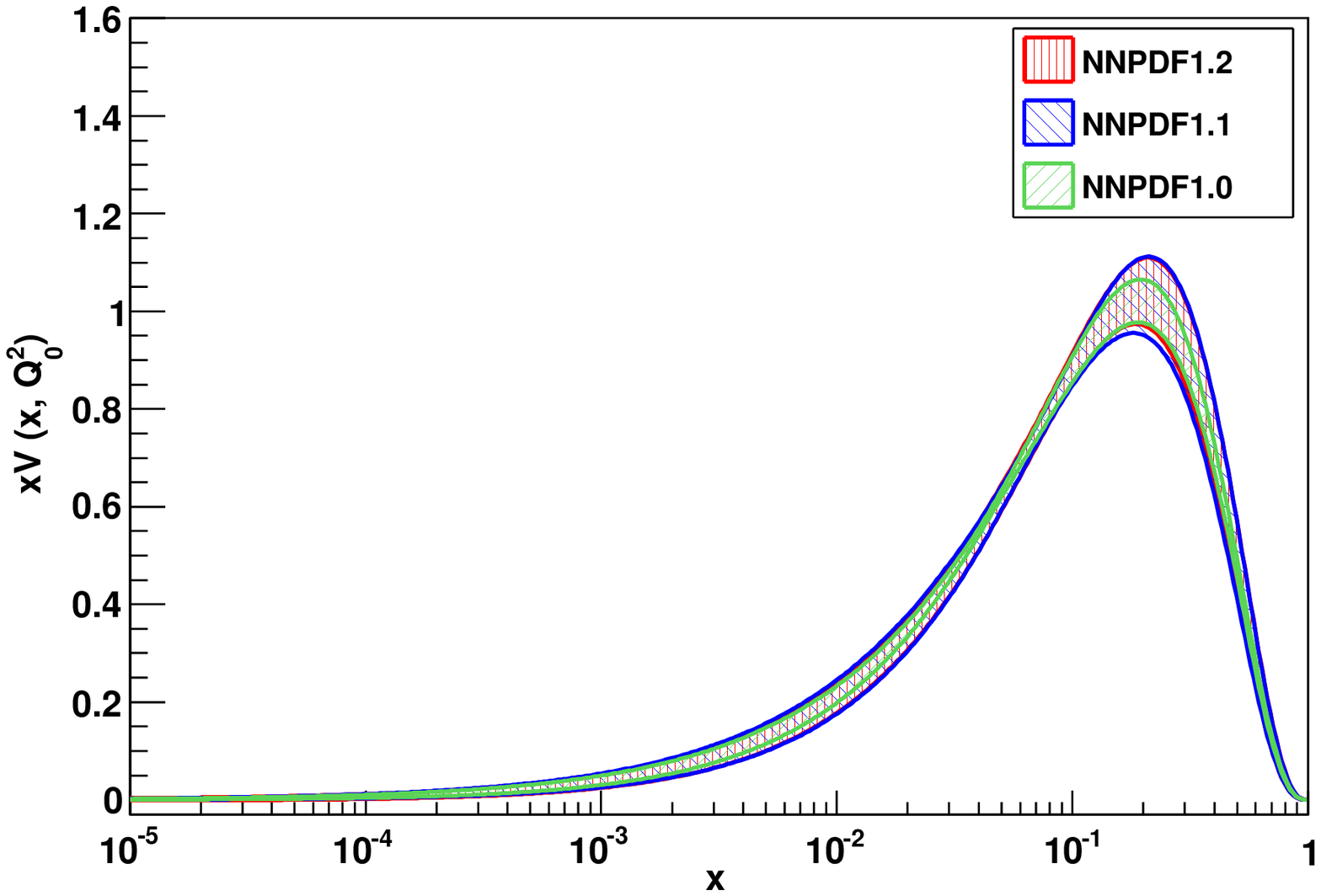}
\epsfig{width=0.48\textwidth,figure=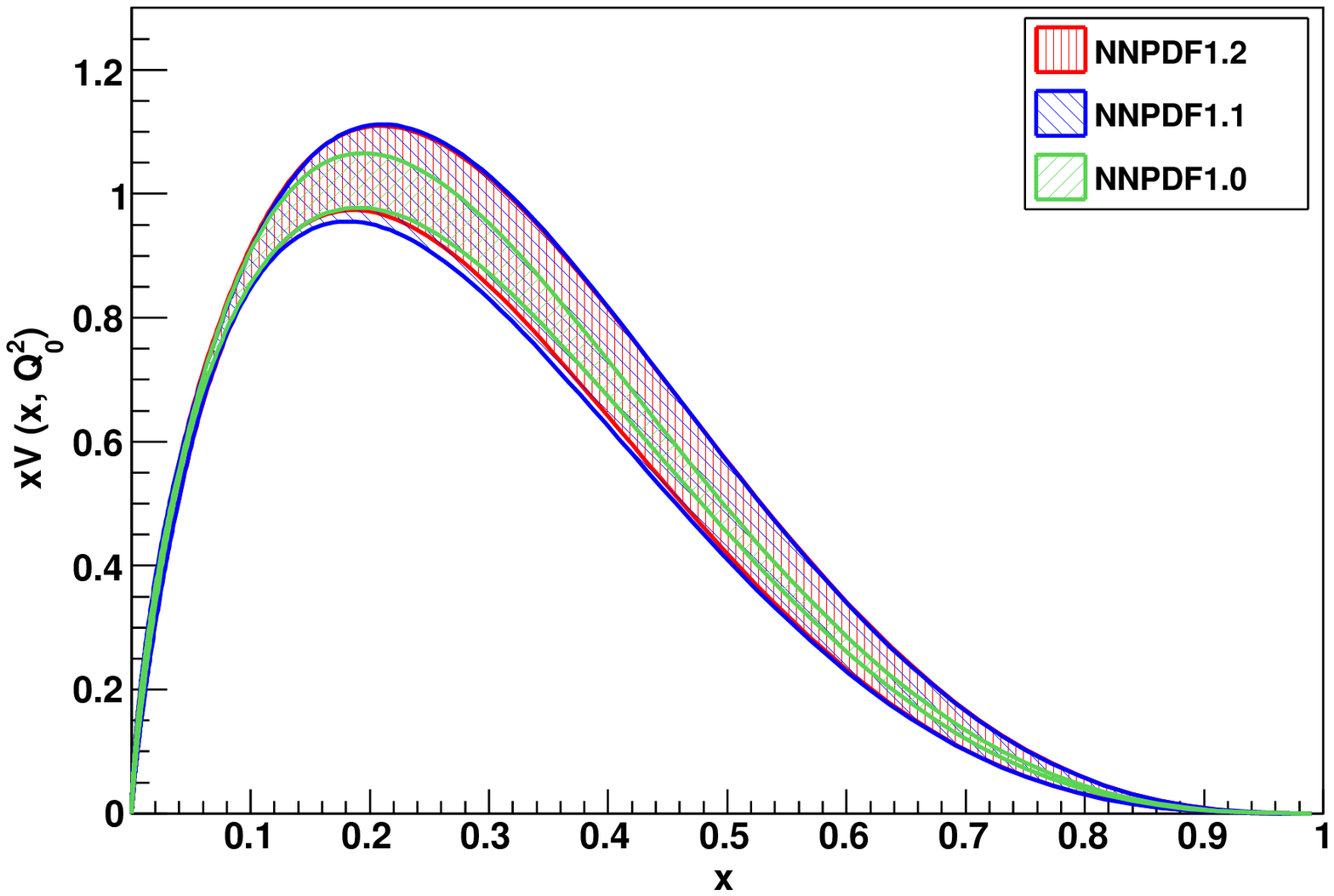}
\epsfig{width=0.48\textwidth,figure=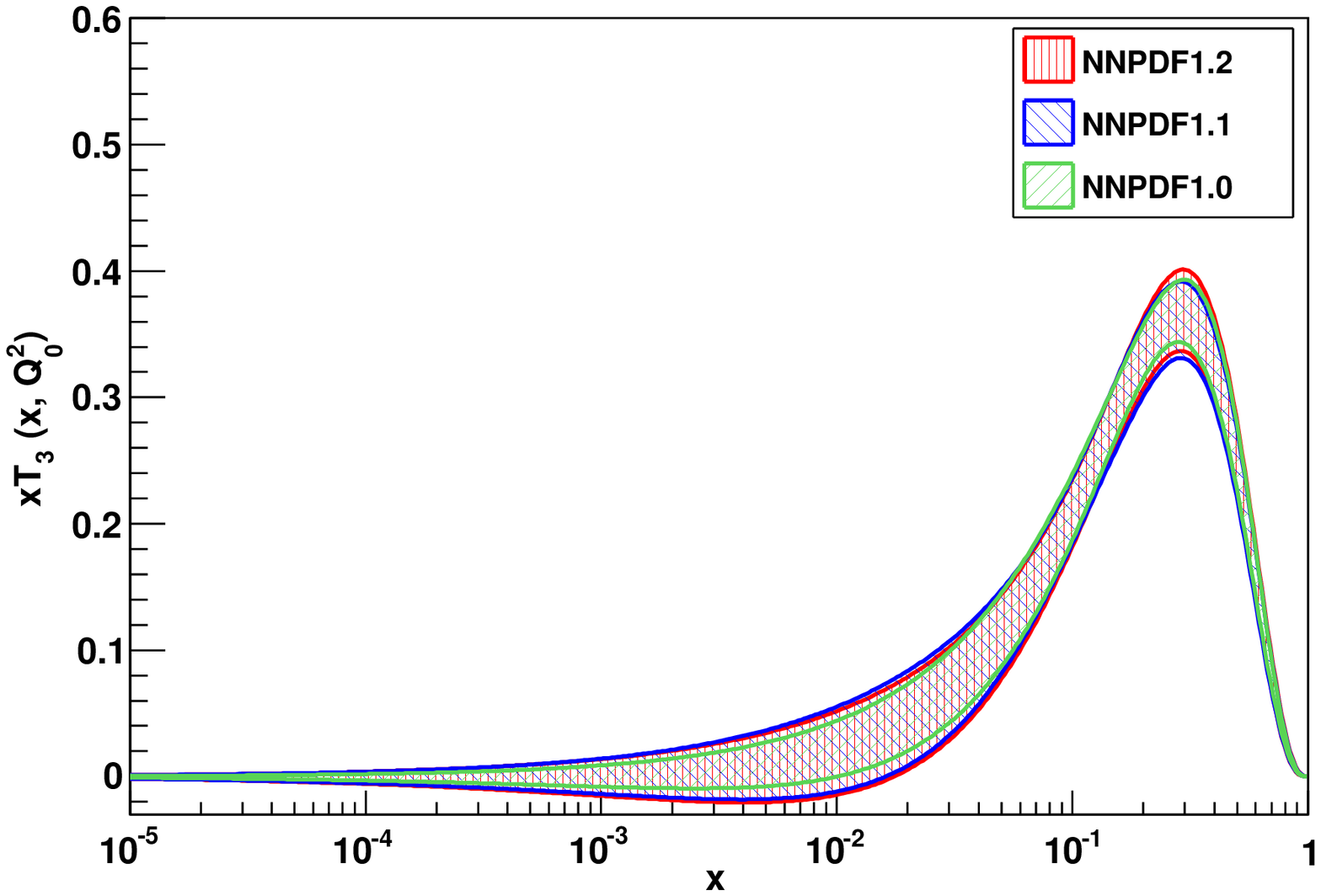}
\epsfig{width=0.48\textwidth,figure=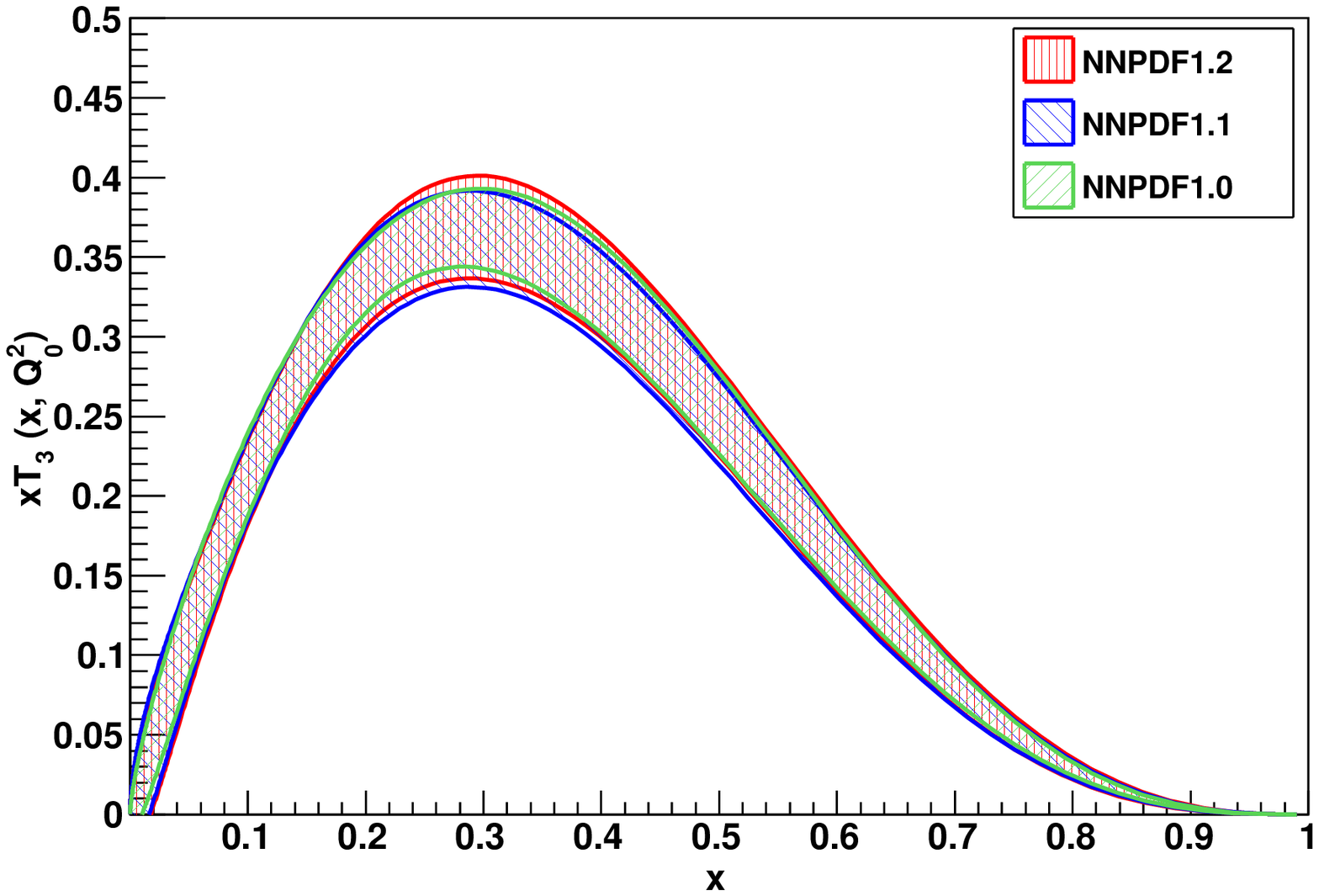}
\epsfig{width=0.48\textwidth,figure=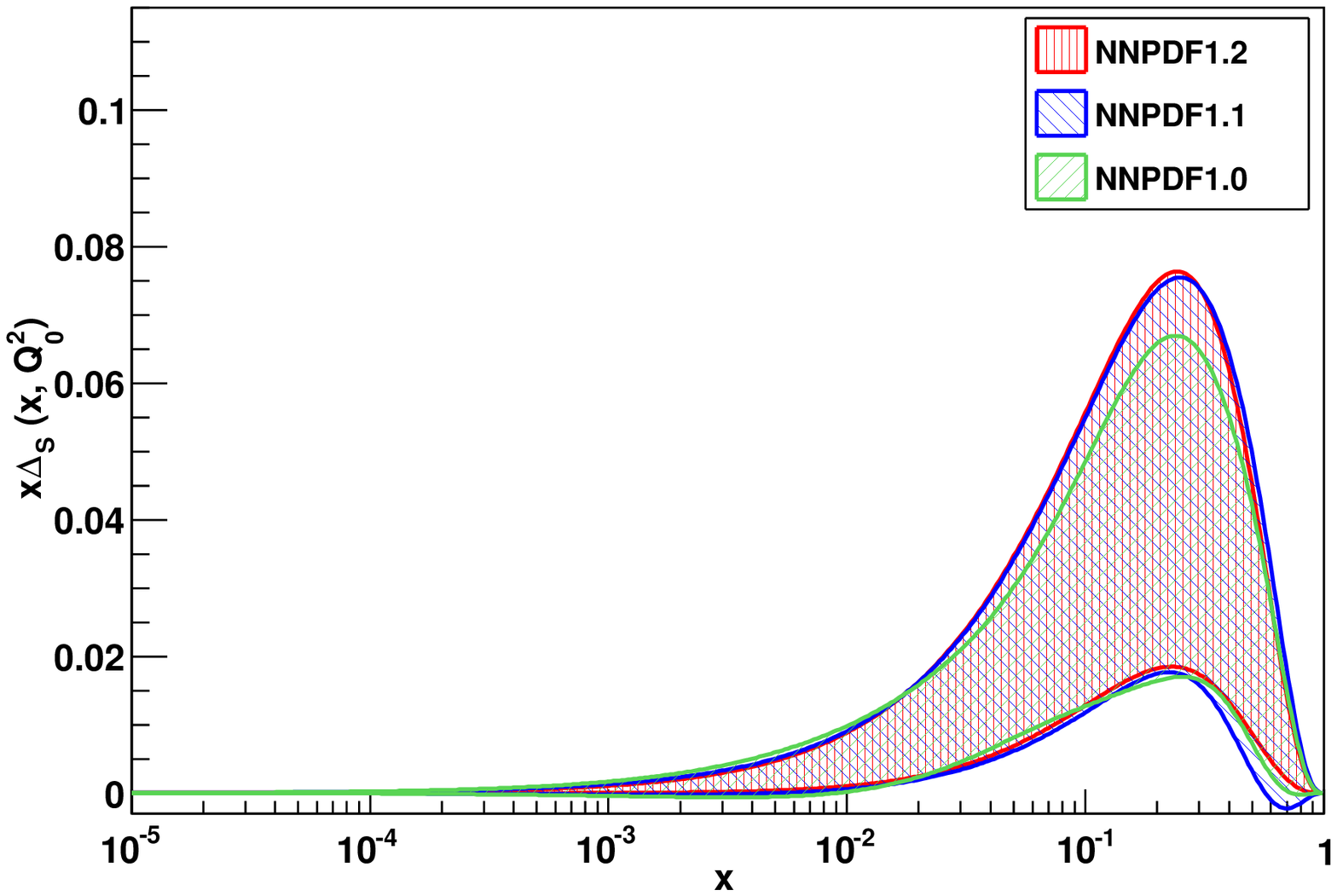}
\epsfig{width=0.48\textwidth,figure=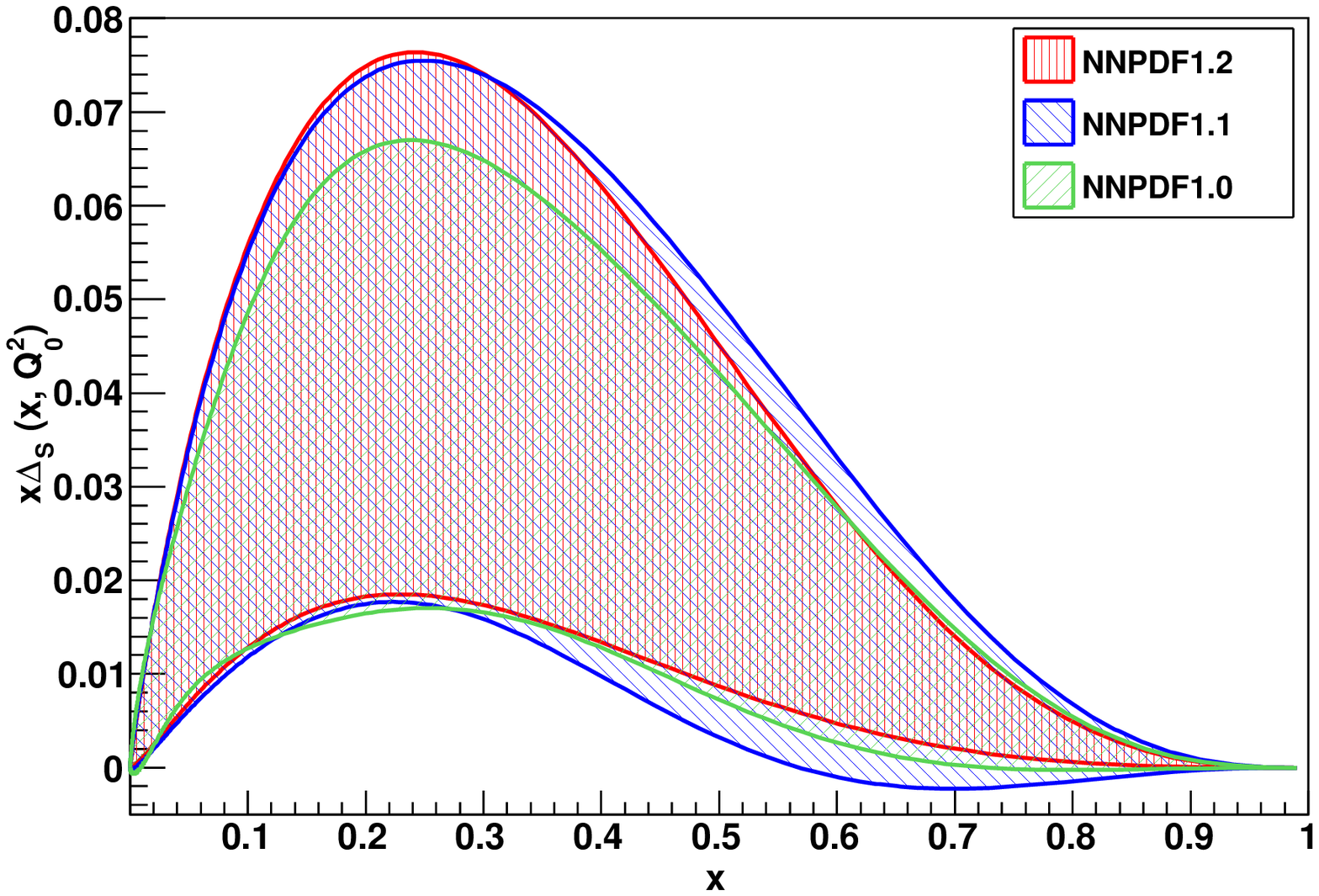}
\end{center}
\label{fig:1p2nsing}
\caption{\small Same as Fig.~\ref{fig:1p2sing}, but for 
the valence and nonsinglet  PDFs.}
\end{figure}

\begin{figure}
\begin{center}
\epsfig{width=0.48\textwidth,figure=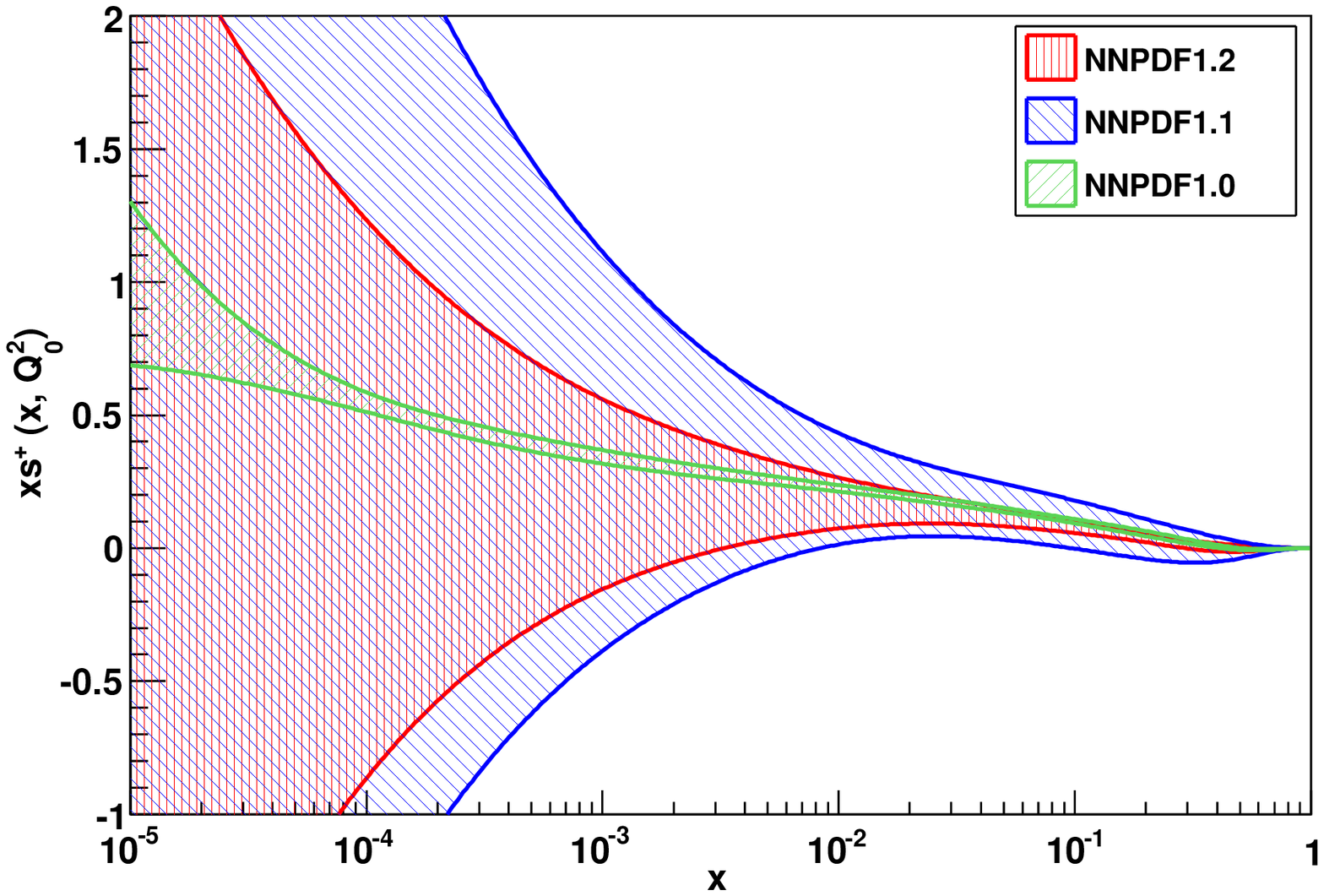}
\epsfig{width=0.48\textwidth,figure=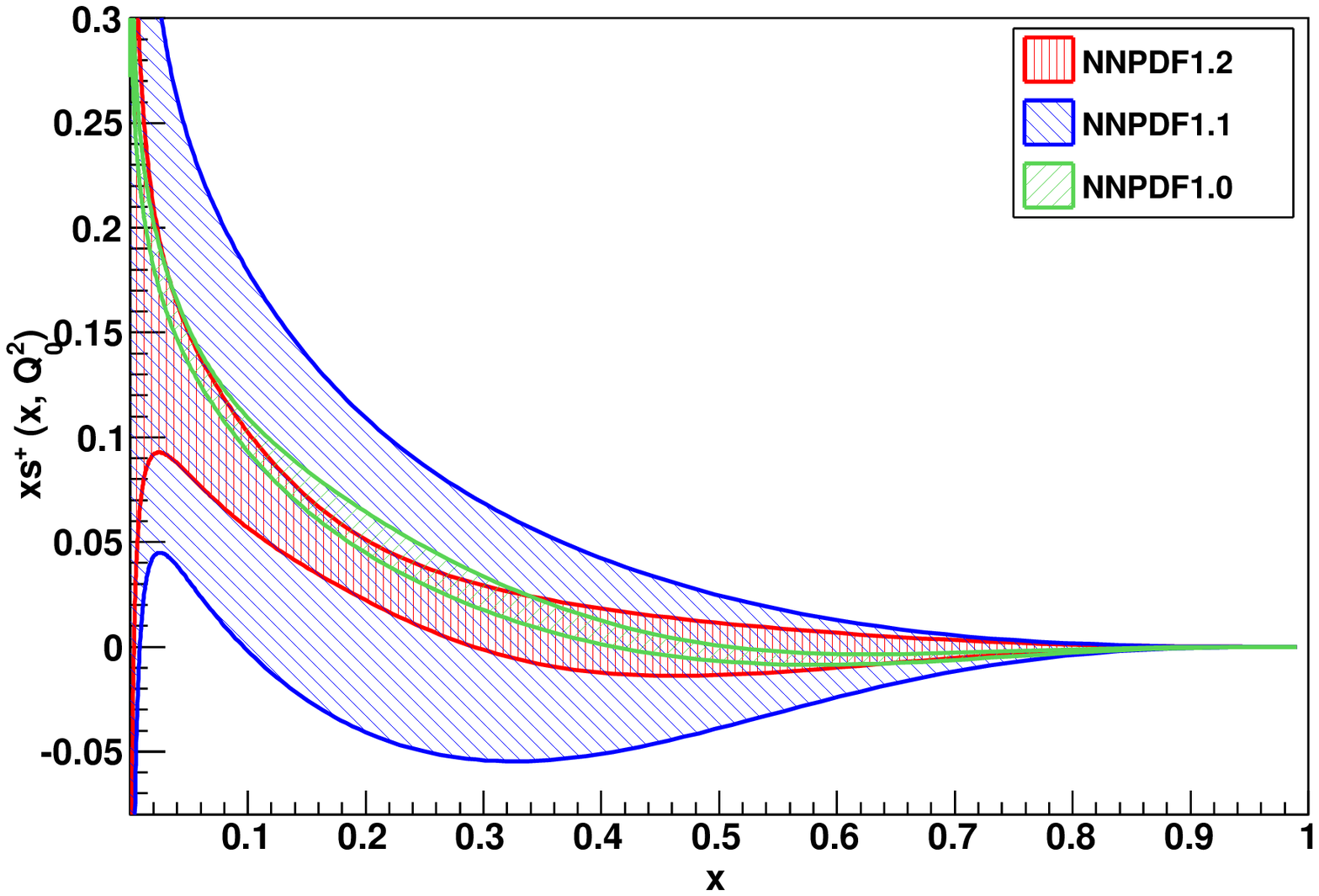}
\epsfig{width=0.48\textwidth,figure=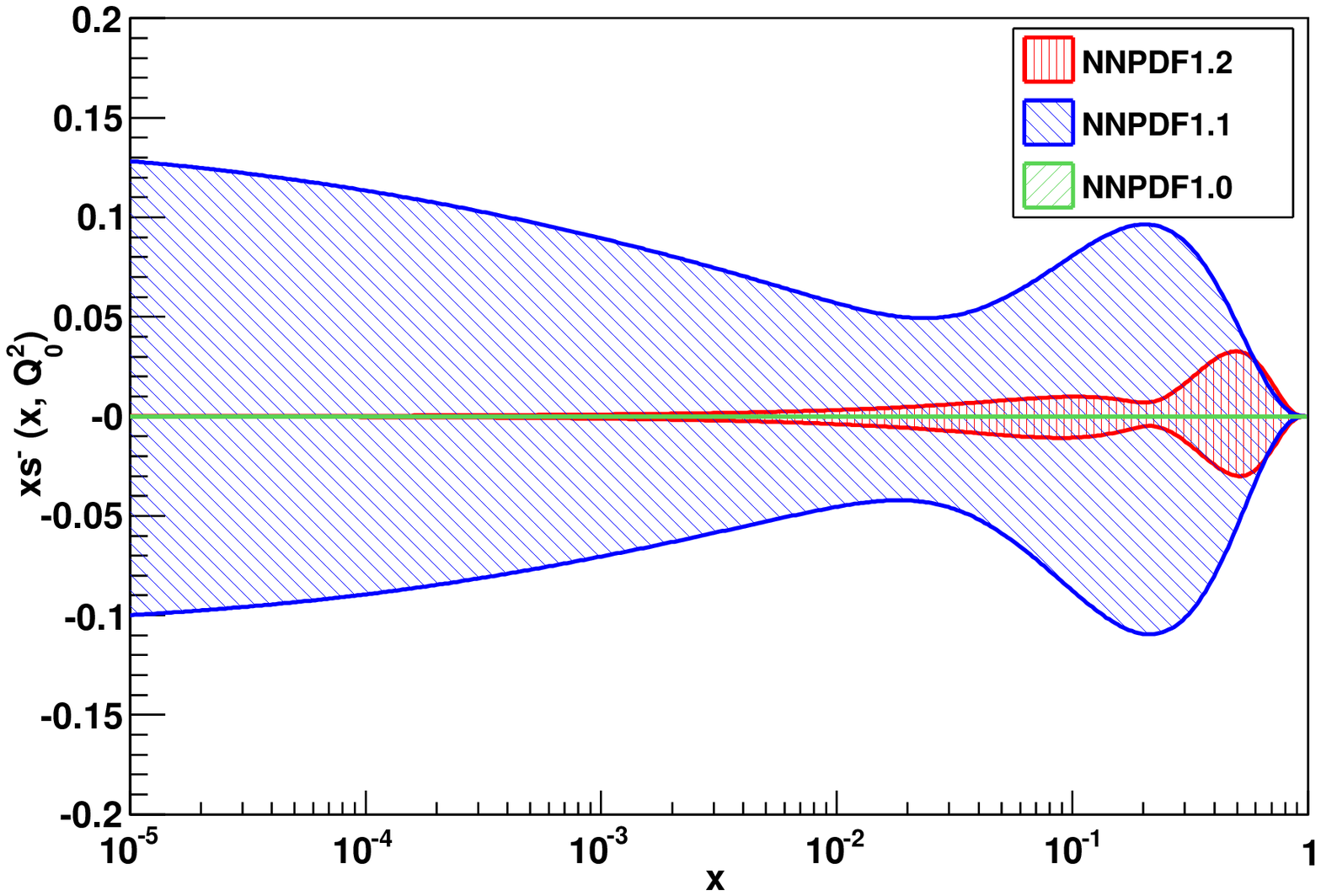}
\epsfig{width=0.48\textwidth,figure=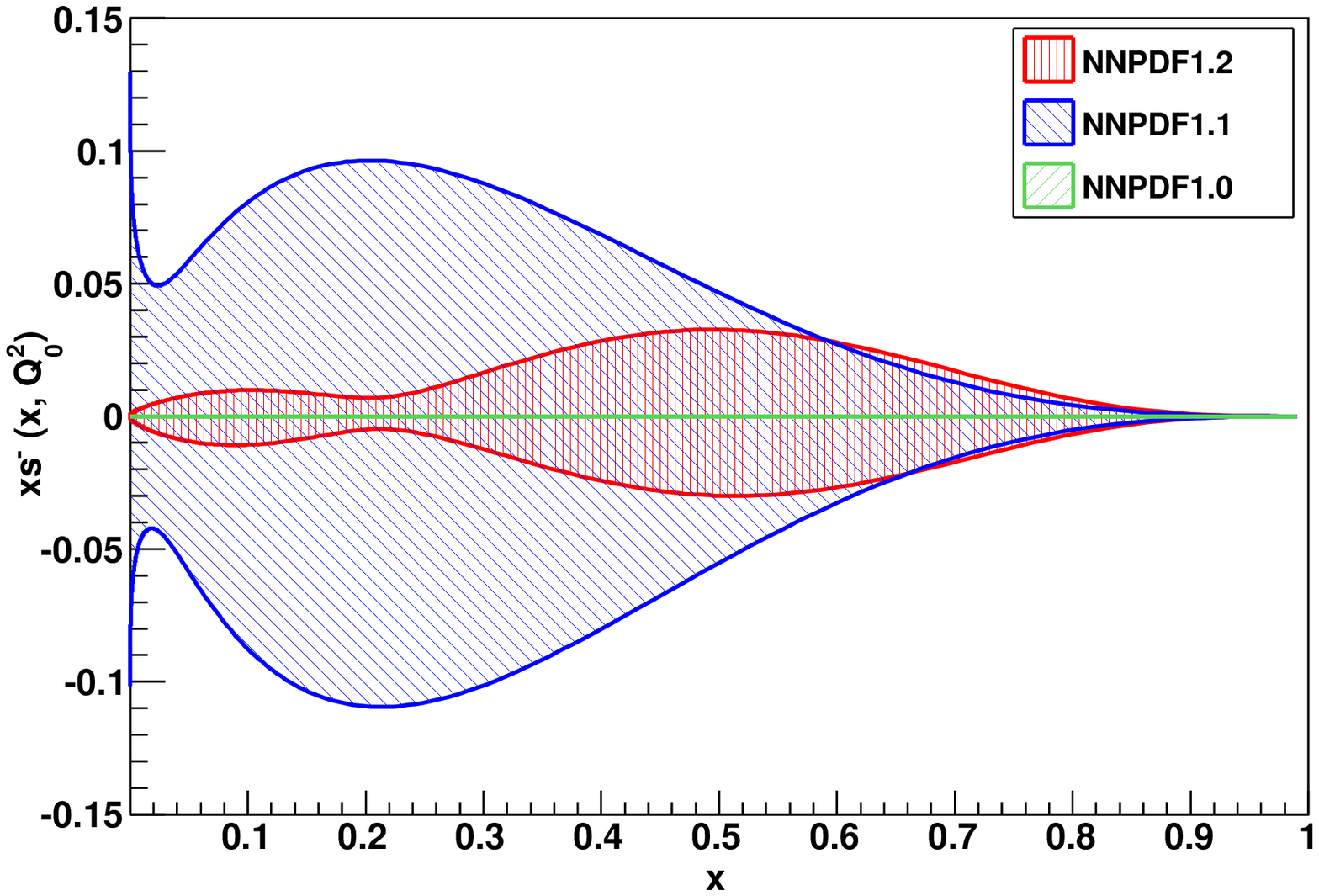}
\end{center}\caption{\small Same as Fig.~\ref{fig:1p2sing}, but for 
the strange sector PDFs. Note that in NNPDF1.0 $ s^\pm$ were  assumed to be
respectively  $ s^+(x,Q_0^2)=\frac{1}{2}\left(\bar u+\bar d\right)$
and $ s^-(x,Q_0^2)=0$. \label{fig:1p2strang}}
\end{figure}

\subsection{The NNPDF1.2 parton set: parton distributions}
\label{sec:comp}
The NNPDF1.2 set of parton distributions at the starting scale
$Q_0^2=2$~GeV$^2$ is displayed in
Figs.~\ref{fig:1p2sing}-\ref{fig:1p2strang}, and compared to the
previous sets
  NNPDF1.0~\cite{Ball:2008by} and NNPDF1.1~\cite{Rojo:2008ke}. The
  distances (defined as in Ref.~\cite{Ball:2008by}) between each pair
  of these three sets are shown in Table~\ref{tab:stabtab-nnpdfrel}.

The general features of this PDF set and its comparison to the
previous NNPDF sets are the following
\begin{itemize}
\item In the singlet sector, there is very little difference in
  central values and uncertainties between the NNPDF1.2 and NNPDF1.0
  parton sets: the distance between the sets is compatible with
  statistical fluctuations. The NNPDF1.1, which had an independent
  parametrization for the strange distribution without any data to
  constrain it displays an increase in the uncertainty of the quark
  singlet due to this unconstrained strange contribution. 
\item The isospin triplet and the sea asymmetry  
are the same in all NNPDF sets within fluctuations. The total valence has
the same central value in all sets within fluctuations, and the same
uncertainty in the NNPDF1.2 and NNPDF1.1 sets, while the uncertainty
on it was somewhat underestimated in NNPDF1.0. This underestimate of
the NNPDF1.0 valence uncertainty was
already singled out based on a statistical stability analysis in
Sect.~5.4 of
Ref.~\cite{Ball:2008by}, where it was suggested that it could be cured
by a randomization of the preprocessing exponents in
Eq.~(\ref{gennn}). This randomization has been implemented in NNPDF1.1
and NNPDF1.2, which indeed have somewhat larger valence uncertainty,
compatible with each other.  This is despite the fact that the strange
contribution to the total valence is affected by a much larger
uncertainty in NNPDF1.1 than in NNPDF1.2.

\item The central value and uncertainty on the
strange distributions Eq.~(\ref{spmdef}) are compatible with
  those of NNPDF1.1, where strangeness was independently parametrized
  but essentially unconstrained by data, whereas they are incompatible
  with those of NNPDF1.0, where strangeness was determined by the
  assumptions $ s^+(x,Q_0^2)=\frac{1}{2}\left(\bar u+\bar d\right)$
and $ s^-(x,Q_0^2)=0$. This means that this simple assumption, though
perhaps not too far off, is insufficient to determine the strange
distribution within its stated accuracy. This conclusion was also
reached recently in Ref.~\cite{Lai:2007dq}.
 The uncertainty on strangeness as we determine
it here turns out to be
rather larger than that induced by the NNPDF1.0 assumption, but much
smaller than that obtained in NNPDF1.1 in the absence of dimuon data. 
 It is thus possible to determine the shape of $s^+$ with reasonable
 accuracy. However, our determination of $s^-$ turns out to be
 compatible with the NNPDF1.0 assumption that $ s^-(x,Q_0^2)=0$.
We shall discuss the features of the strange distribution in greater
detail in Sect.~\ref{sec:strange} below.
\end{itemize}

\begin{table}[t!]
\begin{center}\begin{minipage}{.45\linewidth}
\tiny
\begin{tabular}{|c|c|c|}
\hline
\multicolumn{3}{|c|}{NNPDF1.2 vs. NNPDF1.1} \\
\hline
\hline 
 &  Data  & Extrapolation \\
\hline 
$\Sigma(x,Q_0^2)$ & $5~10^{-4} \le x \le 0.1$  & $10^{-5} \le x \le 10^{-4}$  \\
\hline
$\la d[q]\ra$  & 2.7 &  1.2 \\
$\la d[\sigma]\ra$  & 3.1 & 1.8\\
\hline 
\hline 
$g(x,Q_0^2)$ & $5~10^{-4} \le x \le 0.1$  & $10^{-5} \le x \le 10^{-4}$  \\
\hline
$\la d[q]\ra$  & 2.4 & 2.0 \\
$\la d[\sigma]\ra$  & 1.3& 1.4  \\
\hline 
\hline 
$T_3(x,Q_0^2)$ & $0.05 \le x \le 0.75$  & $10^{-3} \le x \le 10^{-2}$  \\
\hline
$\la d[q]\ra$  & 1.5 &  0.9\\
$\la d[\sigma]\ra$  & 1.1  & 1.2  \\
\hline 
\hline 
$V(x,Q_0^2)$ & $0.1 \le x \le 0.6$  & $3~10^{-3} \le x \le 3~10^{-2}$  \\
\hline
$\la d[q]\ra$  & 1.1 & 1.0 \\
$\la d[\sigma]\ra$  & 1.3  & 1.4  \\
\hline 
\hline 
$\Delta_S(x,Q_0^2)$ & $0.1 \le x \le 0.6$  
& $3~10^{-3} \le x \le 3~10^{-2}$  \\
\hline
$\la d[q]\ra$  & 0.8 & 0.8 \\
$\la d[\sigma]\ra$   & 1.3  & 1.1   \\
\hline 
\hline 
$s^+(x,Q_0^2)$ &  $5~10^{-4} \le x \le 0.1$  & $10^{-5} \le x \le 10^{-4}$ \\
\hline
$\la d[q]\ra$  & 2.0 &  1.6\\
$\la d[\sigma]\ra$  & 4.5  & 1.8   \\
\hline 
\hline 
$s^-(x,Q_0^2)$ & $0.1 \le x \le 0.6$  
& $3~10^{-3} \le x \le 3~10^{-2}$  \\
\hline
$\la d[q]\ra$  & 1.1 &  1.3\\
$\la d[\sigma]\ra$  & 6.1  & 4.6   \\
\hline 
\end{tabular}

\end{minipage}\begin{minipage}{.45\linewidth}
\tiny
\begin{tabular}{|c|c|c|}
\hline
\multicolumn{3}{|c|}{NNPDF1.1 vs. NNPDF1.0} \\
\hline
\hline 
 &  Data  & Extrapolation \\
\hline 
$\Sigma(x,Q_0^2)$ & $5~10^{-4} \le x \le 0.1$  & $10^{-5} \le x \le 10^{-4}$  \\
\hline
$\la d[q]\ra$  & 1.6 &  0.9 \\
$\la d[\sigma]\ra$  & 4.0  & 2.3 \\
\hline 
\hline 
$g(x,Q_0^2)$ & $5~10^{-4} \le x \le 0.1$  & $10^{-5} \le x \le 10^{-4}$  \\
\hline
$\la d[q]\ra$  & 2.3 & 1.7 \\
$\la d[\sigma]\ra$  & 1.6 & 1.2   \\
\hline 
\hline 
$T_3(x,Q_0^2)$ & $0.05 \le x \le 0.75$  & $10^{-3} \le x \le 10^{-2}$  \\
\hline
$\la d[q]\ra$  &  1.6 & 0.8 \\
$\la d[\sigma]\ra$  & 1.8  &3.4   \\
\hline 
\hline 
$V(x,Q_0^2)$ & $0.1 \le x \le 0.6$  & $3~10^{-3} \le x \le 3~10^{-2}$  \\
\hline
$\la d[q]\ra$  & 1.8  &  1.7\\
$\la d[\sigma]\ra$  & 5.3  & 5.2   \\
\hline 
\hline 
$\Delta_S(x,Q_0^2)$ & $0.1 \le x \le 0.6$  
& $3~10^{-3} \le x \le 3~10^{-2}$  \\
\hline
$\la d[q]\ra$  & 1.2 &  1.0\\
$\la d[\sigma]\ra$   & 1.6 & 1.1    \\
\hline 
\hline 
$s^+(x,Q_0^2)$ &  $5~10^{-4} \le x \le 0.1$  & $10^{-5} \le x \le 10^{-4}$ \\
\hline
$\la d[q]\ra$  & 1.0 & 1.0 \\
$\la d[\sigma]\ra$  & 5.4  & 2.3   \\
\hline 
\hline 
$s^-(x,Q_0^2)$ & $0.1 \le x \le 0.6$  
& $3~10^{-3} \le x \le 3~10^{-2}$  \\
\hline
$\la d[q]\ra$  & 1.1  & 1.3\\
$\la d[\sigma]\ra$  & 7.4  & 4.6   \\
\hline 
\end{tabular}

\end{minipage}
\tiny
\begin{tabular}{|c|c|c|}
\hline
\multicolumn{3}{|c|}{ NNPDF1.2 vs. NNPDF1.0} \\
\hline
\hline 
 &  Data  & Extrapolation \\
\hline 
$\Sigma(x,Q_0^2)$ & $5~10^{-4} \le x \le 0.1$  & $10^{-5} \le x \le 10^{-4}$  \\
\hline
$\la d[q]\ra$  & 3.2 & 1.9 \\
$\la d[\sigma]\ra$  &  2.9& 3.3 \\
\hline 
\hline 
$g(x,Q_0^2)$ & $5~10^{-4} \le x \le 0.1$  & $10^{-5} \le x \le 10^{-4}$  \\
\hline
$\la d[q]\ra$  & 1.7  & 0.9 \\
$\la d[\sigma]\ra$  &  1.6& 1.3 \\
\hline 
\hline 
$T_3(x,Q_0^2)$ & $0.05 \le x \le 0.75$  & $10^{-3} \le x \le 10^{-2}$  \\
\hline
$\la d[q]\ra$  &  1.1 &  1.0 \\
$\la d[\sigma]\ra$  & 2.0  & 3.2  \\
\hline 
\hline 
$V(x,Q_0^2)$ & $0.1 \le x \le 0.6$  & $3~10^{-3} \le x \le 3~10^{-2}$  \\
\hline
$\la d[q]\ra$  & 2.6  & 2.4 \\
$\la d[\sigma]\ra$  & 5.3 & 4.9  \\
\hline 
\hline 
$\Delta_S(x,Q_0^2)$ & $0.1 \le x \le 0.6$  
& $3~10^{-3} \le x \le 3~10^{-2}$  \\
\hline
$\la d[q]\ra$  & 1.4 & 0.9 \\
$\la d[\sigma]\ra$  & 1.5  & 1.2   \\
\hline 
\hline 
$s^+(x,Q_0^2)$ &  $5~10^{-4} \le x \le 0.1$  & $10^{-5} \le x \le 10^{-4}$ \\
\hline
$\la d[q]\ra$  & 6.2 & 3.7 \\
$\la d[\sigma]\ra$  & 5.7  & 3.8   \\
\hline 
\hline 
$s^-(x,Q_0^2)$ & $0.1 \le x \le 0.6$  
& $3~10^{-3} \le x \le 3~10^{-2}$  \\
\hline
$\la d[q]\ra$  & 1.3 & 1.2\\
$\la d[\sigma]\ra$  & 6.8 & 6.5   \\
\hline 
\end{tabular}

\end{center}
\caption{\small Distance between
the NNPDF1.0, NNPDF1.1 and NNPDF1.2 parton sets. All distances are
computed from a set of  
$N_{\rm rep}=100$ replicas.
\label{tab:stabtab-nnpdfrel}}
\end{table}

\subsection{Theoretical uncertainties}
\label{sec:thunc}

As discussed in Sects.~\ref{sec:hq}-\ref{sec:nucorr}, dimuon data are
potentially sensitive to the treatment of the quark mass, and neutrino
data in general are potentially sensitive to nuclear corrections. In
order to explore this sensitivity, we have repeated the NNPDF1.2 fit
using also for dimuon data
the ZM-VFN scheme (as in Ref.~\cite{Ball:2008by}) instead of the
improved I-ZM-VFN quark mass treatment discussed in Sect.~\ref{sec:hq}
and used for the default NNPDF1.2 fit (the ZM-VFN is used for all
other data anyway). The distances between results thus obtained are displayed in
Tab.~\ref{tab:stabtab-checks}. It is apparent that there is a
certain change in the central value of the strange $s^+$ distribution
in the region of the data, of
order of about  ten, which, with $100$ replicas, means  that the central
value has moved by about $1.4\sigma$ in units of the standard
deviation. The uncertainty on $s^+$ itself, and the central value
of the singlet distribution in the region of
the data are affected to a lesser extent, while all other PDFs are
unaffected. Thus the charm mass corrections displayed in
Fig.~\ref{fig:izmcomp} have a small but noticeable effect on the
determination of the total strange $s^+$ distribution. Our approximate
treatment will correspondingly be a source of systematics, which we
shall take into account when discussing quantities related to
strangeness. 

In order to study the sensitivity to the nuclear corrections displayed
in Fig.~\ref{fig:nucorr2} we have repeated the NNPDF1.2 fit with
all neutrino data corrected for nuclear effects according to the
models of de~Florian-Sassot~\cite{deFlorian:2003qf} and
HKN07~\cite{Hirai:2007sx}. The distances  tabulated in
Tab.~\ref{tab:stabtab-checks} show that the effect of nuclear
corrections is negligible: fits with or without nuclear corrections
differ by an amount which is compatible with statistical fluctuations.

\begin{table}[t!]
\begin{center}
\scriptsize
\begin{tabular}{|c|c|c|c|c|c|c|}
\hline
& \multicolumn{2}{|c|}{ZM} &\multicolumn{2}{|c|}{De~Florian-Sassot} &
\multicolumn{2}{|c|}{HKN07}  \\
\hline
\hline 
 &  Data  & Extrapolation &  Data  & Extrap. &  Data  & Extrap.
.   \\
\hline 
$\Sigma(x,Q_0^2)$ & $5~10^{-4} \le x \le 0.1$  & $10^{-5} \le x \le 10^{-4}$
& &&&  \\
\hline
$\la d[q]\ra$  &  5.2& 1.0 & 2.3 & 1.4 & 2.3& 0.9\\
$\la d[\sigma]\ra$  & 2.5  & 1.6    & 1.5& 1.2 &1.2& 1.1 \\
\hline 
\hline 
$g(x,Q_0^2)$ & $5~10^{-4} \le x \le 0.1$  & $10^{-5} \le x \le 10^{-4}$
&&&&\\
\hline
$\la d[q]\ra$  &  1.4 &  1.5   & 1.2& 1.0 &1.4& 1.1\\
$\la d[\sigma]\ra$  & 1.8  & 1.5   & 1.2& 1.2 & 1.2& 1.4 \\
\hline 
\hline 
$T_3(x,Q_0^2)$ & $0.05 \le x \le 0.75$  & $10^{-3} \le x \le 10^{-2}$ 
& &&&  \\
\hline
$\la d[q]\ra$  &  1.4 & 2.0     & 1.3& 1.0& 1.0& 1.0\\
$\la d[\sigma]\ra$  & 2.9   & 0.9   & 1.4& 1.5& 1.1& 1.1 \\
\hline 
\hline 
$V(x,Q_0^2)$ & $0.1 \le x \le 0.6$  & $3~10^{-3} \le x \le 3~10^{-2}$
&&&& \\
\hline
$\la d[q]\ra$  & 1.2  & 1.2  & 1.3& 1.2& 0.8& 0.7\\
$\la d[\sigma]\ra$  & 1.5  & 1.1  & 1.3& 1.5& 1.3& 0.9\\
\hline 
\hline 
$\Delta_S(x,Q_0^2)$ & $0.1 \le x \le 0.6$  
& $3~10^{-3} \le x \le 3~10^{-2}$
 & &&&  \\
\hline
$\la d[q]\ra$  & 2.1 & 2.3   & 0.8& 1.0& 1.1& 1.0\\
$\la d[\sigma]\ra$  & 1.1   & 1.1    & 1.2& 1.3& 1.0& 1.3\\
\hline 
\hline 
$s^+(x,Q_0^2)$ &  $5~10^{-4} \le x \le 0.1$  & $10^{-5} \le x \le 10^{-4}$
 & &&&\\
\hline
$\la d[q]\ra$  & 9.4 & 1.1  & 2.1& 1.5& 1.6& 1.1\\
$\la d[\sigma]\ra$  & 3.4   & 1.6   & 1.5& 1.0& 1.5& 1.0 \\
\hline 
\hline 
$s^-(x,Q_0^2)$ & $0.1 \le x \le 0.6$  
& $3~10^{-3} \le x \le 3~10^{-2}$
& &&& \\
\hline
$\la d[q]\ra$  & 0.9 & 0.9  & 1.0& 1.1& 1.3& 1.1 \\
$\la d[\sigma]\ra$  & 1.4  & 1.2    & 1.0& 1.0 & 1.4& 0.9\\
\hline 
\end{tabular}

\end{center}
\caption{\small Distances between PDFs
computed from  a   set of 
$N_{\rm rep}=100$ replicas from the default NNPDF1.2 set,
and 100 replicas obtained using a ZM-VFN scheme instead of the default
I-ZM-VFN scheme of Sect.~\ref{sec:hq}, or introducing nuclear
corrections computed using the  
de~Florian-Sassot~\cite{deFlorian:2003qf} and
HKN07~\cite{Hirai:2007sx} models.
\label{tab:stabtab-checks}}
\end{table}

\subsection{Determination of the strange distribution}
\label{sec:strange}

The determination of the strange and antistrange PDFs is problematic
because of the scarceness of the experimental information on these
quantities, which makes it difficult to separate the genuine
information from theoretical bias, a situation which our methodology
is especially suited to deal with. In previous parton fits, a range
of possible shapes of the strange PDFs was explored by assuming
different functional forms and studying the variation of
results~\cite{Lai:2007dq}. 
%------------------------------------------------------------
\begin{figure}
\begin{center}
\epsfig{width=0.49\textwidth,figure=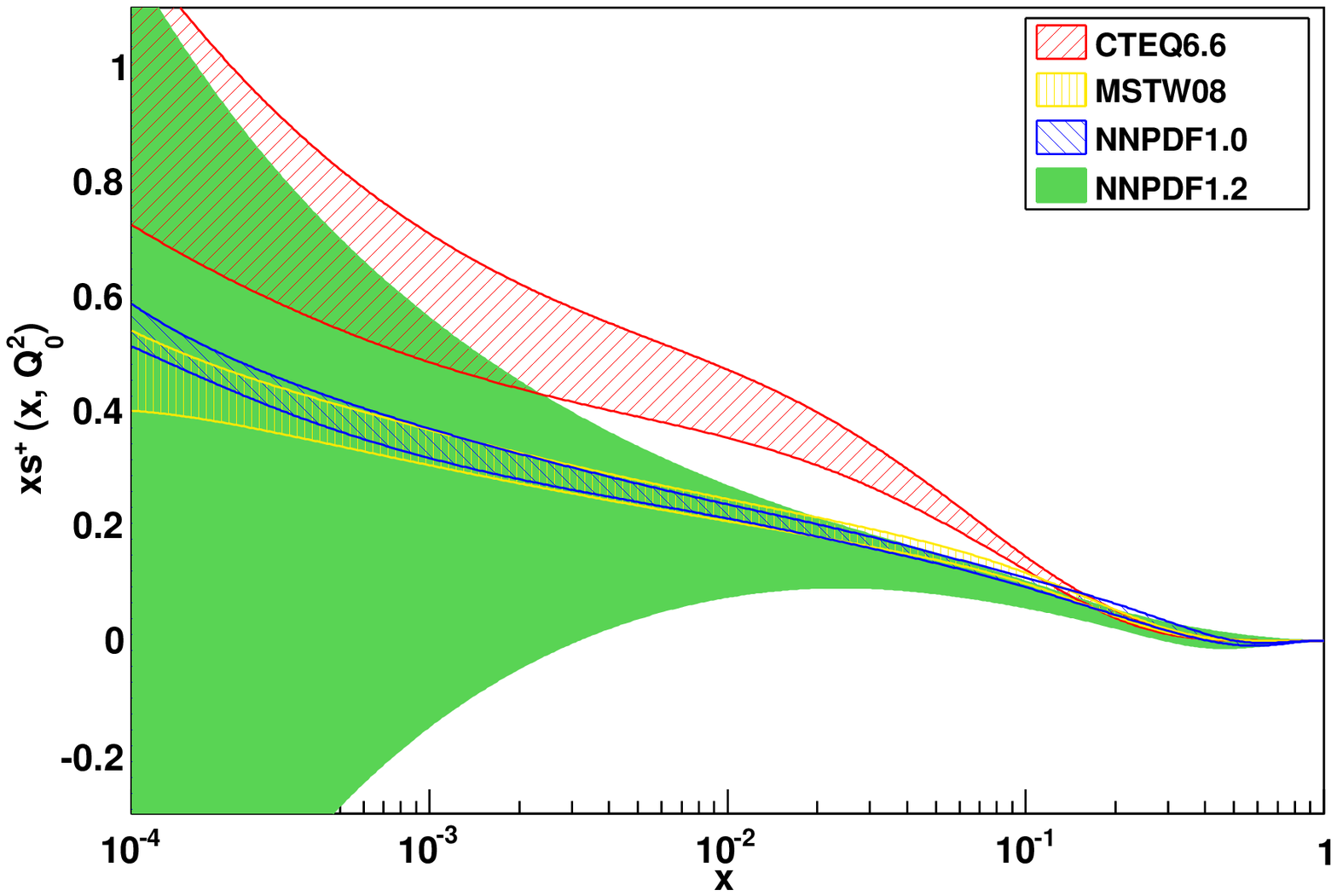}
\epsfig{width=0.49\textwidth,figure=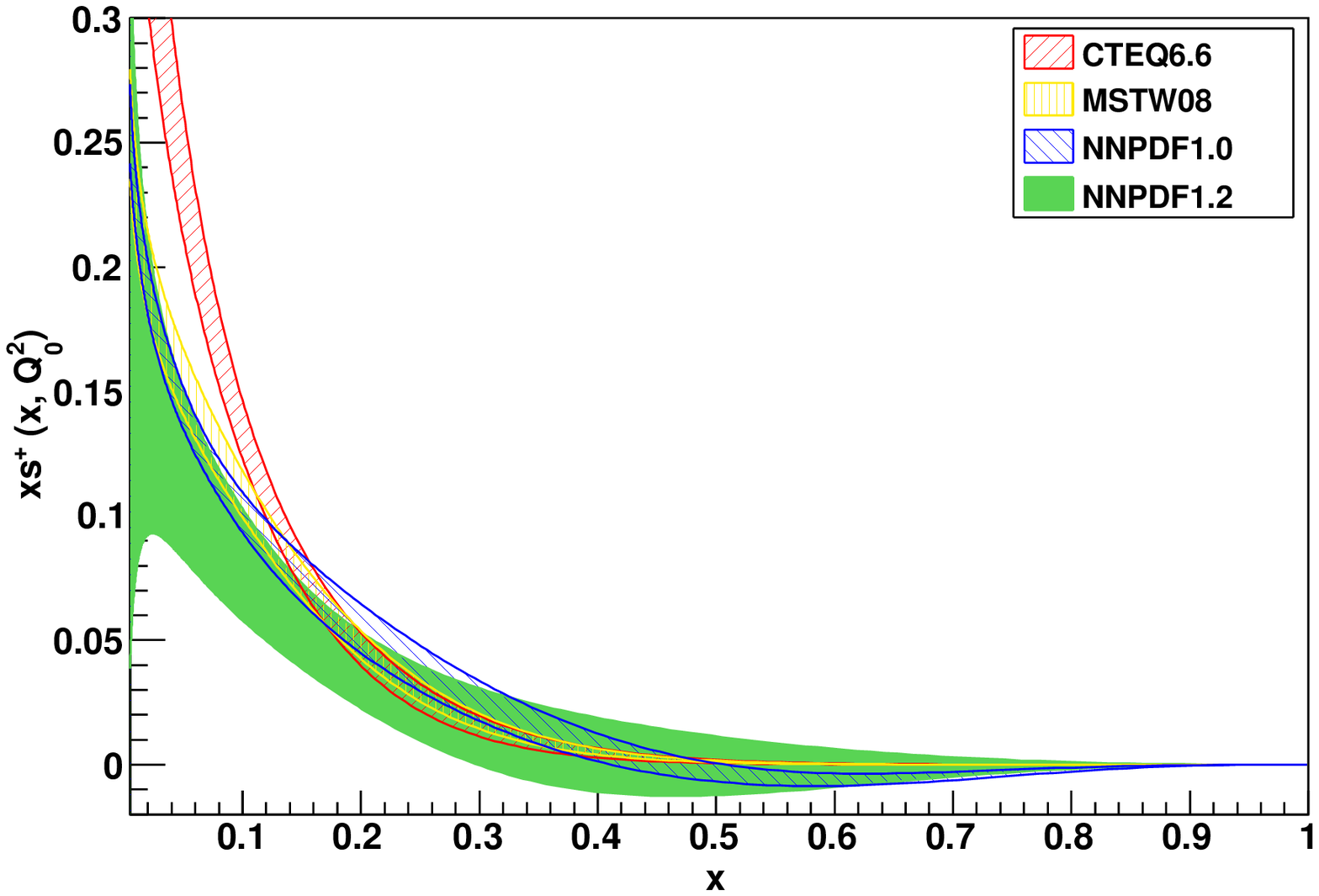} 
\epsfig{width=0.49\textwidth,figure=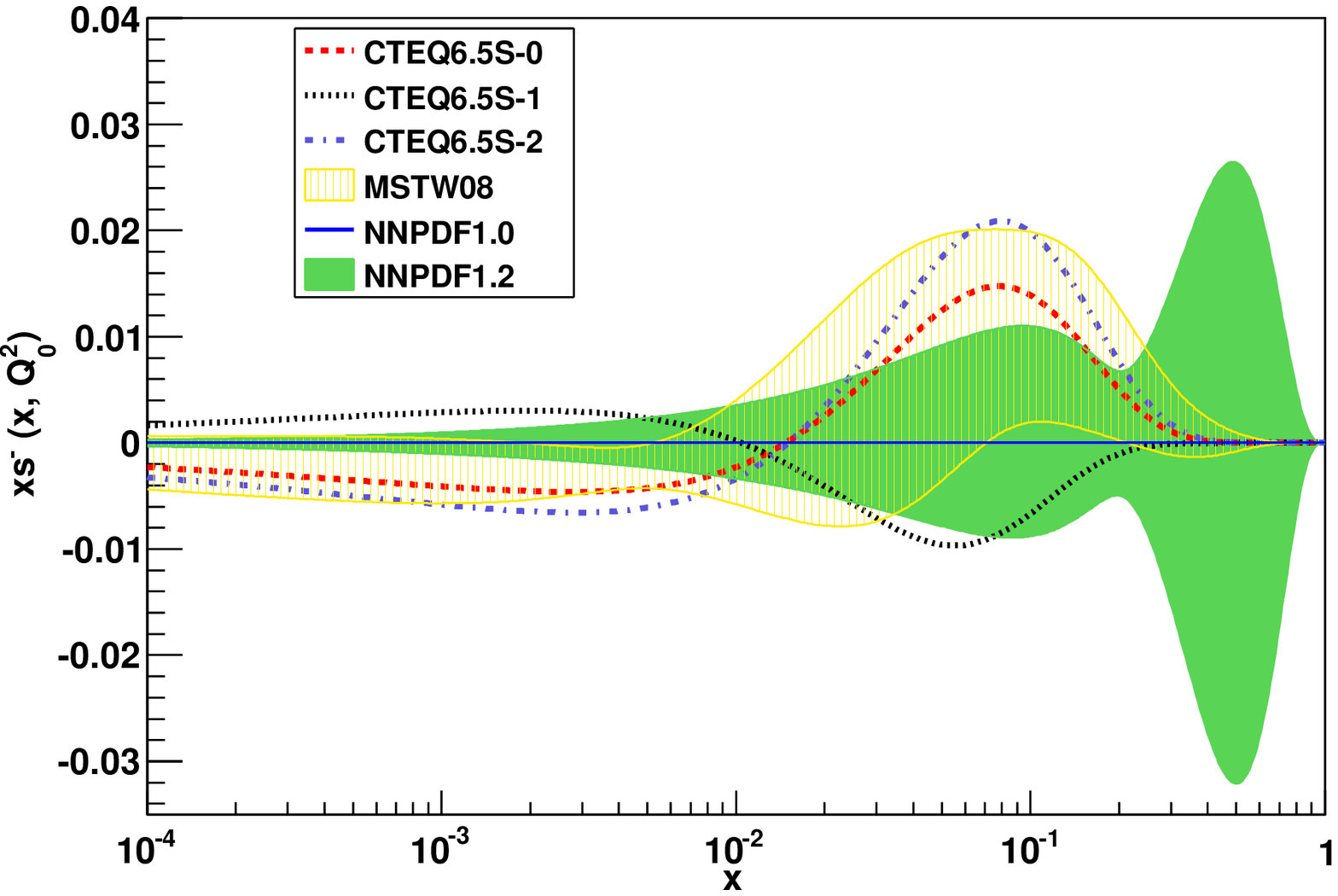}
\epsfig{width=0.49\textwidth,figure=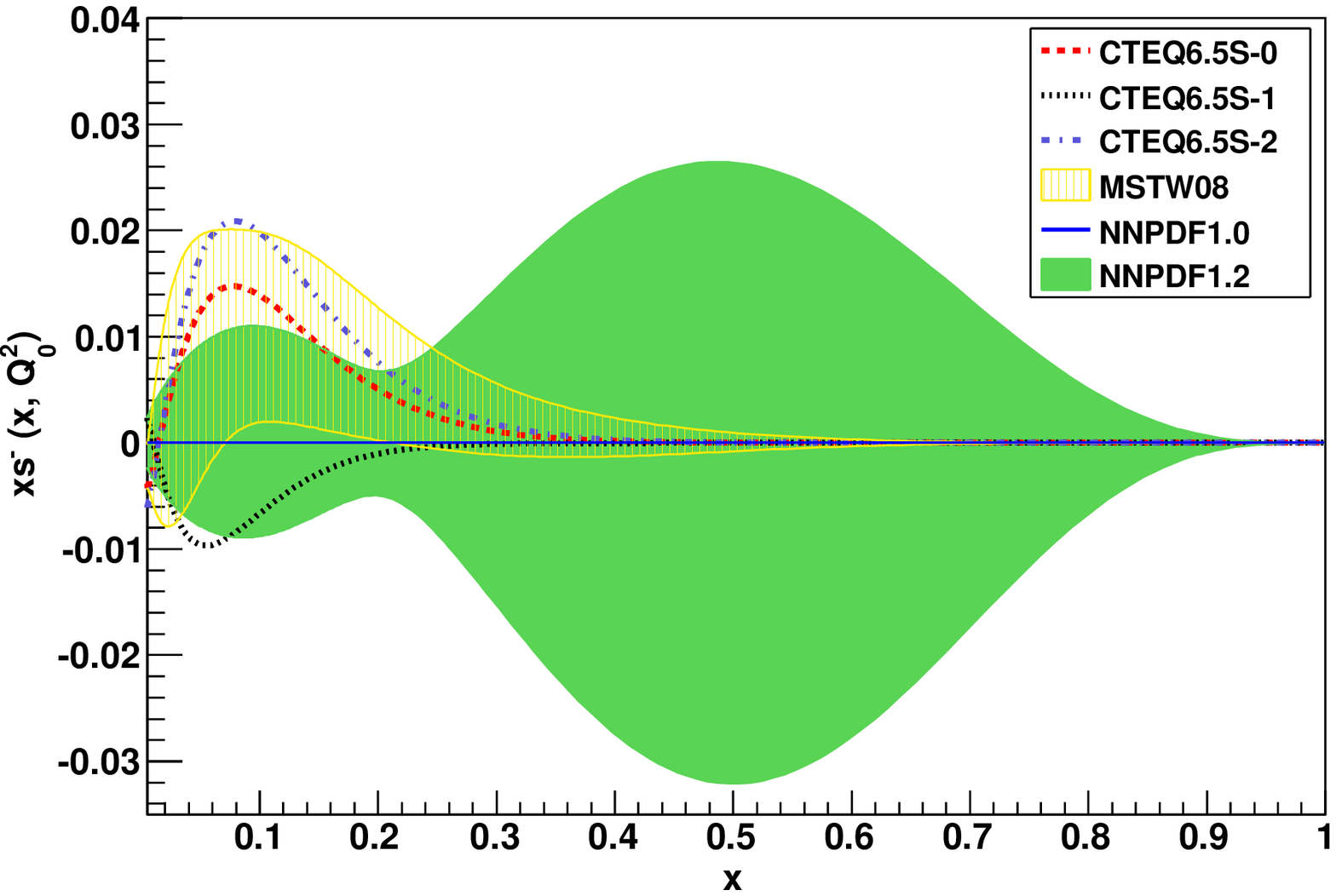} 
\epsfig{width=0.49\textwidth,figure=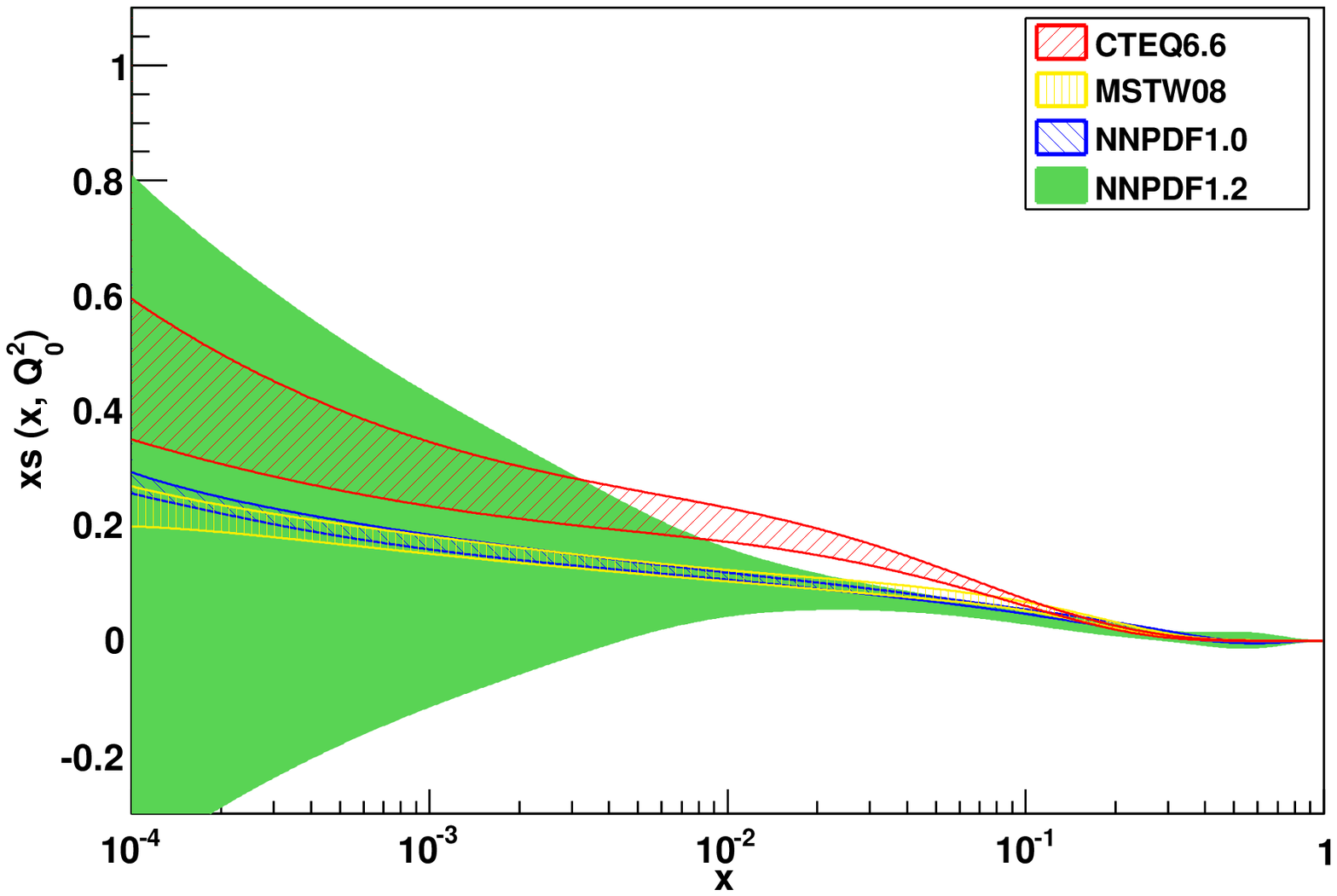}
\epsfig{width=0.49\textwidth,figure=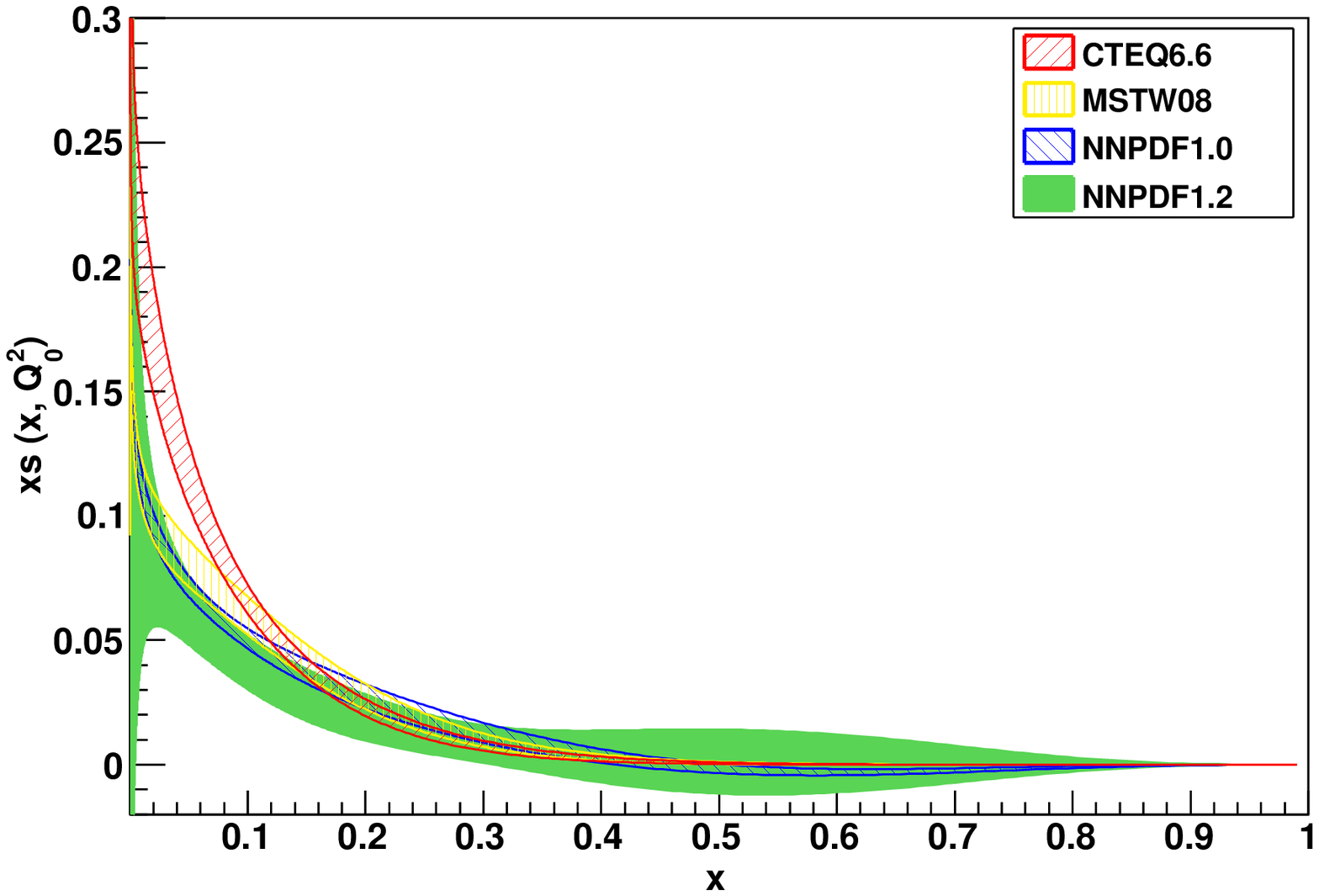} 
\epsfig{width=0.49\textwidth,figure=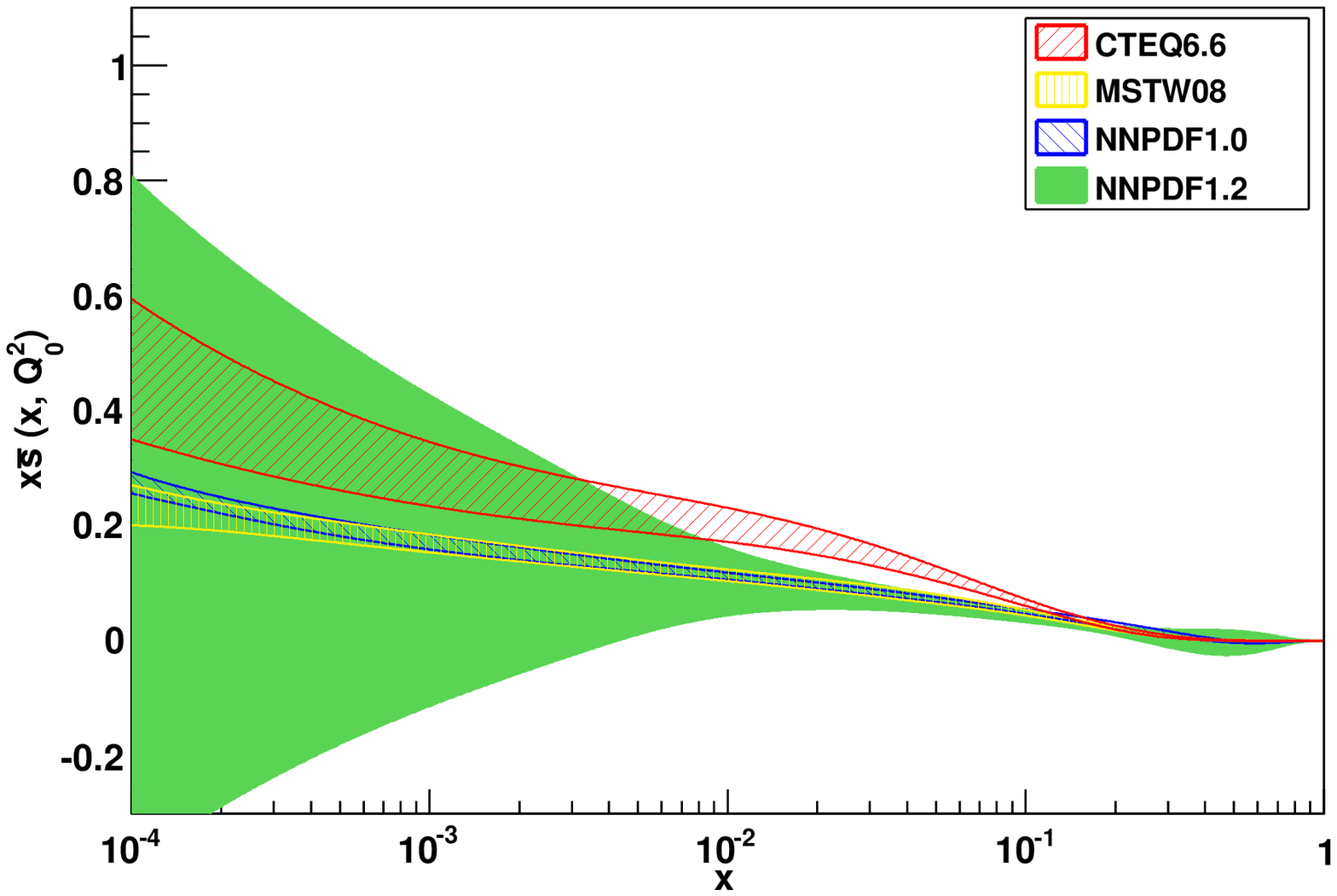}
\epsfig{width=0.49\textwidth,figure=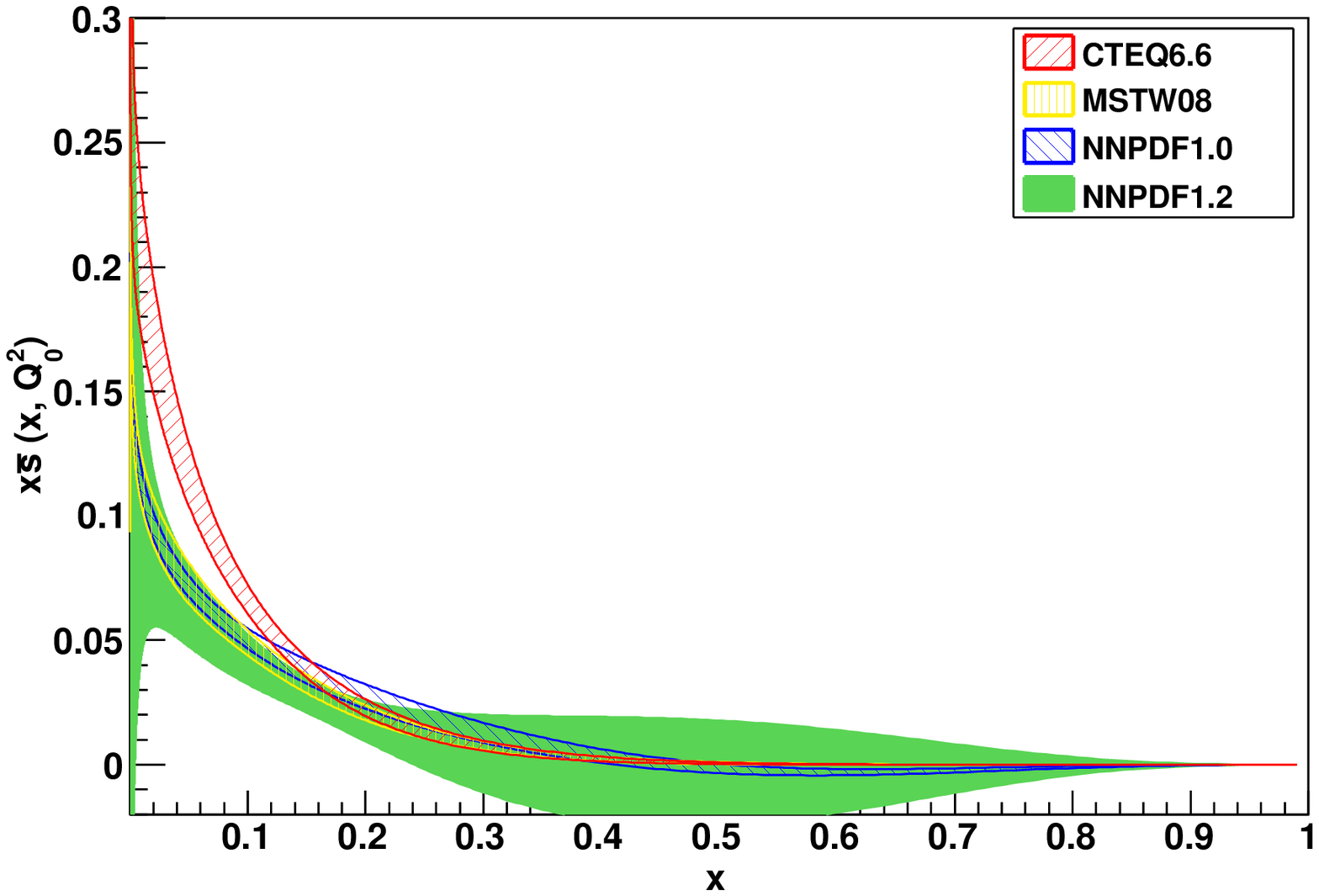} 
\end{center}
\caption{\small From top to bottom,
the strange C-even and C-odd combinations $s^+(x,Q_0^2)$,  $s^-(x,Q_0^2)$ 
Eq.~(\ref{spmdef}) and the corresponding strange  $s(x,Q_0^2)$ and antistrange 
$\bar s(x,Q_0^2)$ PDFs, plotted at the input scale versus $x$ 
on a log (left) or linear(right) scale, computed from the final set of 
$N_{\rm rep}=1000$ replicas.  
The NNPDF1.2 result is compared to the MSTW08~\cite{Martin:2009iq} and 
CTEQ6.6~\cite{Nadolsky:2008zw} global fits. For $s^-$ some of the results 
obtained from the CTEQ6.5s strangeness series~\cite{Lai:2007dq} are also 
shown.}
\label{fig:strangePDFs}
\end{figure}
%------------------------------------------------------------------

\begin{figure}
\begin{center}
\epsfig{width=0.49\textwidth,figure=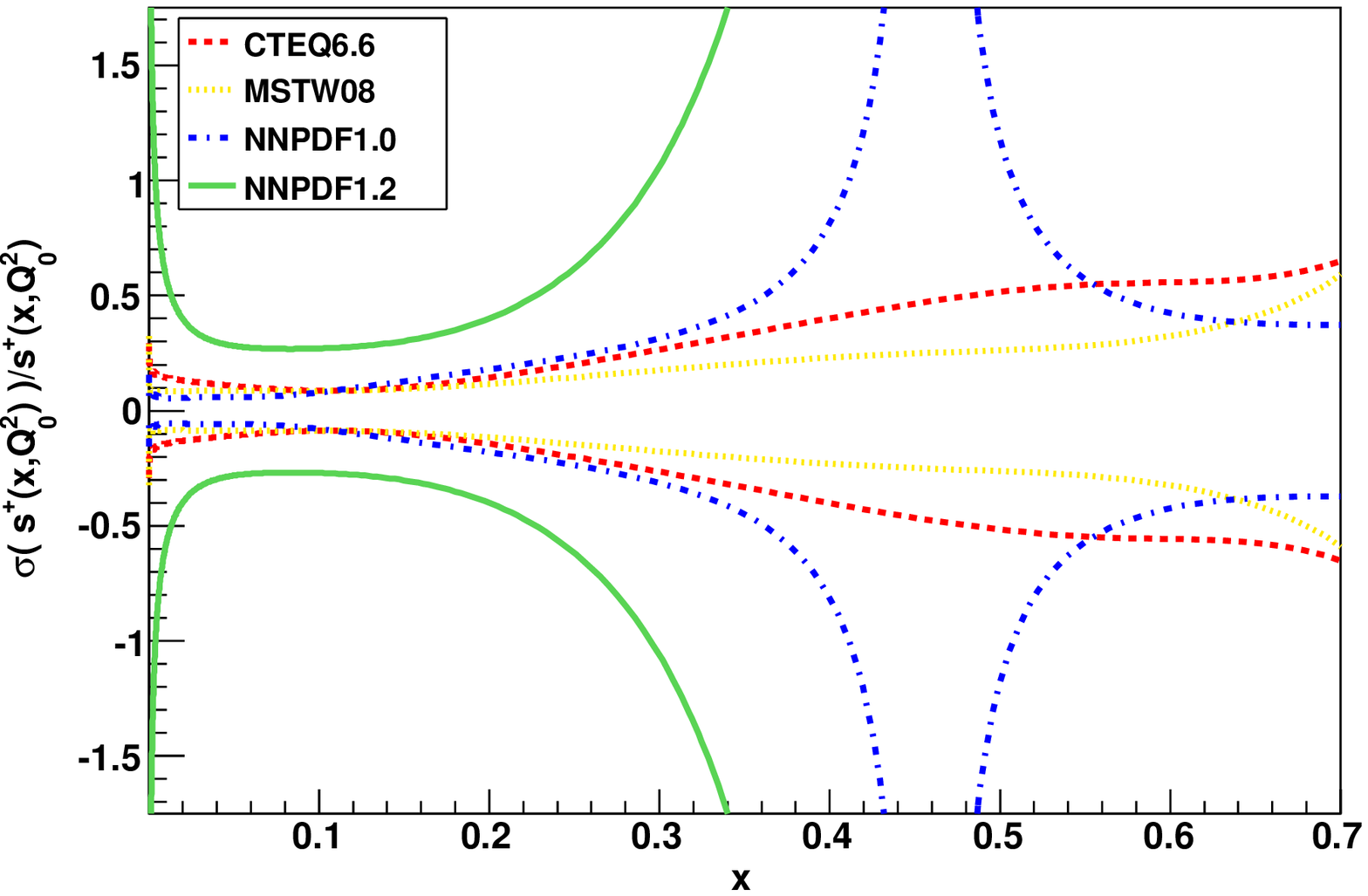}
\epsfig{width=0.49\textwidth,figure=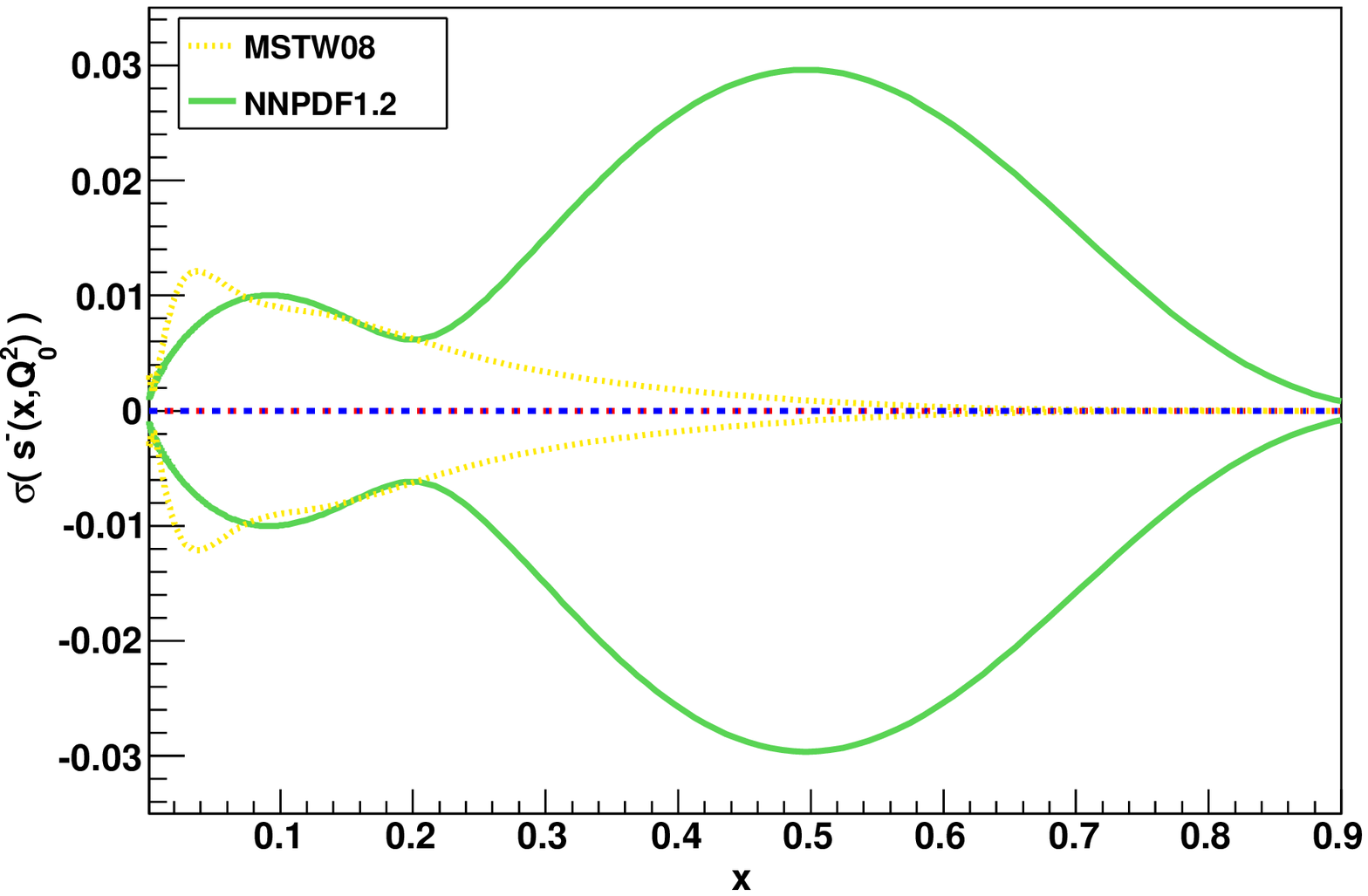} 
\end{center}
\caption{\small The uncertainty on the strange PDFs
  $s^\pm(x,Q_0^2)$
shown in Fig.~\ref{fig:strangePDFs}. All bands correspond to one $\sigma$.
The relative uncertainty is shown for $s^+$ (left) and the absolute
uncertainty for $s^-$ (right).}
\label{fig:strangePDFrel}
\end{figure}

The $s^\pm(x,Q_0^2)$, $s(x,Q_0^2)$ and
$\bar{s}(x,Q_0^2)$ strange PDFs Eq.~(\ref{spmdef}) are shown at the
input scale in Fig.~\ref{fig:strangePDFs}, where they are also
compared to the most recent CTEQ6.6~\cite{Nadolsky:2008zw} and
MSTW08~\cite{Martin:2009iq} sets.  Whereas the CTEQ collaboration has
not performed a full determination of the $s^-$ uncertainty band,
a study of the dependence of the  best-fit  $s^-$ on assumptions on
its functional form was performed in Ref.~\cite{Lai:2007dq}: several of
the corresponding results are also shown in
Fig.~\ref{fig:strangePDFs}. For greater clarity, in
Fig.~\ref{fig:strangePDFrel} we also plot the 
uncertainties on these PDFs.
%------------------------------------------------------------
%------------------------------------------------------------------

\begin{figure}[t!]
\begin{center}
\epsfig{width=0.49\textwidth,figure=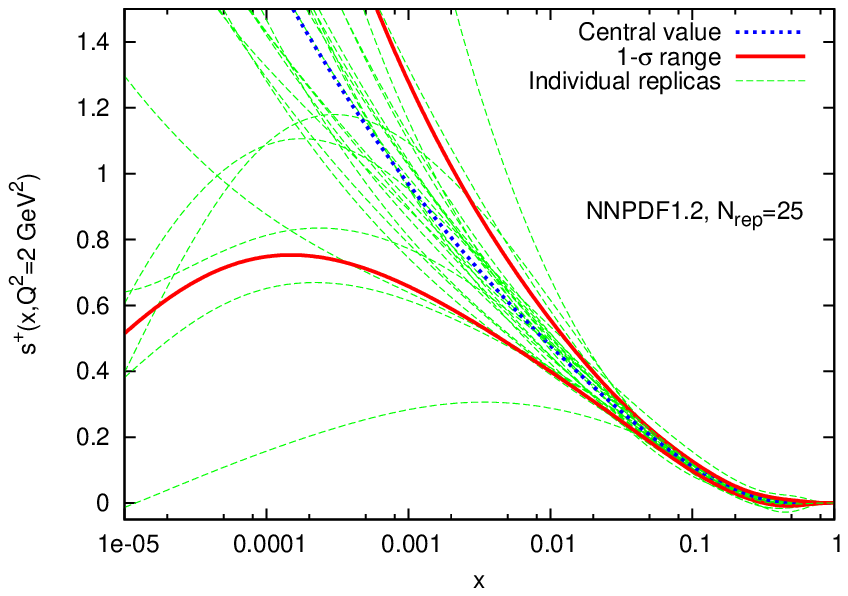} 
\epsfig{width=0.49\textwidth,figure=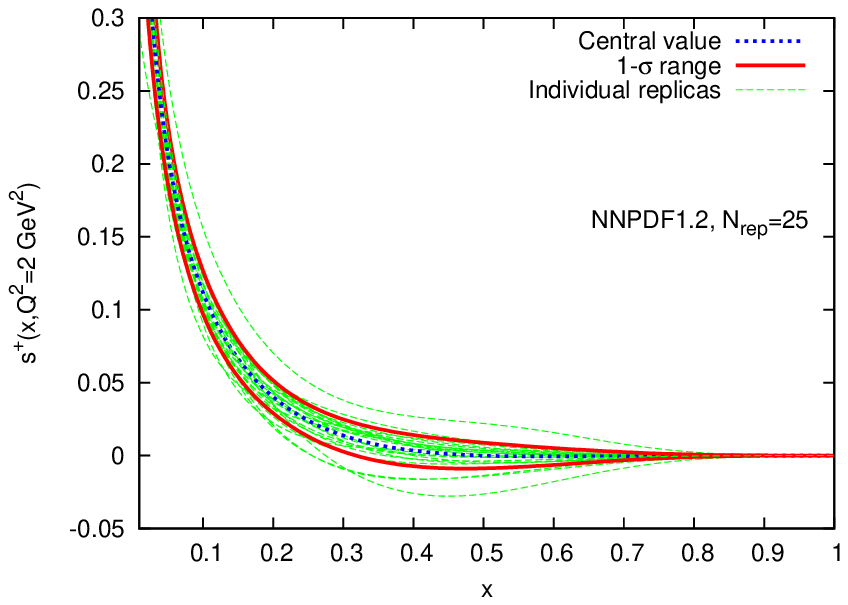} 
\epsfig{width=0.49\textwidth,figure=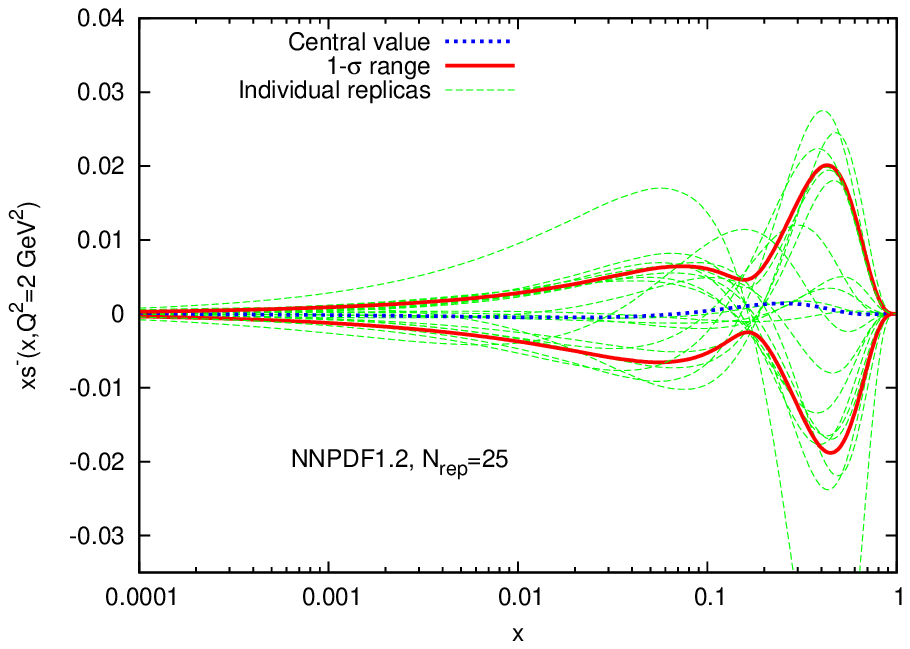} 
\epsfig{width=0.49\textwidth,figure=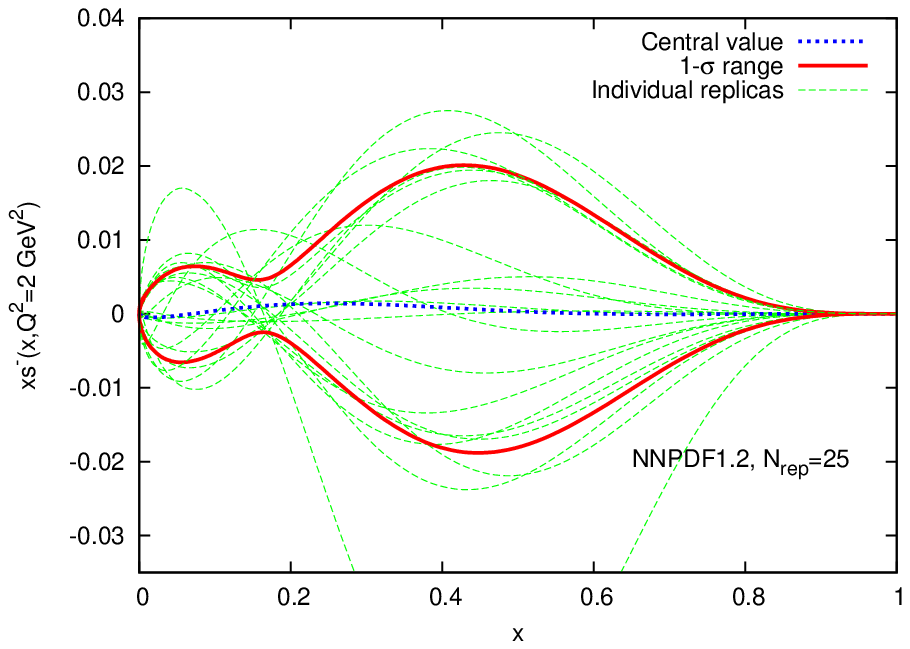} 
\caption{\small A set of randomly chosen $N_{\rep}=25$ replicas of the 
strange PDFs $s^+(x,Q_0^2)$,  $s^-(x,Q_0^2)$
 out of the full set of 
Fig.~\ref{fig:strangePDFs}, and the PDFs computed from them.}
\label{fig:strangePDFsreps}
\end{center}
\end{figure}

\begin{figure}[t!]
\begin{center}
\epsfig{width=0.49\textwidth,figure=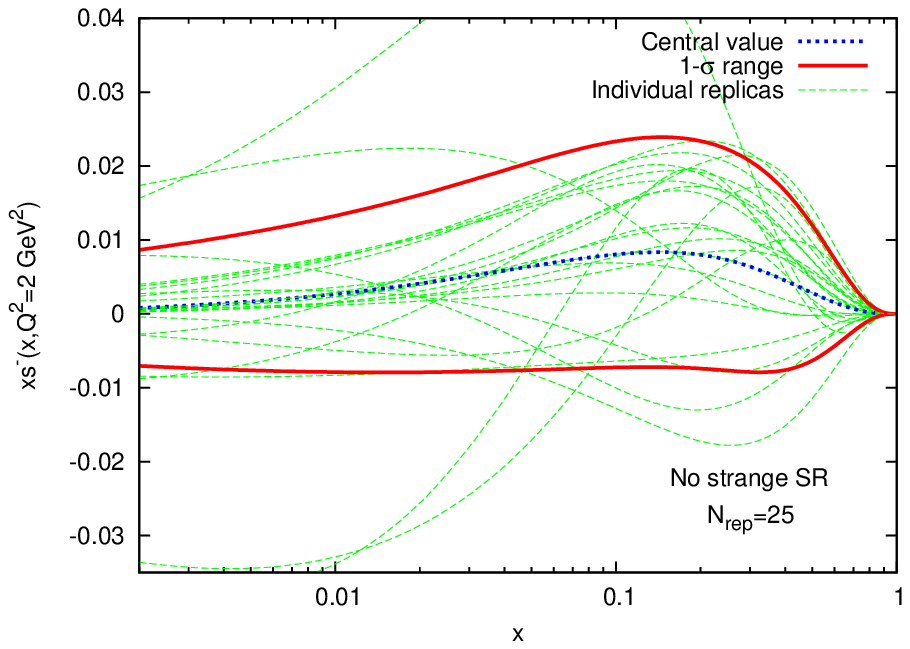} 
\epsfig{width=0.49\textwidth,figure=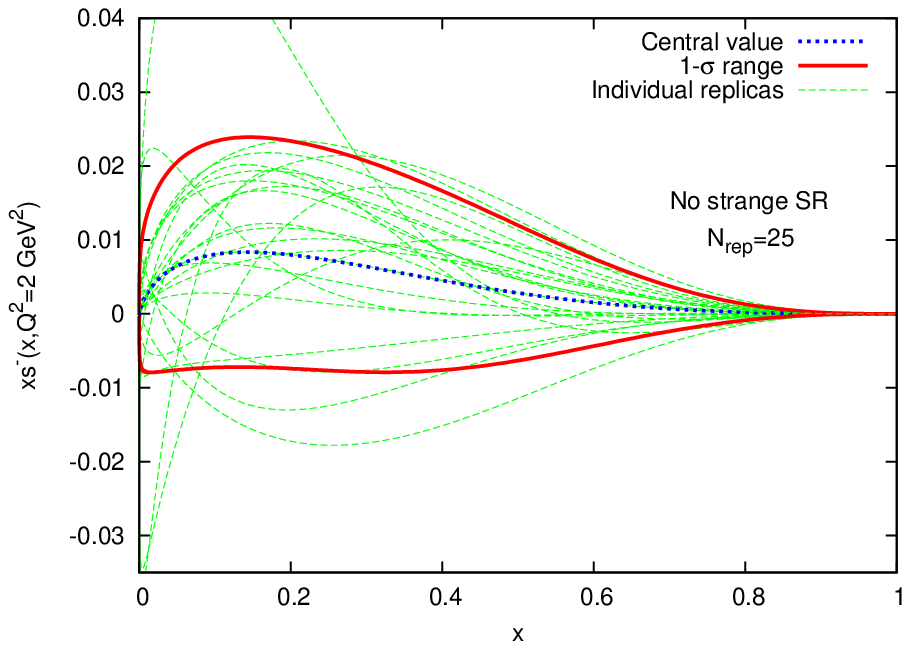} 
\caption{\small Same as the lower row of
  Fig.~\ref{fig:strangePDFsreps} when the sum rule Eq.~(\ref{eq:sr}) is
  not imposed.}
\label{fig:strangePDFnosr}
\end{center}
\end{figure}

In the data region $x\gsim0.03$ all determinations of $s^\pm$ agree, however
the NNPDF1.2 has a much larger uncertainty than other existing
determinations. The origin of this can be understood by looking at
Fig.~\ref{fig:strangePDFsreps}, where we display 25 randomly chosen replicas
out of our full set, and the mean and standard deviation computed from
them: clearly, our large uncertainty is a consequence of the great 
flexibility afforded by the neural network parametrization. This 
is particularly noticeable in the case of  $s^-$,
which must have at least one node because of the sum rule
Eq.~(\ref{eq:sr}): individual replicas cross the $x$--axis in different
places, with different  sign (from positive to negative or
conversely), and some replicas have more than one crossing. 
It is interesting to observe that the ``neck'' in the uncertainty on
$s^-$ around $x\approx 0.1$ corresponds to the value of $x$ at which
the crossing is most likely to occur. The role played by the valence
sum rule Eq.~(\ref{eq:sr}) in determining these features of the
strangeness asymmetry $s^-$ can be elucidated by repeating the fit without
imposing it. The results,  displayed in
Fig.~\ref{fig:strangePDFnosr}, show that even without the sum rule 
constraint many replicas still cross the the $x$--axis.

%------------------------------------------------------------
%------------------------------------------------------------------

\begin{figure}
\begin{center}
\epsfig{width=0.49\textwidth,figure=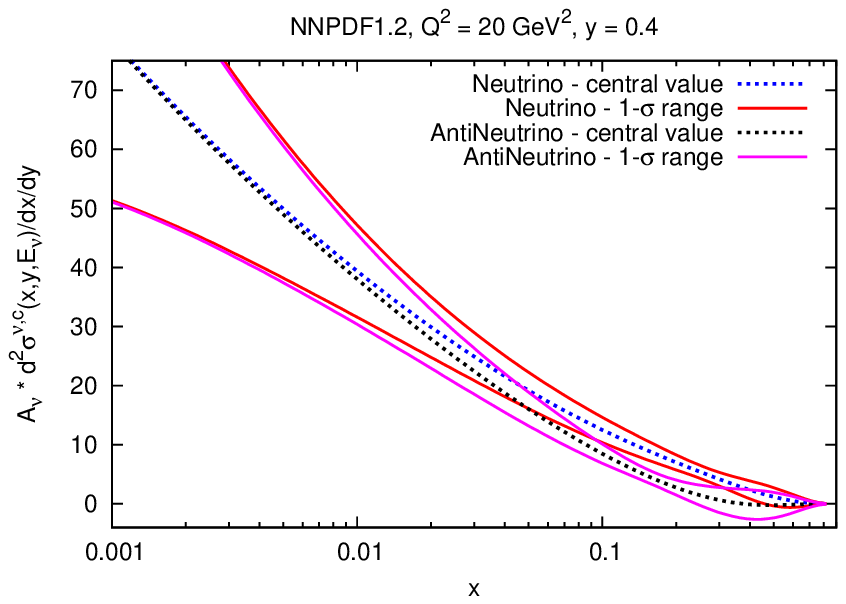}
\epsfig{width=0.49\textwidth,figure=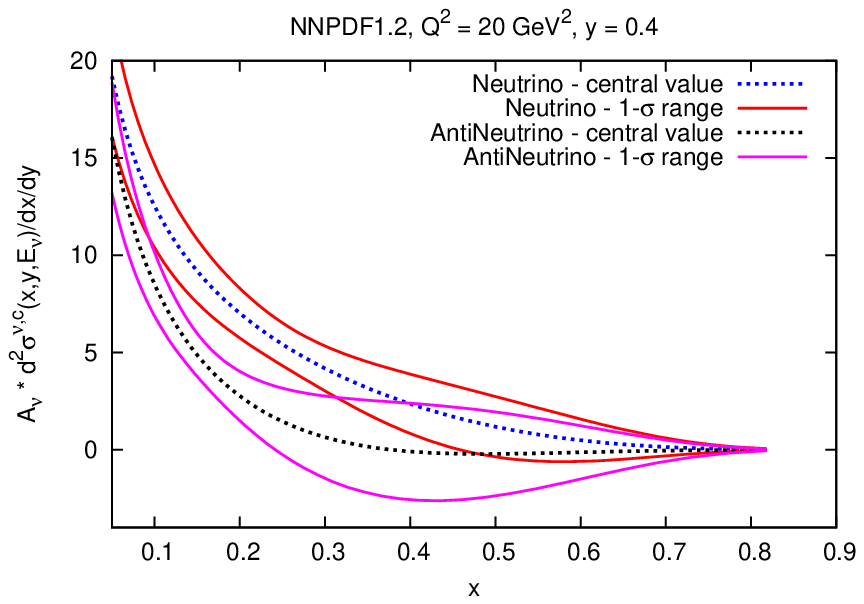} 
\epsfig{width=0.49\textwidth,figure=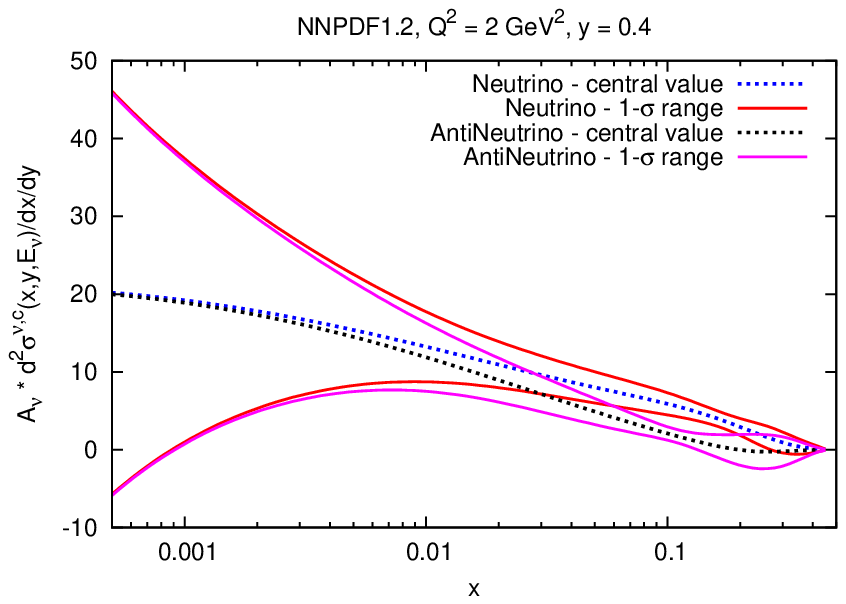}
\epsfig{width=0.49\textwidth,figure=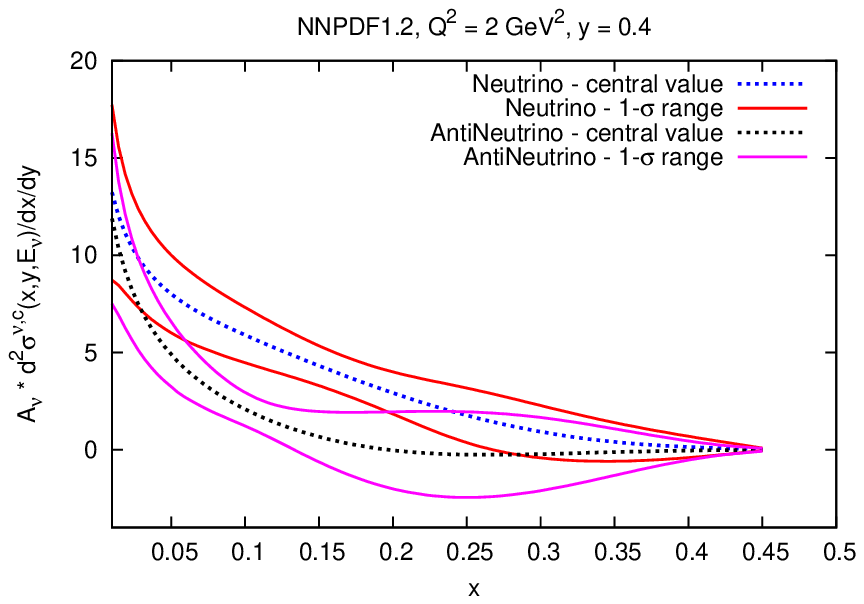} 
\end{center}
\caption{\small The total neutrino and antineutrino 
dimuon cross sections at the
starting scale
  $Q^2_0=2$~GeV$^2$ (lower row) and at the ``NuTeV'' 
scale $Q^2_0=20$~GeV$^2$ (upper row), 
plotted versus $x$ on a log (left) or linear
  (right) scale.}
\label{fig:positivity}
\end{figure}

Another theoretical constraint which may help in reducing
uncertainties is that of positivity of cross sections. For instance,
 as in
Ref.~\cite{Ball:2008by}, in the determination of the NNPDF1.2 PDF set we have imposed positivity of the structure
function $F_L$ at low $x$ and $Q^2$, which helps in reducing the
uncertainty of the gluon distribution at the edge of the HERA data
region. In view of the fact that (see Fig.~\ref{fig:strangePDFs}) 
both $s(x,Q_o^2)$ and $\bar s(x,Q_o^2)$  can turn negative to within one
sigma for $x\lsim 10^{-2}$, and also in the large $x\gsim 0.2$ region,
one may wonder whether imposing positivity of the  dimuon cross
section might likewise help in   reducing the uncertainty on the strange
and antistrange
distributions. In order to test this, in Fig.~\ref{fig:positivity} we display the total
dimuon cross section, both at the initial $Q^2=Q_0^2=2$~GeV$^2$  and at the
typical scale of the NuTeV data $Q^2=20$~GeV$^2$,
 computed using the 
NNPDF1.2 PDFs of Fig.~\ref{fig:strangePDFs}. The cross section only
becomes significantly  negative at low $Q^2$ and
very low $x\lsim10^{-5}$. For antineutrinos, it also become 
somewhat negative at large $x$: at the scale of the large $x$ data
$Q^2\gsim20$~GeV$^2$ for  $x\gsim 0.3$. We conclude that  the
constraint of positivity only affects physical observables quite far
from the data region.  
We have thus not imposed this
constraint in the current fit. It might be worth implementing  it in
future fits which include Drell-Yan data, as these could further
constrain strangeness, especially at large $x$. 

Further constraints
could be based on theoretical expectations: for example, one may
expect the strange PDF to be smaller than the light quark valence
PDFs; indeed, the systematic implementation of theoretical or model
constraints  in parton fits has been advocated e.g. in
Ref.~\cite{Honkanen:2008mb}. However, expectations based on models of
the nucleon have often turned out to be in disagreement with experiment: for
instance, in the polarized case the strange distribution turns out to
be unexpectedly large and in fact larger than the up distribution (see
e.g. Ref.~\cite{deFlorian:2008mr}). To obtain reliable phenomenology,
such as the determination of electroweak parameters to be discussed
below, we prefer therefore to only rely on exact  constraints, such as
the valence sum rule or positivity.

%------------------------------------------------------------
%------------------------------------------------------------------

%------------------------------------------------------------
%------------------------------------------------------------------

The features of the strange distributions which are most interesting
for physics applications (as we shall discuss in more detail in
Section~\ref{sec:ewparam}) are the momentum fractions,
defined as
\be
 \lc S^\pm\rc(Q^2) \equiv \int_0^1 dx x s^\pm(x,Q^2)\ , 
\label{eq:strangmomdef}
\ee
with similar definitions for moments of other PDF combinations,
and in particular their ratio to the light sea or respectively light
valence momentum
fractions:
\bea
\label{eq:mom1}
K_S(Q^2)&\equiv &\frac{\int_0^1 dx~x~
s^+\lp x,Q^2\rp
}{\int_0^1 dx~x\lp \bar{u}\lp x,Q^2\rp+
\bar{d}\lp x,Q^2\rp\rp}=\frac{\lc S^+\rc}{\lc \bar{U}+\bar{D}\rc} 
\ ,\\\label{eq:rdef}
R_S(Q^2)&\equiv&
2 \frac{\int_0^1 dx xs^-(x,Q^2)}
{\int_0^1 dx x\lp u^-(x,Q^2) +  d^-(x,Q^2) \rp}
=2\frac{\lc S^-\rc}{\lc U^-+D^-\rc}  \ .
\eea
In many parton fits, including the NNPDF1.0 fit, 
these quantities are
taken to be fixed at the starting scale: the value of the relative 
total strange momentum (sometimes also
called strange suppression) is, since the earliest
measurements, taken to be~\cite{Abramowicz:1982zr} 
$K_S(Q_0^2)\approx 0.5$, while the strange asymmetry is assumed to vanish, i.e. 
$R_S(Q_0^2)=0$.  

%------------------------------------------------------------
\begin{figure}
\begin{center}
\epsfig{width=0.83\textwidth,figure=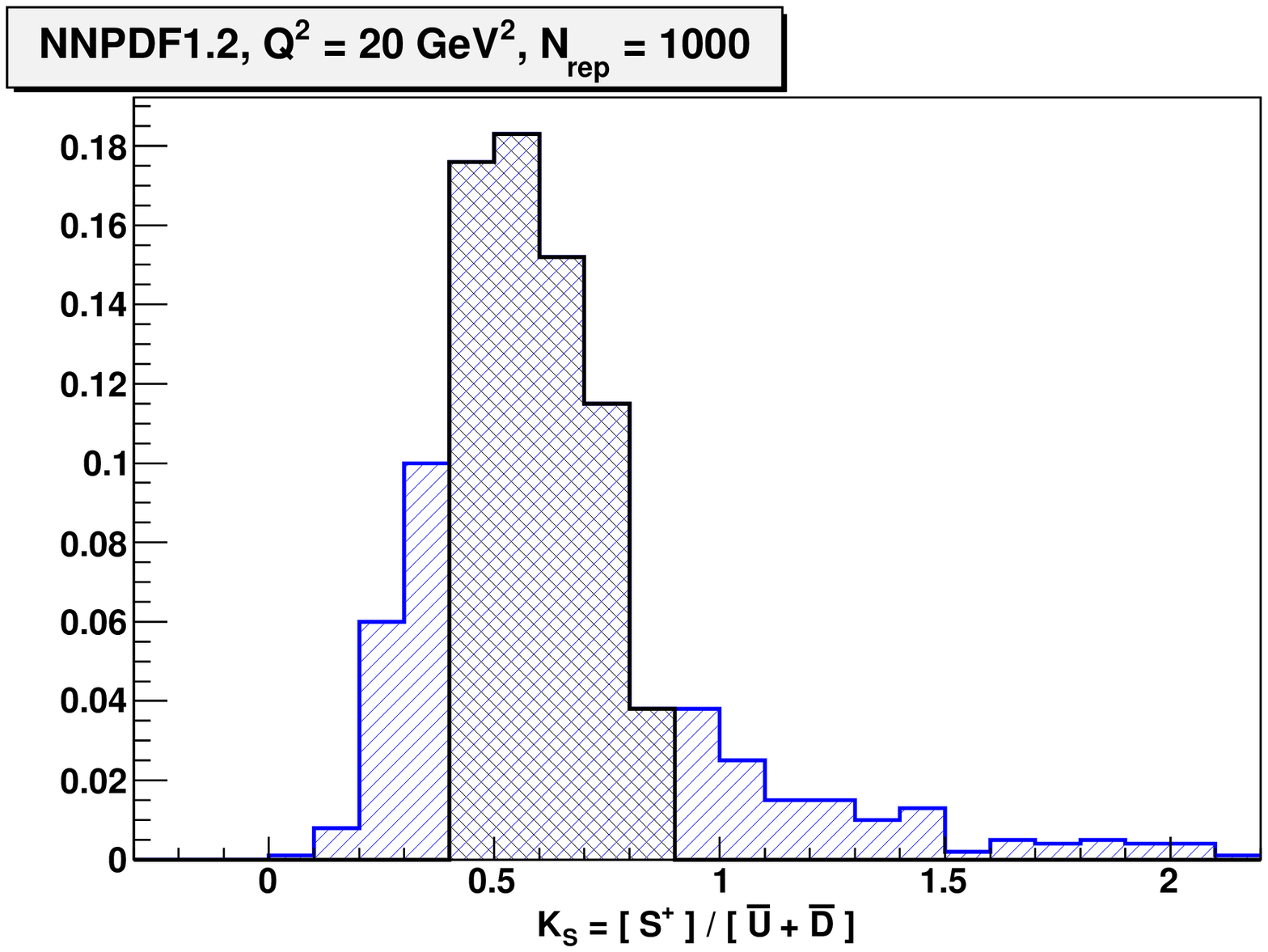} 
\end{center}
\caption{\small Probability distribution of
$K_S$  at $Q^2=20$~GeV$^2$ computed from the reference set of  $N_{\rm
    rep}=1000$ NNPDF1.2 PDF replicas. The central cross-hatched region
 corresponds to the central  68\% confidence 
interval, $K_S\lp Q^2=20 \,{\rm GeV}^2\rp=0.71^{+ 0.19}_{-0.31}{}^{\rm stat}$.
}
\label{fig:strangePDFdists}
\end{figure}
%------------------------------------------------------------------
\begin{table}
 \begin{center}
 \begin{tabular}{|c|c|c|}
  \hline
Analysis & Reference & $K_S\lp Q^2=20 \,{\rm GeV}^2\rp$  \\ \hline
NNPDF1.2 & This work &  $0.71^{+ 0.19}_{-0.31}$ \\
MSTW08 & \cite{Martin:2009iq} & $0.56\pm 0.03$ \\
CTEQ66 & \cite{Nadolsky:2008zw} & $0.72\pm 0.05$ \\
AKP08 & \cite{Alekhin:2008mb} & $0.59\pm 0.08 $\\
 \hline
 \end{tabular}
\end{center}
\caption{\small The relative strange momentum fraction
$K_S(Q^2)$ Eq.~(\ref{eq:mom1}), as determined from various parton sets.
All uncertainties   correspond to 68\% confidence levels.
\label{tab:srat}}
 \end{table}

The value and uncertainty on these quantities can be determined from
the NNPDF1.2 set by performing averages over
replica PDFs~\cite{Ball:2008by}, which for $K_S$ and $R_S$ 
will not necessarily coincide with the ratio of average PDFs, because
Eqs.~(\ref{eq:mom1}-\ref{eq:rdef}) are not linear in the PDFs.
In fact, because the denominator in Eq.~(\ref{eq:mom1}) can assume
rather small values, we expect that the distribution of values of the total
strange fraction  $K_S$ can be
rather asymmetric and non-gaussian. The probability distribution of
$K_S$ at $Q^2=20$~GeV$^2$ is shown in Fig.~\ref{fig:strangePDFdists},
and turns out to be indeed quite far from gaussian. Therefore, we
compute the one-$\sigma$ uncertainty as a central 68\% confidence
integral, namely requiring the two outer tails of the
probability distribution  (lighter blue region in
Fig.~\ref{fig:strangePDFdists}) to each correspond to 16\%
probability, with the central value still given by the average. 
The result we thus obtain for the expected $K_S$ and its uncertainty
are shown in
Table~\ref{tab:srat}, along with the results found using other
parton sets. The
median of the probability distribution is equal to $K_S^{\rm
  med}=0.59$, significantly different from the average because of the
asymmetry.  
%(which are all symmetric since they correspond to Hessian
%uncertainties~\cite{Pumplin:2001ct,Martin:2002aw}).
The NNPDF1.2 uncertainty is much larger than that found in other
fits, for the reasons discussed above. Note that, however, all values are
essentially consistent with the simple assumption $K_S=0.5$ used in
older parton fits.

%------------------------------------------------------------
\begin{figure}
\begin{center}
\epsfig{width=0.80\textwidth,figure=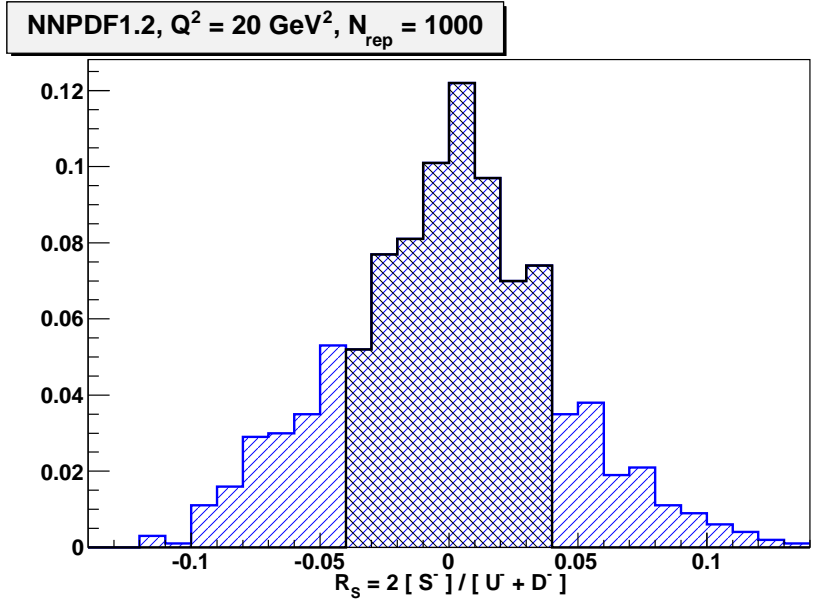} 
\end{center}
\caption{\small Probability distribution of
$R_S$  at $Q^2=20$~GeV$^2$ computed from the reference set of  $N_{\rm
    rep}=1000$ NNPDF1.2 PDF replicas.  The central cross-hatched region
corresponds to the central  68\% confidence 
interval, $R_S\lp Q^2=20 \, {\rm GeV}^2\rp=0.006\pm 0.045^{\rm stat}$.}
\label{fig:weinsrdists}
\end{figure}
%------------------------------------------------------------------

%--------------------------------------------------
 \begin{table}
 \begin{center}
 \begin{tabular}{|c|c|c|c|c|}
  \hline
  Analysis & Reference & $\lc S^-\rc(Q^2)\cdot 10^3$ & $Q^2$ [GeV$^2$] & 
$\lc S^-\rc\lp Q^2_{\rm ref}=20~{\rm GeV}^2\rp \cdot 10^3$ \\ \hline
 NNPDF1.2 & This work &  $0.5\pm 8.6$ & $20$ & $0.5\pm 8.6$  \\
MSTW08 & \cite{Martin:2009iq} & $2.4 \pm 2.0$ & 1 & $1.7\pm 1.4$ \\
CTEQ6.5s & \cite{Lai:2007dq} & $2.0\pm 1.8$ 
& 1.69 & $1.6\pm 1.4$ \\
CTEQ6.1s & \cite{Olness:2003wz} & $1.5 \pm 1.5$ & 1.69 & $1.2\pm 1.2$ \\
AKP08 & \cite{Alekhin:2008mb} & $1.0\pm 1.3 $ & 20 &
$1.0\pm 1.3 $ \\
NuTeV07 & \cite{Mason:2007zz} & $2.2 \pm 1.3 $ &
16 & $2.2 \pm 1.3$\\
BPPZ03 & \cite{Barone:2006xj}& $ 1.8 \pm 3.8$ & 20 &
$ 1.8 \pm 3.8$\\
 \hline
 \end{tabular}
\end{center}
\caption{\small 
Comparison of various determinations of strangeness momentum asymmetry
$\lc S^-\rc(Q^2)$ Eq.~(\ref{eq:strangmomdef}).
All uncertainties   correspond to 68\% confidence levels. Both the
published value is given, and the value obtained evolving to
$Q^2_{\rm ref}=20~{\rm GeV}^2$ through NLO perturbative evolution.
\label{tab:sasymm}}
 \end{table}

In the case of the strange momentum asymmetry $R_S$
Eq.~(\ref{eq:rdef}) the denominator is fixed by knowledge of the valence
content of the nucleon, which is known quite accurately: hence we
expect the uncertainty to be symmetric and dominated by uncertainty
of the numerator. Indeed, the probability distribution for $R_S$,
shown in
Fig.~\ref{fig:weinsrdists}, turns out to be approximately gaussian so
that the uncertainty computed from the central  68\% confidence essentially
coincides with the standard deviation of the distribution, while
central value and uncertainty for $R_S$ are
essentially proportional to those of the strangeness asymmetry
$\lc S^-\rc$. This latter quantity has been determined by various groups, at
various scales: several of these results are collected in
Table~\ref{tab:sasymm} and compared to our own. Results are given both
at the scale at which they were determined, and then also when evolved to 
a common scale, exploiting the fact that at NLO (though not at
NNLO~\cite{Catani:2004nc}) $\lc S^-\rc$ evolves multiplicatively. In
  this case, too, the NNPDF1.2 uncertainty is much larger than that
  obtained in other
fits: while for all other groups  there is an indication that a
positive value of $\lc S^-\rc$ is favored (all results being
nevertheless compatible with zero), this indication looses its
significance in our analysis due to the very large uncertainty.

In the next section we will see that, surprisingly, even with such
large uncertainties it is possible to exploit our determination of
$K_S$ and $R_S$ for a determination of electroweak parameters. In view
of this, it is useful to also study possible sources of systematic
uncertainty on these quantities. Possible significant sources of
systematics are the following:

\begin{itemize}
\item Heavy quark mass effects. The treatment of heavy quark mass
  effects entails various ambiguities related to the prescription
    used to deal with subleading terms~\cite{Thorne:2008xf}. 
In our case, a further source of systematics is due to the fact that
the charm quark mass is treated
  approximately, using the I-ZM-VFN scheme as discussed in
  Sect.~\ref{sec:hq}, and then only for dimuon data. The corresponding
  uncertainty is conservatively estimated by repeating the fit in a
  pure ZM-VFN scheme. 
\item Nuclear corrections. Their effect is estimated by repeating the
  fit with CHORUS and NuTeV data corrected using  
the de~Florian-Sassot~\cite{deFlorian:2003qf} and
HKN07~\cite{Hirai:2007sx} models.
\item Higher order QCD corrections. These are very conservatively
  estimated by repeating the fit at LO.
\end{itemize}

 \begin{table}
 \begin{center}
 \begin{tabular}{|c|c|c|c|}
  \hline
    &  $ K_S$ (mean) & $R_S$ 
  \\ \hline
 Reference & $0.71^{+ 0.19}_{-0.31}$  &  
 $\lp 6 \pm 45\rp\cdot 10^{-3}$  \\\hline
ZM-VFN & $0.47^{+ 0.10}_{-0.20}$ &
 $\lp 8 \pm 39\rp\cdot 10^{-3}$ \\
Nuclear - dFS03 & $0.74^{+ 0.21}_{-0.40}$ 
 & $\lp 12 \pm 48\rp\cdot 10^{-3}$  \\
Nuclear  - HKN07 & $0.68^{+ 0.24}_{-0.29}$
 & $\lp 0 \pm 40\rp\cdot 10^{-3}$  \\
LO  &  $0.61^{+ 0.33}_{-0.22}$ & 
 $\lp 1 \pm 38\rp\cdot 10^{-3}$     \\\hline
No strange SR &  $0.62^{+ 0.20}_{-0.21}$ & 
 $\lp 17 \pm 32\rp\cdot 10^{-3}$ \\
 \hline
 \end{tabular}
\end{center}
\caption{\small The strange relative total and 
valence momentum fractions  $K_S$
  and $R_S$, Eqs.~(\ref{eq:mom1},\ref{eq:rdef}),
 at the scale  $Q^2=20$ GeV$^2$. The first row gives the
  value computed from the reference NNPDF1.2 set of $N_{\rm rep}=1000$
  replicas, while the other rows give results from sets of $N_{\rm
    rep}=100$ replicas each obtained from alternative fits discussed
  in the text. All uncertainties are one-$\sigma$ or 68\% central
  confidence intervals.  
\label{tab:strangecheck}}
 \end{table}

%------------------------------------------------------------
\begin{figure}
\begin{center}
\epsfig{width=0.98\textwidth,figure=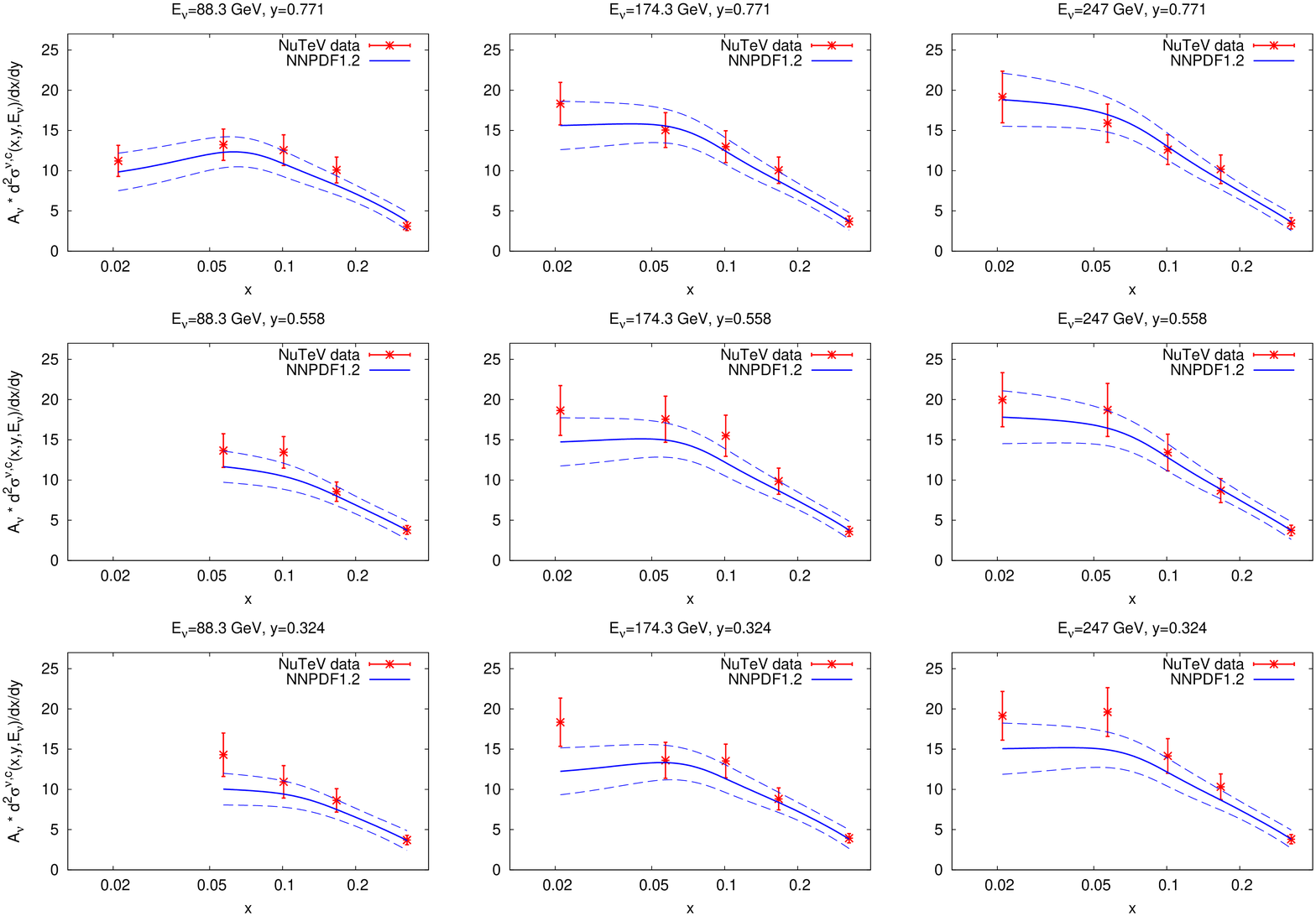} 
\epsfig{width=0.98\textwidth,figure=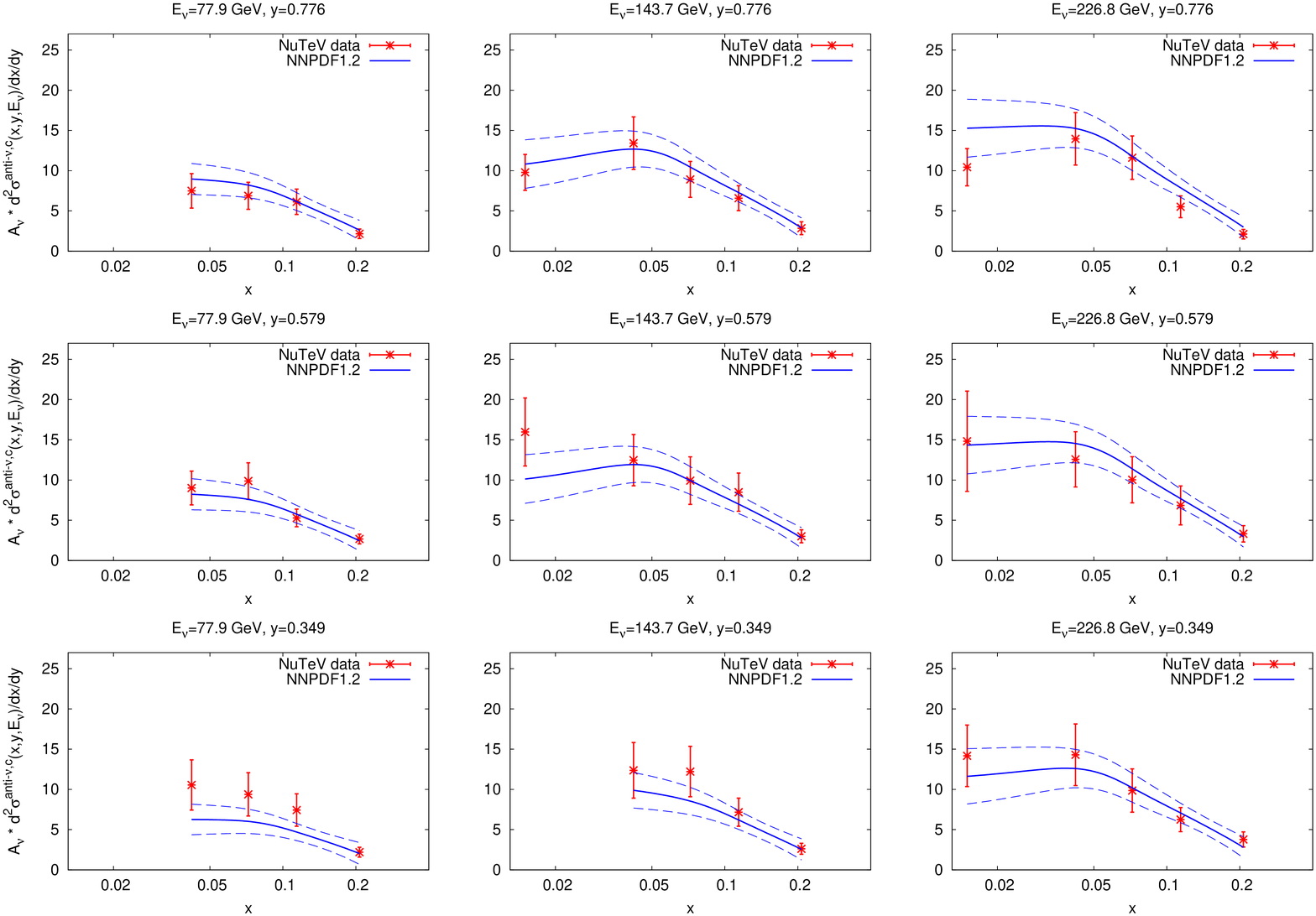}
 \end{center}
\caption{\small Comparison between the NuTeV data and the NNPDF1.2
  theoretical predictions for neutrino (upper three rows) and
  anti-neutrino (lower three rows) dimuon production. 
All cross section in the plots are
rescaled by a factor $A_{\nu}=\frac{1}{E_{\nu}}
\frac{10^2}{G_F^2M_NE_{\nu}}$. The neutrino kinematics parameters $(E_{\nu},y)$
are related to $x$ and $Q^2$ by Eq.~(\ref{eq:nukin}).
The solid  line is the central NNPDF1.2 prediction and the
dashed lines the 1-$\sigma$ interval.}
\label{fig:dimuon}
\end{figure}
%------------------------------------------------------------------
The results from $K_S$ and $R_S$ obtained in each of these cases are
compared in Table~\ref{tab:strangecheck} to the reference NNPDF1.2
result, all at the scale $Q^2=20$~GeV$^2$. It is apparent that the
effect of any of these systematics is rather moderate, even if
very conservatively estimated. In the same table we also show the
result of a fit in which the sum rule Eq.~(\ref{eq:sr}) is not imposed:
even in this case the result changes very little. 

Estimating the effect of the systematics from the sum in quadrature of
the shift of central values due to the four central rows of
Table~\ref{tab:strangecheck} we get, at $Q^2=20$~GeV$^2$
\bea
\label{eq:ksval}
K_S&=& 0.71 {{}^{+0.19}_{-0.31}}^{\rm stat}\pm 0.26^{\rm  syst},\\
R_S&=& 0.006\pm0.045^{\rm stat} \pm 0.010^{\rm syst}.
\label{eq:rsval}
\eea
The systematics on $R_S$ is thus negligible, and mostly due to nuclear
effects. The systematics on $K_S$ is not quite negligible, and
almost entirely due to 
the treatment of the heavy quark mass: this is an aspect of our
analysis which could be improved in the future within a more accurate
treatment of quark mass effects.

\subsection{Comparison with experimental data}

%------------------------------------------------------------
\begin{figure}[t!]
\begin{center}
\epsfig{width=0.80\textwidth,figure=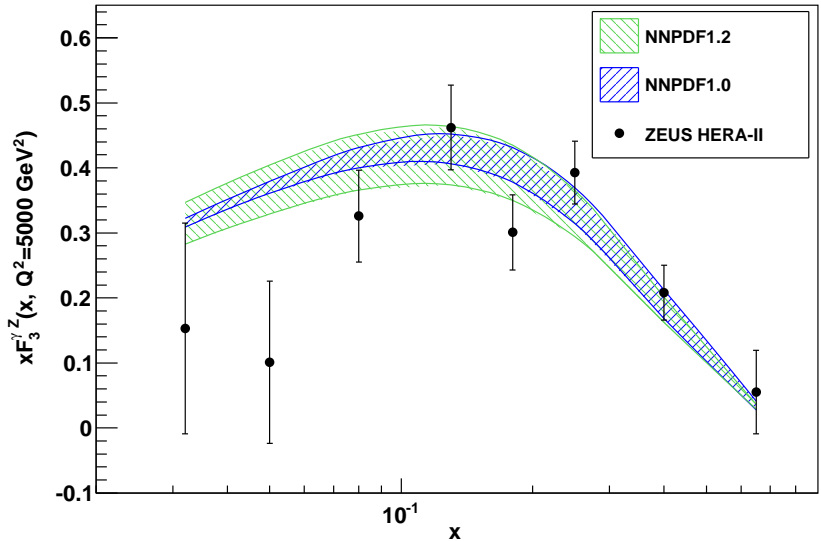}
\caption{\small Comparison with the experimental
  determination~\cite{Chekanov:2009gm} of the interference  
structure function $xF_{3}^{\gamma Z}(x,Q^2)$at $Q^2=5000$~GeV$^2$
with the NLO prediction obtained using the NNPDF1.0 and NNPDF1.2
parton sets. \label{fig:xf3}} 
\end{center}
\end{figure}
%------------------------------------------------------------

The results obtained from a NLO computation of the dimuon cross
section from the reference set of NNPDF1.2 parton distributions with
$N_{\rm rep}=1000$ replicas are compared in Fig.~\ref{fig:dimuon} to
the NuTeV experimental data. The agreement is clearly excellent in all
bins except for the lowest $Q^2$ values (bottom left plot), where the
approximate treatment of the quark mass leads to a deterioration in
quality of the fit.

In Ref.~\cite{Chekanov:2009gm}, an extraction of the interference
parity--violating structure function $xF_3^{\gamma Z}$ evolved
to a common
scale $Q^2=5000$~GeV$^2$ was also presented. This extraction is based
on data already included in our fit, so these data do not provide any
extra information. However, it is interesting to
compare directly to it, because this structure function is directly
sensitive to the flavour and valence/sea decomposition of PDFs
(specifically to strangeness), which
 is difficult
to probe directly (see e.g. Ref.~\cite{Ball:2008by}, appendix A.2). 
Because the  contribution of $xF_3^{\gamma Z}$
to the total reduced cross--section is small and only relevant
in a limited region of
phase space, the agreement between data and theory for this quantity
could in principle be poor without this being significantly reflected in the
quality of the global fit. 
\begin{table}
  \centering
  {\scriptsize
  \begin{tabular}{|c|c|c|c|c|}
    \hline
      &   & $\sigma(W^+){\rm Br}\lp W^+ \to l^+\nu_l\rp\,$ 
          & $\sigma(W^-){\rm Br}\lp W^- \to l^+\nu_l\rp\,$ 
          & $\sigma(Z^0){\rm Br}\lp Z^0 \to l^+l^-\rp\,$ \\
    \hline
    \multirow{2}{*}{NNPDF 1.0} & $10$ TeV &  8.49 $\pm$ 0.18 & 5.81 $\pm$ 0.13 & 1.36 $\pm$ 0.02 \\
                               & $14$ TeV & 11.83 $\pm$ 0.26 & 8.41 $\pm$ 0.20 & 1.95 $\pm$ 0.04 \\
    \hline
    \multirow{2}{*}{NNPDF 1.1} & $10$ TeV &  8.52 $\pm$ 0.33 & 5.79 $\pm$ 0.28 & 1.36 $\pm$ 0.04 \\
                               & $14$ TeV & 11.86 $\pm$ 0.46 & 8.38 $\pm$ 0.39 & 1.95 $\pm$ 0.06 \\
    \hline
    \multirow{2}{*}{NNPDF 1.2} & $10$ TeV &  8.61 $\pm$ 0.25 & 5.85 $\pm$ 0.15 & 1.37 $\pm$ 0.03 \\
                               & $14$ TeV & 11.99 $\pm$ 0.34 & 8.47 $\pm$ 0.21 & 1.97 $\pm$ 0.04 \\
    \hline
  \end{tabular}}
  \caption{\small Cross sections for gauge boson production at the
    LHC. All quantities have been computed at NLO using
    MCFM~\cite{Campbell:2000bg,Campbell:2002tg,Campbell:2004ch,MCFMurl}
    and NNPDF partons.\label{tab:LHCobs}}
\end{table}

A comparison of these data with the NLO
prediction obtained using the NNPDF1.0 and NNPDF1.2
parton sets
is shown in
Fig.~\ref{fig:xf3}, and shows good agreement: $\chi^2=1.53$ for NNPDF1.2, and
$\chi^2=1.55$ for NNPDF1.0, comparable to the value for other data in
the valence region 
(despite the fact that for the NNPDF1.0 fit neither the data of
Ref.~\cite{Chekanov:2009gm} nor dimuon data were used).
The widening of the uncertainty band when going from
NNPDF1.0 to NNPDF1.2 is a consequence of the sensitivity of this
structure function to valence combinations, and strangeness in
particular: very precise measurements of it could greatly improve
flavour separation of PDFs.

A detailed study of the phenomenological implications of our
reassessment of the strangeness uncertainty for LHC observables is
beyond the scope of this work. However, in Table~\ref{tab:LHCobs} we
collect the total cross section for $W$ and $Z$ production computed at
NLO with
MCFM~\cite{Campbell:2000bg,Campbell:2002tg,Campbell:2004ch,MCFMurl}:
results obtained with the NNPDF1.2 and NNPDF1.1 parton sets are
compared to those found using NNPDF1.0. already discussed in
Ref.~\cite{Ball:2008by}. Because of the increased uncertainty on the
strange distribution, the uncertainty in the cross section is larger
in NNPDF1.1 and NNPDF1.2, though less so in NNPDF1.2 due to the 
constraint
from dimuon data.

\section{Precision determination of electroweak parameters}
\label{sec:ewparam}

Neutrino DIS data, and especially dimuon data, can be used to perform direct
measurements of electroweak
parameters~\cite{CSB:1997xy,Mangano:2001mj}. However the potential precision
of these measurements can be spoiled by PDF uncertainties. Indeed, we have seen
in Sect.~\ref{sec:results} that the uncertainties we obtain on the
strange distributions are quite large, typically larger by almost one order of
magnitude than those found in previous global fits.

The CKM matrix elements control the strength of the coupling of various
partons to neutrinos according to Eqs.~(\ref{fnexpr},\ref{fnbexpr}).
In spite of the large PDF uncertainties in the strange sector,
we shall provide here the most precise
direct determination up to date of the  CKM matrix element  $|V_{cs}|$ 
within a single experiment. We will also provide
a determination of $|V_{cd}|$ with an accuracy consistent with
previous results from neutrino data. These 
remarkable results are possible
because  PDF uncertainties are 
free from parametrization bias, thus they may be 
disentangled  from the uncertainty on the physical
parameters. 

We will then turn to a study of the impact of PDF
uncertainties on the extraction of the electroweak
mixing angle $\sin^2\theta_{\rm W}$ from the Paschos-Wolfenstein ratio:
we will show that once PDF uncertainties are properly taken into account, 
the NuTeV measurement of this ratio~\cite{Zeller:2001hh} is in full agreement
with the standard model prediction.

\subsection{Determination of $|V_{cs}|$ and
$|V_{cd}|$.}

Since the pioneering CDHS studies~\cite{Abramowicz:1982zr}, neutrino DIS
has been used as a means to  directly determine CKM matrix elements: 
the parton--model expressions for the
neutrino and anti-neutrino dimuon production Eqs.~(\ref{fnexpr},\ref{fnbexpr})
provide two equations which relate two experimentally measurable
cross sections to the two unknowns $|V_{cd}|$ and $|V_{cs}|$. 

However, these
equations also contain as unknowns the second moments  of the light
quark PDFs (the total cross section is proportional to the second
moment of the PDF). 
The standard
lore~\cite{Abramowicz:1982zr,Bolton:1997pq,Amsler:2008zzb}  
is then that if one
assumes that $S^-\approx0$, the linear combination
$F_2^{\nu,c}-F_2^{\bar\nu,c}$ only depends on the $|V_{cd}|$ and the
$u$ and $d$ valence
components, which are well measured by other experiments, so it can be
used to determine $|V_{cd}|$. On the other
hand, the orthogonal combination $F_2^{\nu,c}+F_2^{\bar\nu,c}$ depends
on the $|V_{cs}|/|V_{cd}|$ ratio, but also on $K_S$ Eq.~(\ref{eq:mom1}),
and thus it can only be used to determine the combination
$|V_{cs}|K_S$.  Indeed, the PDG~\cite{Amsler:2008zzb} 
quotes a value of $|V_{cd}|=0.23\pm 0.11$ obtained 
from the average neutrino dimuon experiments as the best current
direct determination. 
Only the bound $|V_{cs}|\ge 0.74$ at 90\% confidence
level~\cite{Bolton:1997pq} was quoted in previous
PDG~\cite{Groom:2000in} editions, but this is now superseded by a direct 
determination $|V_{cs}| = 1.04\pm0.06$ from $D$ decays (for a recent
update, see Ref.~\cite{Narison:2008bc}).  Of course,
the values obtained from the current global CKM
fits~\cite{UTFIT,CKMFITTER,Amsler:2008zzb}  
are much more precise than these direct
determinations (see Table~\ref{tab:CKMtab} below).

In the NNPDF1.2 reference fit, $|V_{cd}|$, $|V_{cs}|$, and $|V_{cb}|$ are 
each fixed to the current PDG value~\cite{Amsler:2008zzb},
obtained from the global CKM unitarity fit. 
We now show that, thanks to the fact that we are free of bias related
to the parametrization of strangeness, we can extract both $|V_{cs}|$
and $|V_{cd}|$ from the fit. In order to do this,
we perform a scan over the values of $|V_{cs}|$ and $|V_{cd}|$ used 
in the fit, holding $|V_{cb}|$ fixed, but relaxing the unitarity constraint
(in practice, because of its smallness, 
the precise value chosen for $|V_{cb}|$ 
is inconsequential). The best--fit value and uncertainty for
the CKM  parameters are then determined in the standard way by maximum
likelihood from the $\chi^2$ profile. 

The $\chi^2$ determined from a
set of $N_{\rm dat}$ data points fluctuates, with a standard deviation equal to
$\sigma_{\chi^2}=\sqrt{2 N_{\rm dat}}$. In order to determine 
the $\chi^2$ profile as
the underlying parameters are varied, these fluctuations must be
kept under control. Within our Monte Carlo approach, this could
be done by using the same set of data replicas each time the $\chi^2$
is recomputed with different values of the underlying parameters. This
might however bias the result in a random way depending on the
particular set of replicas which has been chosen in the first
place. We prefer thus to vary randomly the set of replicas which is
used for different parameter values: fluctuations are then kept under
control by using a sufficiently large set of replicas, given 
the fluctuation of the $\chi^2$ computed from a replica average has a
standard deviation equal to $\sigma_{\chi^2}/\sqrt{N_{\rm  rep}}$. 
Because only dimuon  data are sensitive to the CKM matrix
  elements, we can determine their values
 from the dependence of
the $\chi^2$ of the fit to these data
  only, rather than for that of the fit to the global  dataset. 
Because we have (see Tab.~\ref{tab:expsets}) $84$ dimuon
  data points, a set of $N_{\rm rep}=500$ replicas is 
sufficient to guarantee that point-by-point
  fluctuations are smaller than $\Delta\chi^2=1$.

%------------------------------------------------------------
\begin{figure}
\begin{center}
\epsfig{width=0.81\textwidth,figure=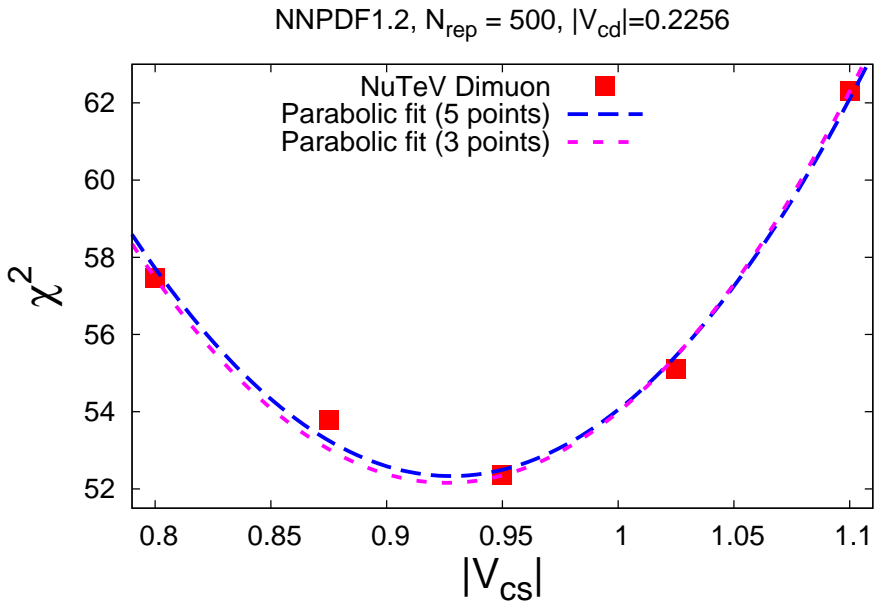} 
\end{center}
\caption{\small The $\chi^2$ of the NuTeV dimuon
data as a function of $|V_{cs}|$ when $|V_{cd}|$ is kept
fixed at its best unitarity fit value. The long-dashed curve is the 
parabolic fit from which
the central value and one-$\sigma$ uncertainty Eq.~(\ref{eq:vcs}) are obtained; the
short-dashed  curve is a parabolic fit to the central and two outer points
only.}
\label{fig:Vcsscan}
\end{figure}

\begin{figure}
\begin{center}
\epsfig{width=0.81\textwidth,figure=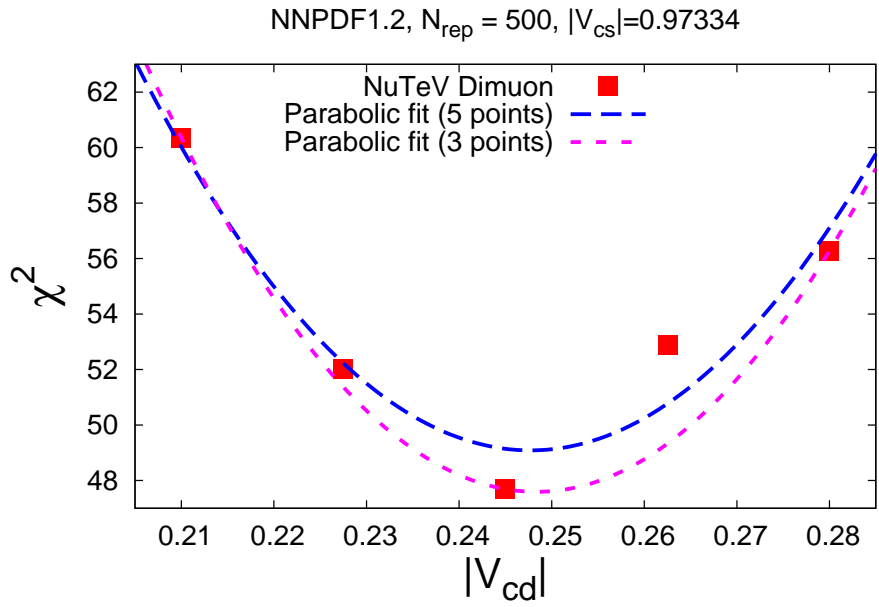} 
\end{center}
\caption{\small 
The $\chi^2$ of the NuTeV dimuon
data as a function of $|V_{cd}|$ when $|V_{cs}|$ is kept
fixed at its best unitarity fit value. The log-dashed curve is the 
parabolic fit from which
the central value and one-$\sigma$ uncertainty Eq.~(\ref{eq:vcd}) are obtained; the
short-dashed  curve is a parabolic fit to the central and two outer points
only.}
\label{fig:Vcdscan}
\end{figure}
%------------------------------------------------------------------

\begin{figure}
\begin{center}
\epsfig{width=0.78\textwidth,figure=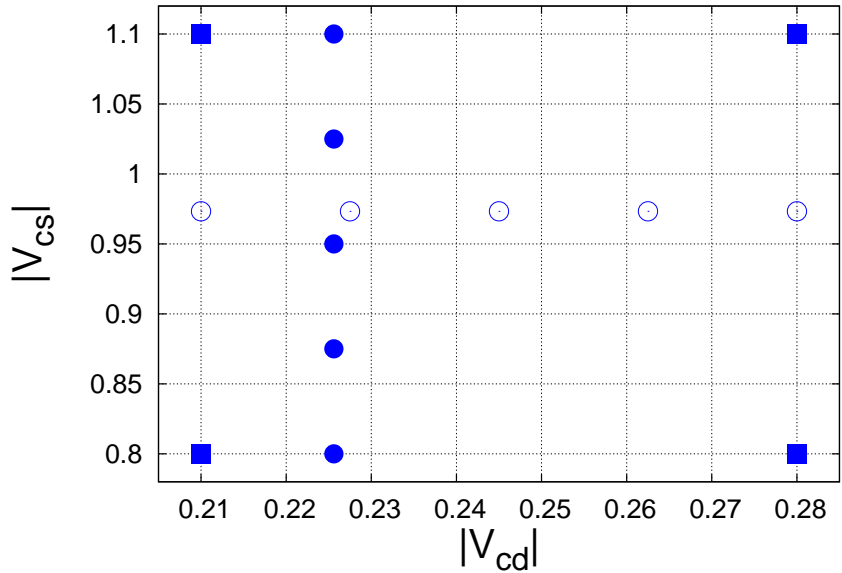} 
\end{center}
\caption{\small The grid of points used in the determination of
the CKM matrix elements $|V_{cs}|$ and $|V_{cd}|$. Open circles 
denote points used for the determination of $|V_{cd}|$ Eq.~(\ref{eq:vcd}),
and full circles points used for the determination of $|V_{cs}|$
Eq.~(\ref{eq:vcs}). All points are used in the joint determination
Eq.~(\ref{eq:finalCKM}).}
\label{fig:CKMfit-grid}
\end{figure}

First, we vary independently  each of the two CKM matrix
elements, keeping the other fixed at its central 
value in the CKM unitarity fit. The 
$\chi^2$ profile is computed for five equally spaced values
of the parameter which is being varied. The values have been chosen  
on the basis of a
preliminary exploration of the space of parameters based on fits with
a small number of replicas; they are displayed in
Fig.~\ref{fig:CKMfit-grid}. The ensuing $\chi^2$ profile is displayed
in Fig.~\ref{fig:Vcsscan} for  $|V_{cs}|$ 
and in Fig.~\ref{fig:Vcdscan} for  $|V_{cd}|$.
We observe well-defined minima in both cases. A parabolic fit leads to
\bea
\label{eq:vcs}
|V_{cs}| &=&0.93\pm 0.06,
\\
\label{eq:vcd}
|V_{cd}|&=&0.248\pm 0.012,
\eea
where the one-$\sigma$ uncertainty is obtained from the 
condition $\Delta\chi^2=1$.  The fit is
quite stable upon the choice of different subsets of the
five available points: if it is repeated by only retaining the central
and two outer points neither the central values nor the uncertainties
Eqs.~(\ref{eq:vcs}-\ref{eq:vcd}) vary significantly. This confirms that
the number of replicas used to compute the $\chi^2$ is sufficiently
large for the result not to be biased by statistical
fluctuations. Both fits are shown in
Figs.~\ref{fig:Vcsscan}-\ref{fig:Vcdscan}. 

\begin{figure}
\begin{center}
\epsfig{width=0.78\textwidth,figure=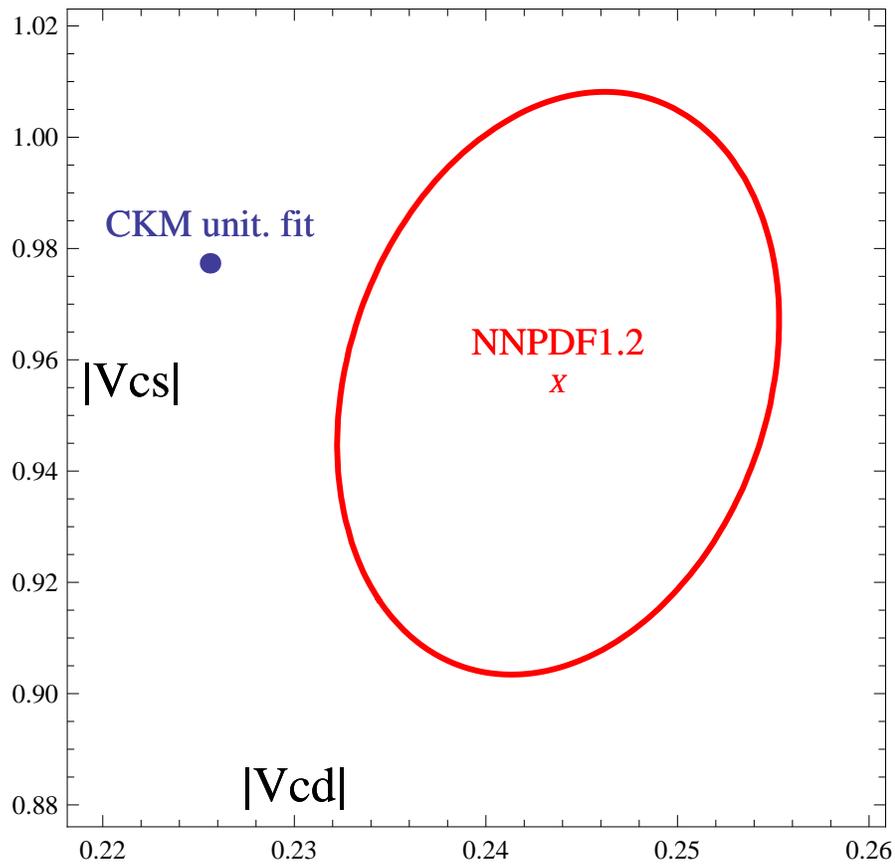} 
\end{center}
\caption{\small Location of the best-fit point and one-$\sigma$
 (statistical $\Delta\chi^2=1$ uncertainty) 
ellipse in the ($|V_{cd}|$,~$|V_{cs}|$) plane
for the best-fit $\chi^2$ paraboloid obtained from the $\chi^2$
computed at the points displayed in Fig.~\ref{fig:CKMfit-grid}.
%Note that the ellipse is computed with statistical uncertainties
%only.
The best unitarity fit result  \cite{Amsler:2008zzb} is also shown for
comparison.
}
\label{fig:CKMellipse-total}
\end{figure}

This shows that either CKM matrix element can be determined from our
data, with comparable uncertainty, by taking the other fixed. 
We can thus perform a simultaneous determination of both these
CKM matrix elements. In order to improve the accuracy of this
determination, we compute the $\chi^2$ at four more points in the
($|V_{cd}|$,~$|V_{cs}|$) plane, denoted by squares in
Fig.~\ref{fig:CKMfit-grid}.  
The $\chi^2$ in these additional points is computed from a smaller set
of $N_{\rm rep}=100$ replicas.
The result of the combined fit is then
\bea
|V_{cs}|&=&0.96\pm 0.05,\\
|V_{cd}|&=&0.244\pm 0.012.
\label{eq:finalCKM}
\eea
The uncertainties turn out to be almost identical to the diagonal
uncertainties, and the correlation coefficient is relatively small
$\rho=0.21$, reflected in a moderate shift in central values in
comparison to the separate fits Eqs.~(\ref{eq:vcs}-\ref{eq:vcd}).
The location of the best-fit point and one-$\sigma$
 ($\Delta\chi^2=1$) ellipse in the ($|V_{cd}|$,~$|V_{cs}|$) plane
for the best-fit $\chi^2$ paraboloid is shown in Fig.
\ref{fig:CKMellipse-total}.

%------------------------------------------------------------
%------------------------------------------------------------------
%------------------------------------------------------------
%------------------------------------------------------------------

\begin{table}
\begin{center}
\small
\begin{tabular}{|c|c|c|}
\hline
&    $|V_{cd}|$ & $|V_{cs}|$ \\
\hline
\hline
Statistical & $\pm 0.012$ & $\pm 0.05$\\
\hline
Mass effects  & $\pm 0.007 $  &  $\pm 0.02 $\\
Higher order QCD  & $\pm 0.010$   & $\pm 0.03$ \\
Nuclear corrections  & $\pm 0.008$   &   $\pm 0.03$ \\
\hline
 Total systematic uncertainty & $\pm 0.014 $   & $\pm 0.05 $ \\
\hline
\hline
Total uncertainty &  $\pm 0.019 $ & $\pm 0.07 $ \\
\hline
\end{tabular}
\end{center}
\caption{\small Summary of statistical and systematic uncertainties
in the present
determination of $|V_{cs}|$ and $|V_{cd}|$.
\label{tab:CKMsys} }
\end{table}
This determination Eq.~(\ref{eq:finalCKM}) is affected by the same
systematics that we examined in Sect.~\ref{sec:thunc}, namely,
higher order QCD corrections, treatment of heavy quark effects and modeling
of nuclear
corrections. In order to
assess their impact in the CKM element
determination, we have repeated the determination of each of the two
parameters as the other is kept fixed, Eqs.~(\ref{eq:vcs}-\ref{eq:vcd}),
by recomputing the $\chi^2$ for a smaller set of $N_{\rm rep}=100$
replicas along the points denoted as circles in
Fig.~\ref{fig:CKMfit-grid}, with each of these  three effects varied in turn
as we did in Sect.~\ref{sec:thunc}. We then take the shift in central
value as an estimate of the corresponding uncertainty.
The results are summarized in Table~\ref{tab:CKMsys}. 
Putting everything together, we find
\bea
\label{finalCKMsys1}
|V_{cs}|&=&0.96\pm 0.07^{\rm tot}\ , \\
\label{finalCKMsys}
|V_{cd}|&=&0.244\pm 0.019^{\rm tot}.
\eea

In Table~\ref{tab:CKMtab} we compare our final results
Eqs.~(\ref{finalCKMsys1}-\ref{finalCKMsys}) with the best CKM unitarity
fit results and with other direct determinations.  Our determination
of $|V_{cd}|$ is consistent with other direct determinations, and of
comparable accuracy, though one should bear in mind that previous
determinations from dimuon data were based on fits with a fixed
functional form, and thus subject to potentially large systematics
bias.  Our determination of $|V_{cs}|$ is rather more accurate that
any other direct determination from dimuon data, more accurate than
any single direct determination, and of comparable accuracy to the PDF
average of determinations from $D$ decays. 
% In view of this, it might
%be interesting to consider the possibility of including our
%determinations of $|V_{cd}|$ and $|V_{cs}|$ in the input to the CKM
%fits, as for example in Ref.~\cite{Charles:2004jd,Bona:2005vz}.

%-----------------------------------------------------
 \begin{table}
 \begin{center}
\small
 \begin{tabular}{|c|c|c|c|}
  \hline
 Analysis & Description & Reference & $|V_{cs}|$ \\ \hline
NNPDF1.2 & {\footnotesize Direct determination from global PDF analysis} 
& This work & $0.96\pm 0.07^{\rm tot}$\\ 
CDHS & {\footnotesize LO determination from $\nu N\to \mu^+\mu^- X$} &
 \cite{Abramowicz:1982zr} & $\ge 0.59$ (90\% C.L.) \\
CCFR & 
{\footnotesize NLO determination from $\nu N\to \mu^+\mu^- X$} & 
 \cite{Bazarko:1994tt,Bolton:1997pq}  & $\ge 0.74$ (90\% C.L.) \\
PDG08   & 
{\footnotesize Averages of determinations from
 $D$ decays}&  \cite{Amsler:2008zzb}  &$1.04 \pm 0.06$  \\
Hocker    & 
{\footnotesize Averages of determinations from
$\nu N\to \mu^+\mu^- X$ }&  \cite{Hocker:2001xe}  &$1.04 \pm 0.16$  \\
DELPHI   & 
{\footnotesize Direct measurement from
$W^+\to c\bar{s}$ decays}&  \cite{Abreu:1998ap}  &$0.94^{+0.32}_{-0.26} 
\pm 0.13$  \\
PDG08 &  
{\footnotesize CKM unitarity fit}  & \cite{Amsler:2008zzb}  & 
$0.97334\pm 0.00023$\\
 \hline
 \end{tabular}
 \begin{tabular}{|c|c|c|c|}
  \hline
 Analysis & Description & Reference & $|V_{cd}|$ \\ \hline
NNPDF1.2 & {\footnotesize Direct determination from global PDF analysis} 
& This work & $0.244\pm 0.019^{\rm tot}$ \\ 
CDHS & 
{\footnotesize LO determination from $\nu N\to \mu^+\mu^- X$} & 
 \cite{Abramowicz:1982zr}  &$0.24 \pm 0.03$  \\
CCFR & 
{\footnotesize NLO determination from $\nu N\to \mu^+\mu^- X$} & 
 \cite{Bolton:1997pq}  &$0.232^{+0.017}_{-0.019}$  \\
PDG08   & 
{\footnotesize Averages of direct determinations
from  $\nu N\to \mu^+\mu^- X$}&  \cite{Amsler:2008zzb}  &$0.230 \pm 0.011$  \\
PDG08 & 
{\footnotesize Average of determinations from $D\to K/\pi l\nu$ decays} & 
 \cite{Amsler:2008zzb}  &$0.218 \pm 0.023$ \\
PDG08 & {\footnotesize CKM unitarity fit}  & \cite{Amsler:2008zzb}  & $0.2256\pm 0.0010$  \\
 \hline
 \end{tabular}
\end{center}
\caption{\small Comparison of the present determination
of the CKM matrix elements $|V_{cs}|$ (upper table) 
and $|V_{cd}|$ (lower table) with
other available direct measurements, averages and
CKM constrained fits.
\label{tab:CKMtab}}
 \end{table}
% ------------------------------

\clearpage

\subsection{PDF corrections to the Paschos-Wolfenstein ratio}

The successful determination of the CKM matrix elements which control
charged current scattering suggests that we might use our parton set for a
reliable
reassessment of the determination  
of the coupling which controls neutral current neutrino DIS. As is well
known~\cite{Paschos:1972kj}, this coupling depends on the electroweak
mixing angle, which can thus be extracted from its experimental
measurement. Specifically, in the parton model one has
\bea
\label{eq:PW}
R_{\rm PW} &\equiv&
\frac{\sigma(\nu {\cal N}\to \nu X)-\sigma(\bar\nu {\cal N}\to
\bar\nu X)}{\sigma(\nu {\cal N}\to \ell X) - \sigma(\bar{\nu}{\cal
    N}\to \bar{\ell}X)} \nonumber \\
&=&\frac{1}{2}-  \sin^2 \theta_{\rm W}+\left[\frac{( \lc U^-\rc  -
 \lc D^-\rc )+( \lc C^-\rc - \lc S^-\rc)}{  \lc{\cal Q}^-\rc }
\frac{1}{6}\left(3-7\sin^2 \theta_{\rm W}\right)\right]\ ,
\eea
where $ \theta_{\rm W}$ is the electroweak mixing angle, 
$\lc S^-\rc $ is  the strange valence momentum fraction
Eq.~(\ref{eq:strangmomdef}), $\lc U^-\rc$, $\lc D^-\rc$  and
 $\lc C^-\rc$ 
the valence momentum
fractions of other quark flavors, and
$ \lc {\cal Q}^-\rc 
\equiv ( \lc U^-\rc  +
\lc D^-\rc)/{2}$.

The recent experimental determination~\cite{Zeller:2001hh}
\be
\sin^2\theta_{\rm W}\Big|_{\rm NuTeV} = 0.2277 \pm 0.0014^{\rm stat} \pm
0.0009^{\rm sys}= 0.2277\pm 0.0017^{\rm tot} \ ,  
\label{eq:nutevtheta}
\ee
is obtained using Eq.~(\ref{eq:PW}) under the assumption that for an
isoscalar nucleon target $\lc U^-\rc$-$\lc D^-\rc$=$\lc C^-\rc$=$\lc S^-\rc$=$0$, so the term in square
brackets in Eq.~(\ref{eq:PW}) vanishes. Of course, the NuTeV iron target is
not exactly isoscalar; however, the corresponding correction can be
computed~\cite{Zeller:2001hh} with small uncertainty~\cite{Davidson:2001ji}.
The result Eq.~(\ref{eq:nutevtheta}) disagrees at the three-$\sigma$ level with
the value determined in global 
precision electroweak
fits, such as~\cite{Flacher:2008zq,Heinemeyer:2004gx}
\be
\sin^2\theta_{\rm W}\Big|_{\rm EWfit} = 0.2223 \pm 0.0003 \ .
\label{eq:ewfit}
\ee
Possible explanations for this include nuclear effects, electroweak
corrections, QCD corrections, and physics beyond the standard
model~\cite{Davidson:2001ji} (see e.g. \cite{Londergan:2007zza} for an
updated list of references). However, one may
also\cite{Davidson:2001ji} 
question the
validity of the assumption of the vanishing of
the contribution in square brackets in
Eq.~(\ref{eq:PW}). The possibility that $\lc U^-\rc -\lc D^-\rc
\not=0$ even for an
isoscalar target due to isospin violation induced by
QED evolution effects was discussed in
Ref.~\cite{Martin:2004dh}: it could easily explain about a third of
the observed discrepancy.

In our fit, isospin symmetry is assumed, and furthermore 
$\lc C^-\rc=0$. We are then left with
the correction
\be
\label{eq:maps2W}
\delta_s \sin^2 \theta_{\rm W} 
=-R_S \frac{1}{6}\left( 3-7\sin^2 \theta_{\rm W}\right),
\ee
with $R_S$ defined in Eq.~(\ref{eq:rdef}). 
Using the value of $R_S$ Eq.~(\ref{eq:rsval}), obtained at the typical
scale $Q^2=20$~GeV$^2$ of the NuTeV data (and whose scale dependence
is very small anyway~\cite{Catani:2004nc}) we obtain
\be
\label{eq:maps2W2}
\delta_s \sin^2 \theta_{\rm W} 
= -0.001\pm 0.011^{\rm PDFs} \pm 0.002^{\rm th},
\ee
where the theoretical uncertainty comes from the effects
discussed above in Sect.~\ref{sec:thunc}, and it is not to be confused with the
experimental systematics in the NuTeV measurement
Eq.~(\ref{eq:nutevtheta}).

\begin{figure}
\begin{center}
\epsfig{width=0.65\textwidth,figure=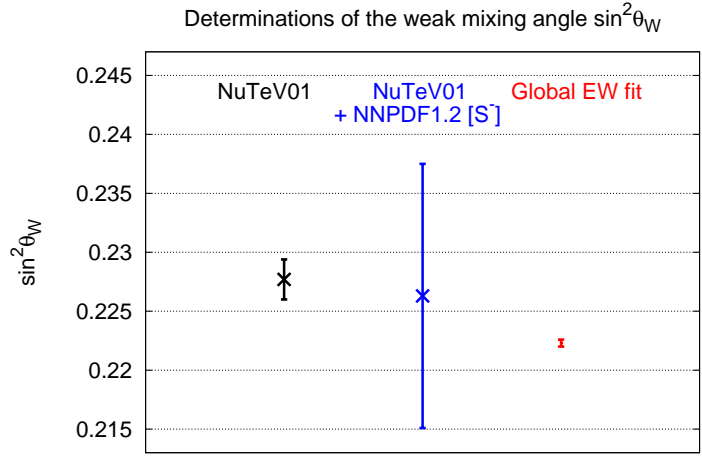} 
\end{center}
\caption{\small Comparison between the 
NuTeV determination of $\sin^2\theta_{\rm W}$, Eq.~(\ref{eq:nutevtheta}),
the result from the global electroweak fit, Eq.~(\ref{eq:ewfit}), and 
the NuTeV result after the  correction due to the uncertainty on $S^-$ 
Eq.~(\ref{eq:nutevcorr}).
}
\label{fig:nutev}
\end{figure}

Even neglecting these theoretical uncertainties (which we estimated
very conservatively), the additional PDF  uncertainty due to
strangeness alone  
is thus about twice the observed discrepancy in $\sin^2 \theta_{\rm W}$.
We must conclude therefore that the apparent inconsistency between the NuTeV
measurement and the global electroweak fit disappears once the uncertainty
on the strange distribution is properly taken into account.
Applying the correction Eq.~(\ref{eq:maps2W2}) the NuTeV result becomes
\be
\sin^2\theta_{\rm W}\Big|_{\rm NuTeV} = 0.2263 \pm 0.0014^{\rm stat} \pm
0.0009^{\rm sys}\pm0.0107^{\rm PDFs}.  
\label{eq:nutevcorr}
\ee
We recommend that the corrected result Eq.~(\ref{eq:nutevcorr}) be used,
for instance in  global electroweak fits. This corrected result is
compared graphically in Fig.~\ref{fig:nutev}
to the original NuTeV result Eq.~(\ref{eq:nutevtheta}) and the 
result from the global
electroweak fit Eq.~(\ref{eq:ewfit}).

%------------------------------------------------------------
%------------------------------------------------------------------

% ------------------------------------------------

%\include{sec-lhcpheno}

% --------------------------------------------
%
%\section{Conclusions and outlook}
%
%-------------------------------------------------
%-----------------------------------------------

\section{Conclusions and outlook}
\label{sec:conclusions}

We have presented an upgrade of the NNPDF1.0 parton set, which now
includes an independent parametrization for the strange distributions,
and the inclusion of dimuon data which constrain them.
Besides being an intermediate step towards a fully global fit
including hadronic data, our results are interesting as a test of the
NNPDF methodology, as a state-of-the art determination of the strange
PDFs, and as a determination of electroweak parameters.

We have shown that the NNPDF approach has no difficulty in dealing with
situations where experimental information is scarce and only
provides loose constraints on the form of parton distributions. Within
our approach, this does not require the introduction of theoretical
assumptions or constraints in order to obtain stable results. We can
thus provide reliable estimates of uncertainties,
free of bias induced by  theoretical assumptions.

We have obtained a determination of the strange momentum fraction
and of the strangeness valence component, which, though in agreement
with previous determinations, turn out to be affected by
uncertainties which are sizably larger than those found by other
groups. 

Nevertheless, we have shown that, with the uncertainty on the
strange PDF carefully estimated, the dimuon data can be used to
provide a good determination of the CKM matrix elements $|V_{cd}|$ and
 $|V_{cs}|$. In particular, our determination of $|V_{cs}|$ is the most
accurate ever obtained from neutrino deep-inelastic scattering data,
and it is also more accurate than any individual direct
determination from $D$ decays. We have also shown that once PDF
uncertainties are estimated reliably, the value of the electroweak
mixing angle extracted from NuTeV inclusive data is in agreement with standard
model expectations.

The main defect of our results is that they are still based on an
approximate treatment of the charm mass. Within the context of the
present work, the only significant implication of this is a slight
increase in the systematic uncertainty on our determination of
$|V_{cd}|$. However, this also entails a further small but
non-negligible systematic uncertainty in our determination 
of PDFs~\cite{Ball:2008by}.

It will be interesting to study the implications for LHC observables
of this reassessment of the uncertainty on the strange distribution.
The NNPDF1.2 release is available from the webpage of the
NNPDF Collaboration {\tt http://sophia.ecm.ub.es/nnpdf/}.
\bigskip

{\bf Acknowledgments \\}
This work was partly supported by grants 
PRIN-2006 (Italy), 
MEC FIS2004-05639-C02-01
(Spain) and by the European network HEPTOOLS under contract
MRTN-CT-2006-035505. L.D.D. is funded by an STFC Advanced Fellowship and
M.U. by a SUPA graduate studentship.
We acknowledge discussions with  S.~Alekhin, P.~Nadolsky, P.~Nason and
A.~Vicini.
We are especially grateful to D. Mason for providing
us with the NuTeV data and acceptances, to F.~Olness for information
on NLO acceptances, to R.~Sassot and R.~Petti for 
providing us with their nuclear PDF sets and to A.~Tapper and
K.~Nagano
for help with the HERA-II data. J.R.~acknowledges the hospitality
of the CERN TH Division where part of this work was completed.

%------------------------------------------------------

\appendix

%----------------------------------------------------------------

\section{Kernels for Physical Observables}
\label{sec:kernels}
\def\nn{\nonumber}
\def\GS{\Gamma_{\rm S}}
\def\GNS{\Gamma_{\rm NS}}
\def\half{\smallfrac{1}{2}}
\def\ESp{E_{\rm S}^+} 
\def\ESm{E_{\rm S}^-} 
\def\ENSp{E_{\rm NS}^+} 
\def\ENSm{E_{\rm NS}^-} 

In this appendix we expand the physical observables
for dimuon production in the evolution basis of the PDFs, and derive 
expressions for the kernels, in the same way and using the same notation as 
in Appendix A of Ref. \cite{Ball:2008by}. All
convolutions may be performed either in the ZM-VFNS or in the 
I-ZM-VFN scheme, as discussed 
in Sect. \ref{sec:hq}.

The cross-section for charm production in
neutrino scattering off an isoscalar nucleon
is given by Eq.~(\ref{eq:nuxsecdimuon}), which we write as
\be
\label{nuxsecx}
\widetilde
\sigma^{\nu(\bar{\nu}),c}
=\kappa[\widetilde{Y}_+ F_2^{\nu(\bar{\nu}),c}
-y^2 F_L^{\nu(\bar{\nu},c)}\pm \,Y_-\,xF_3^{\nu(\bar{\nu},c)}],
\ee
where  
\be
\label{Ytilplus}
\kappa= \frac{G_F^2M_N}{2\pi(1+Q^2/M_W^2)^2},\qquad
\widetilde{Y}_+ =\bigg(Y_+ - \frac{2M^2_Nx^2y^2}{Q^2} - y^2\bigg)
\bigg(1+\frac{m_c^2}{Q^2}\bigg)+y^2.
\ee

Taking into account a possible non-isoscalar component of the 
nuclear target by defining $\tau \equiv 1-2Z/A$,
in the quark model we have
 \begin{eqnarray}
\label{fnu_chamrm_qm}
    &&F_2^{\nu,c}=F_L^{\nu,c}=x\,F_3^{\nu,c}=
    x\, \lp |V_{cd}|^2((1+\tau)u+(1-\tau)d)\,
+\,2|V_{cs}|^2s +\,2|V_{cb}|^2 b \rp,\\    
  &&F_2^{\bar{\nu},c}=F_L^{\bar{\nu},c}=-x\,F_3^{\bar{\nu},c}=
    x\, \lp |V_{cd}|^2((1+\tau)\bar{u}+(1-\tau)\bar{d})\,
+\,2|V_{cs}|^2\bar{s}+\,2|V_{cb}|^2\bar{b}\rp,    
  \end{eqnarray}
where all explicit dependence on $x$ and $Q^2$ has been dropped. 
In terms of the PDF evolution eigenstates we then have
\begin{eqnarray}
    \label{fnu_charm_qb}
    &&F_2^{\nu(\bar{\nu}),c}=F_L^{\nu(\bar{\nu}),c}
=\pm\, x\,F_3^{\nu(\bar{\nu}),c}=
x\,\big\{\smallfrac{1}{6}w_0(\Sigma\pm V)
    +\half\tau w_3 (T_{3}\pm V_3)\nn\\
&&\quad
+\smallfrac{1}{6}w_8(T_{8}\pm V_8)
+\smallfrac{1}{12}w_{15}(T_{15}\pm V_{15}) 
+\smallfrac{1}{20}w_{24}(T_{24}\pm V_{24})
+\smallfrac{1}{30}w_0(T_{35}\pm V_{35})
\big\},
     \end{eqnarray}
where the +(-) sign corresponds to neutrino (anti-neutrino) scattering, and
the CKM factors are
\begin{eqnarray}
    \label{wckm}
&&w_0\equiv |V_{cd}|^2+|V_{cs}|^2 +|V_{cb}|^2,
\qquad w_3\equiv |V_{cd}|^2,\qquad 
w_8\equiv |V_{cd}|^2-2|V_{cs}|^2,\nn\\
&&w_{15}\equiv |V_{cd}|^2+|V_{cs}|^2,\qquad
w_{24}\equiv |V_{cd}|^2+|V_{cs}|^2 -4 |V_{cb}|^2.
\end{eqnarray}
Unitarity of the CKM matrix is imposed setting $w_0=1$; in 
the CKM determination in Sec. 5.2 it is however left unconstrained.  
Below $b$ threshold $V_{cb}=0$, so  $w_0=w_{15}=w_{24}$.

In perturbative QCD the charm production neutrino structure functions
thus take the form 
\begin{eqnarray}
    \label{fnu_charm_qcd}
    &&F_i^{\nu(\bar{\nu}),c}= C_{i,q}^s\otimes
\smallfrac{1}{6}w_0\Sigma
+C_{i,g}\otimes
\smallfrac{1}{n_f}w_0 g\pm C_{i,q}^s\otimes\smallfrac{1}{6}w_0V
+C_{i,q}\otimes\big\{
\half\tau w_3 (T_{3}\pm V_3)\nn\\
&&\quad
+\smallfrac{1}{6}w_8(T_{8}\pm V_8)
+\smallfrac{1}{12}w_{15}(T_{15}\pm V_{15}) 
+\smallfrac{1}{20}w_{24}(T_{24}\pm V_{24})
+\smallfrac{1}{30}w_0(T_{35}\pm V_{35})
\big\},\\
    &&F_3^{\nu(\bar{\nu}),c}= \pm C_{3,q}^s\otimes
\smallfrac{1}{6}w_0\Sigma
+ C_{i,q}^s\otimes\smallfrac{1}{6}w_0V
\pm C_{i,q}\otimes\big\{
\half\tau w_3 (T_{3}\pm V_3)\nn\\
&&\quad
+\smallfrac{1}{6}w_8(T_{8}\pm V_8)
+\smallfrac{1}{12}w_{15}(T_{15}\pm V_{15}) 
+\smallfrac{1}{20}w_{24}(T_{24}\pm V_{24})
+\smallfrac{1}{30}w_0(T_{35}\pm V_{35})
\big\},
\end{eqnarray}
where $i=2,L$ and $n_f$ is the number of active flavours. 
We can thus write the charm production neutrino cross-sections as 
%\begin{eqnarray}
%    \label{fnu_charm_qcd}
%    &&\widetilde{\sigma}^{\nu(\bar{\nu}),c}=
%\kappa x\,\big\{(\widetilde{Y}_+ C_{2,q}^s-y^2C_{L,q}+ Y_- C_{3,q})\otimes
%\smallfrac{1}{6}w_0\Sigma
%+(\widetilde{Y}_+ C_{2,g}^s-y^2C_{L,g})\otimes
%\smallfrac{1}{n_f}w_0 g\nn\\
%&& + (\widetilde{Y}_+ C_{2,q}-y^2C_{L,q}+ Y_- C_{3,q})\otimes 
%\big(\pm\smallfrac{1}{6}w_0V
%+\half\tau w_3 (T_{3}\pm V_3)
%+\smallfrac{1}{6}w_8(T_{8}\pm V_8)\nn\\
%&&\qquad\qquad
%+\smallfrac{1}{12}w_{15}(T_{15}\pm V_{15}) 
%+\smallfrac{1}{20}w_{24}(T_{24}\pm V_{24})
%+\smallfrac{1}{30}w_0(T_{35}\pm V_{35})\big)
%\big\},
%\end{eqnarray}
\bea
\label{Knu}
&&\widetilde{\sigma}^{\nu(\bar{\nu}),c} = \kappa\, x \lbrace 
K^{\nu(\bar{\nu}),c}_{\Sigma}
\otimes\Sigma_0
+ K^{\nu(\bar{\nu}),c}_g\otimes g_0 
\pm K^{\nu(\bar{\nu}),c}_{V} 
\otimes V_0\nn\\
&&\quad   
+ K^{\nu(\bar{\nu}),c}_{+}
\otimes (\half\tau w_3 T_{3,0}
+\smallfrac{1}{6}w_8 T_{8,0}
+\smallfrac{1}{12}w_{15}T_{15,0})\nn\\
&&\quad\pm K^{\nu(\bar{\nu}),c}_{-} 
\otimes (\half\tau w_3 V_{3,0}
+\smallfrac{1}{6}w_8 V_{8,0}
+\smallfrac{1}{12}w_{15}V_{15,0})\rbrace,
\eea  
where in Mellin space the kernels are 
\bea
\label{KCCS}
K^{\nu(\bar{\nu}),c}_{\Sigma} &=& 
(\widetilde{Y}_+C^s_{2,q}-y^2C^s_{L,q}+ Y_- C_{3,q}^s)
(\smallfrac{1}{6}w_0\GS^{qq}
 +\smallfrac{1}{20}w_{24}\GS^{24,q}
+\smallfrac{1}{30}w_{0}\GS^{35,q})\nn\\
&&\qquad\qquad\qquad 
+(\widetilde{Y}_+C_{2,g}-y^2C_{L,g})\smallfrac{1}{n_f}w_{0}
\GS^{gq},\\
\label{KCCg}
K^{\nu(\bar{\nu}),c}_{g} &=& 
(\widetilde{Y}_+C^s_{2,q}-y^2C^s_{L,q}+ Y_- C_{3,q}^s)
(\smallfrac{1}{6}w_0\GS^{qg}
 +\smallfrac{1}{20}w_{24}\GS^{24,g}
 +\smallfrac{1}{30}w_0\GS^{35,g})\nn\\
 &&\qquad\qquad\qquad 
+(\widetilde{Y}_+C_{2,g}-y^2C_{L,g})\smallfrac{1}{n_f}w_{0}
\GS^{gg},\\
\label{KCCV}
K^{\nu(\bar{\nu}),c}_{V} &=& 
(\widetilde{Y}_+C^s_{2,q}-y^2C^s_{L,q}+ Y_- C_{3,q}^s)
(\smallfrac{1}{6}w_0\GNS^{v} 
+\smallfrac{1}{20}w_{24}\GNS^{24}
+\smallfrac{1}{30}w_{0}\GNS^{35}),\\
\label{KCC+}
K^{\nu(\bar{\nu}),c}_{+} &=&
(\widetilde{Y}_+C_{2,q}-y^2C_{L,q}+ Y_- C_{3,q})
\GNS^{+},\\
\label{KCC-}
K^{\nu(\bar{\nu}),c}_{-} &=& 
(\widetilde{Y}_+C_{2,q}-y^2C_{L,q}+ Y_- C_{3,q})
\GNS^{-}.
\eea
Below the $b$ threshold the singlet kernels simplify to
\bea
\label{KCCSs}
K^{\nu(\bar{\nu}),c}_{\Sigma} &=&\smallfrac{1}{4}w_{15} 
\big[(\widetilde{Y}_+C^s_{2,q}-y^2C^s_{L,q}+ Y_- C^s_{3,q})
\GS^{qq} 
+(\widetilde{Y}_+C_{2,g}-y^2C_{L,g})
\GS^{gq}\big],\\
\label{KCCgs}
K^{\nu(\bar{\nu}),c}_{g} &=& \smallfrac{1}{4}w_{15}\big[
(\widetilde{Y}_+C^s_{2,q}-y^2C^s_{L,q}+ Y_- C^s_{3,q})
\GS^{qg} 
+(\widetilde{Y}_+C_{2,g}-y^2C_{L,g})
\GS^{gg}\big],\\
\label{KCCVs}
K^{\nu(\bar{\nu}),c}_{V} &=& \smallfrac{1}{4}w_{15}
(\widetilde{Y}_+C^s_{2,q}-y^2C^s_{L,q}+ Y_- C^s_{3,q})
\GNS^{v}.
\eea

%----------------------------------------------------

%\bibliography{nnstrange}

\end{document}